\documentclass[onecolumn,11pt]{article}
\usepackage[utf8]{inputenc}

\newcommand{\monthyear}{\ifcase \month \or January\or February\or March\or %
April\or May \or June\or July\or August\or September\or October\or November\or %
December\fi, \number \year}

\newcommand{\dotp}{\ensuremath{\boldsymbol{\cdot}}}

\def\-{\scalebox{0.4}[1.0]{\(-\)}}

\newcommand{\mathcalOM}[1]{\ensuremath{\mathcal{O}(M^{#1})}}

\newcommand{\ord}[2]{\ensuremath{#1^{(#2)}}}

\newcommand{\facesum}{\sum^l_{\mathcal{V}(i)}}

\newcommand{\uuline}[1]{\underline{\underline{#1}}}

\newcommand{\sk}{\ensuremath{S_k}}

\newcommand{\skt}{\ensuremath{S_{k/2}}}
\newcommand{\ckt}{\ensuremath{C_{k/2}}}

\usepackage{verbatim}

\usepackage{setspace}

\usepackage{amsmath}
\usepackage{amssymb}

\usepackage{mathtools}

\usepackage[hidelinks]{hyperref}

\usepackage[doi=false,url=false,isbn=false,giveninits=true]{biblatex}
\AtEveryBibitem{%
    \clearfield{note}%
    \clearlist{language}%
    \clearfield{month}%
    \clearfield{publisher}%
    \clearfield{pages}%
}
\renewbibmacro{in:}{}
\usepackage[svgnames]{xcolor}

\usepackage[a4paper,margin=1.5cm]{geometry}

\usepackage{cancel}

\usepackage{stmaryrd}

\usepackage{graphicx}
\usepackage{graphics}
\usepackage{epstopdf}

\usepackage{subcaption}

\usepackage{svg}

\usepackage{multirow}
\usepackage{tabularx}

\usepackage{appendix}

\usepackage[perpage,symbol*]{footmisc}
\DefineFNsymbols*{lamportnostar}[math]{\dagger\ddagger\S\P\|{\dagger\dagger}{\ddagger\ddagger}}
\setfnsymbol{lamportnostar}

\usepackage{lineno}

\begin{document}

\title{Artificial diffusion for convective and acoustic low Mach number flows I: Analysis of the modified equations, and application to Roe-type schemes}

\author{Joshua Hope-Collins\footnote{Oxford Thermofluids Institute, University of Oxford, Osney Mead, Oxford, OX2 0ES, England, United Kingdom} \footnote{Department of Mathematics, Imperial College London, London, SW7 2AZ, England, United Kingdom} \footnote{Corresponding author. \textit{Email address}: joshua.hope-collins@eng.ox.ac.uk}\\
Luca di Mare$^{\dagger}$}
\date{\monthyear}

\maketitle
\abstract{

\noindent Three asymptotic limits exist for the Euler equations at low Mach number - purely convective, purely acoustic, and mixed convective-acoustic.
Standard collocated density-based numerical schemes for compressible flow are known to fail at low Mach number due to the incorrect asymptotic scaling of the artificial diffusion.
Previous studies of this class of schemes have shown a variety of behaviours across the different limits and proposed guidelines for the design of low-Mach schemes.
However, these studies have primarily focused on specific discretisations and/or only the convective limit.

In this paper, we review the low-Mach behaviour using the modified equations - the continuous Euler equations augmented with artificial diffusion terms - which are representative of a wide range of schemes in this class.
By considering both convective and acoustic effects, we show that three diffusion scalings naturally arise.
Single- and multiple-scale asymptotic analysis of these scalings shows that many of the important low-Mach features of this class of schemes can be reproduced in a straightforward manner in the continuous setting.

As an example, we show that many existing low-Mach Roe-type finite-volume schemes match one of these three scalings.
Our analysis corroborates previous analysis of these schemes, and we are able to refine previous guidelines on the design of low-Mach schemes by including both convective and acoustic effects.
Discrete analysis and numerical examples demonstrate the behaviour of minimal Roe-type schemes with each of the three scalings for convective, acoustic, and mixed flows.\\


\noindent Highlights:
\begin{itemize}
    \item Asymptotic analysis of numerical schemes at low Mach number
    \item Considers the convective, acoustic, and mixed convective-acoustic low Mach limits
    \item Derivation of required asymptotic scaling of artificial diffusion at each limit
    \item Use of modified equations applies to finite-volume or finite-difference schemes
    \item Application to Roe schemes shows excellent agreement with previous literature
\end{itemize}
}

\newpage
\tableofcontents
\newpage

\graphicspath{{./figures/}}

\section*{Introduction}\label{sec:intro}
Low Mach number flows with non-zero divergence - distinct from properly incompressible divergence-free flows - are of interest in a wide range of applications.
Flows with significant heat transfer or external body forces experience compressibility and buoyancy effects, for example environmental flows \cite{chassignet_buoyancy-driven_2012} and cooling systems for electronic devices and turbomachinery \cite{sohel_murshed_critical_2017,coletti_turbulent_2014}.
Low Mach number acoustic phenomena are also of interest \cite{inoue_sound_2002}, for example noise prediction for wind turbines or aeroplanes during take-off and landing.
Some flows experience both of these effects, such as low Mach number combustion, where heat-release induced compressibility and acoustic interactions both play important roles in governing the flow evolution \cite{silva_assessment_2013,daviller_generalized_2019}.
Some flows contain regions of both low and high Mach number in the same domain, for example aerofoils at high angle of attack where the background flow has low Mach number but regions of higher Mach numbers occur in the vicinity of the aerofoil \cite{potsdam_unsteady_2007}.
To properly simulate any of these cases requires a numerical scheme which is accurate at low Mach number.

There are three common approaches for designing numerical methods for low Mach number flow.
The first uses asymptotic analysis and other assumptions to simplify the full governing equations (either Euler or Navier-Stokes) into a reduced set of equations which can then be solved using specialised solution strategies \cite{muller_low_1999}, such as the Boussinesq buoyancy equations \cite{rehm_equations_1978,majda_derivation_1985} or multiple pressure variable (MPV) methods \cite{klein_semi-implicit_1995,roller_calculation_2005}.
This approach leads to schemes which are highly specialised and often very efficient, however they are restricted to the regimes where the simplifying assumptions hold.
The second approach extends pressure based methods for incompressible flows - such as SIMPLE or fractional step - to the low Mach number regime.
These methods have proved particularly useful for low Mach number aeroacoustics \cite{wall_semi-implicit_2002,yu_fast_2021}.
These methods can leverage the large existing research and code bases for incompressible methods, however they may only handle mixed Mach number flow with moderate, not high Mach number.
The third approach is the one considered in this paper: extending density based methods for compressible flows into the low Mach number regime.
These methods face efficiency and accuracy problems in this regime, but Turkel 1993 \cite{turkel_review_1993} states three reasons justifying the effort placed into overcoming these issues which are still valid today.
Firstly, these methods can handle mixed Mach number flows, a major advantage over the other approaches.
Secondly, these methods handle very naturally flows with large thermal effects which cause large density variations and strong coupling of the energy equation.
Lastly, there is a large body of research and software built upon these methods which can be taken advantage of, including effective acceleration techniques and handling of complex geometries \cite{blazek_computational_2015}.

Collocated density based compressible numerical methods face two substantial problems in the low Mach number regime: efficiency and accuracy.
The efficiency problem is easily understood by considering the wavespeeds of the Euler equations: as $M=u/a\to0$, the condition number $(u+a)/u\to1/M$, and the convective wavespeed $u$ evolves very slowly relative to the acoustic wavespeed $a$.
If the step size of an iterative scheme (for time-accurate evolution or for steady-state convergence) is chosen according to the acoustic wavespeed,
then a large number of steps are required to resolve the convective features.
The common solutions to this problem are semi-implicit methods which remove the acoustic CFL limit \cite{wall_semi-implicit_2002}, or low Mach preconditioners which improve the condition number and restore efficiency \cite{turkel_preconditioning_1994,turkel_preconditioning_1999}.
The design and analysis of low Mach number preconditioning is a substantial research topic in itself, but is not the focus of the current work - although the impact of the artificial diffusion on efficiency will be discussed.
See Turkel 1999 \cite{turkel_preconditioning_1999} for a review of low Mach number preconditioning.

The accuracy problem is less easily demonstrated, and is the main focus of this paper.
Using asymptotic analysis, three limits for the Euler equations can be found at low Mach number.
The limit most often of interest is the convective single-scale limit, where only convective features exist, and solutions approach the incompressible limit as $M\to0$ \cite{klainerman_compressible_1982}.
At the acoustic single-scale limit, only acoustic features exist, and the solutions approach those of the wave equations for linear acoustics.
Finally, a single-space-scale multiple-time-scale (or multiple-space-scale single-time-scale) analysis produces a mixed convective-acoustic limit with convective features on a slow timescale (short space-scale) co-existing with acoustic features on a fast timescale (long space-scale) \cite{muller_low_1999}.
The accuracy problem describes the inability of some schemes to produce solutions which match those of the desired limit (usually the convective limit).

Volpe 1993 \cite{volpe_performance_1993} and Godfrey et al. 1993 \cite{godfrey_preconditioning_1993} showed that a na{\"i}ve application of transonic numerical schemes to flows at the convective low Mach number limit can produce inaccurate, physically inconsistent results.
Godfrey et al. showed that calculating the artificial diffusion based on the preconditioned system instead of the original Euler system produces vastly improved results.
Since then, many papers have been published presenting different methods for modifying the artificial diffusion to achieve accurate low Mach number results.
These use a variety of strategies, building on central schemes \cite{turkel_review_1993,venkateswaran_evaluation_1998,venkateswaran_efficiency_2000,venkateswaran_artficial_2003}, flux-difference \cite{godfrey_preconditioning_1993,weiss_preconditioning_1995,guillard_behaviour_1999,guillard_behavior_2004,guillard_chapter_2017,mary_large_2002,thornber_improved_2008,dellacherie_analysis_2010,rieper_low-mach_2011,oswald_l2roe_2016,potsdam_unsteady_2007,bruel_low_2019,li_all-speed_2008,li_development_2009,li_momentum_2010,li_mechanism_2013,fillion_flica-ovap_2011}, and flux-vector splitting \cite{edwards_low-diffusion_1998,liou_numerical_1999,liou_sequel_2006,shima_parameter-free_2011,sachdev_improved_2012,lin_density_2018} methods.
The majority of studies consider only the convective limit, although a number also address flows with acoustic features \cite{venkateswaran_evaluation_1998,venkateswaran_efficiency_2000,venkateswaran_artficial_2003,potsdam_unsteady_2007,bruel_low_2019,shima_parameter-free_2011,sachdev_improved_2012,merkle_use_1998}.
There have been a number of excellent review and analysis articles published on artificial diffusion at low Mach number, covering the discrete equations of Roe-type schemes at the different limits \cite{guillard_behaviour_1999,guillard_behavior_2004,guillard_chapter_2017,li_mechanism_2013,bruel_low_2019}, the modified equations \cite{dellacherie_analysis_2010}, and the relationship to preconditioning \cite{merkle_use_1998,venkateswaran_evaluation_1998,venkateswaran_efficiency_2000,venkateswaran_artficial_2003}.
We cover the literature on low Mach number artificial diffusion schemes and their analysis in more detail in section \ref{sec:litrev}.\\

In this paper we review the role of the artificial diffusion in the class of collocated, density based schemes at low Mach number.
We have endeavoured to find a formulation of the problem which is as simple as possible whilst still demonstrating the most important findings across this research area.
We hope that this approach will help highlight the root causes of these findings, and enable it to act as an introduction for those with little experience in this field.

To achieve this, we consider all three low Mach number limits - the purely convective, purely acoustic, and mixed convective-acoustic - which are introduced in section \ref{sec:theory}.
After this, we can review the previous literature in section \ref{sec:litrev}.
Previous studies usually begin by showing that a classical transonic scheme is inaccurate for the convective limit, then proceed to `fix' this scheme.
Instead, in section \ref{sec:design} we use the continuous modified equations in the entropy variables to design artificial diffusion schemes which naturally match each of the three limits independently of the specific discretisation.
In section \ref{sec:continuous} we apply single- and multiple-scale asymptotic expansions to the modified equations of each scheme.
In section \ref{sec:discrete} we transform the artificial diffusion to the conservative variables to find the equivalent finite-volume flux functions.
We compare to previous works, and proceed to repeat the asymptotic expansions on the discrete equations.
Finally, in section \ref{sec:examples} we show a series of numerical examples which verify the results of the previous sections.
In a sequel paper\footnote{\emph{Artificial diffusion for convective and acoustic low Mach number flows II: Application to Liou-Steffen, Zha-Bilgen and Toro-Vasquez flux splitting schemes, J. Hope-Collins \& L. di Mare}, in progress.} we extend this to three families of convection-pressure flux splittings: AUSM \cite{liou_new_1993}, Zha-Bilgen \cite{zha_numerical_1993} and Toro-Vasquez \cite{toro_flux_2012}.

\section{Asymptotic expansion of the Euler equations at low Mach number}\label{sec:theory}
We review the theoretical results for the asymptotic solutions of the Euler equations with an ideal gas at low Mach number using single- and multiple-timescale analysis.
We present only the most relevant results, and work in the entropy variables $\underline{w}=(p,u,v,s)^T$.
For a more extensive analysis in the conserved variables, see M\"uller 1999 \cite{muller_low_1999}, and for a multiple space scale analysis see \cite{klein_semi-implicit_1995}.
First, all variables are non-dimensionalised:
\begin{equation}\label{eq:non-dimensionalisation}
\begin{gathered}
    \rho = \frac{\tilde{\rho}}{\rho_{\infty}},
    \quad
    \underline{u} = \frac{\underline{\tilde{u}}}{u_{\infty}},
    \quad
    \underline{x} = \frac{\underline{\tilde{x}}}{L_{\infty}},
    \quad
    E = \frac{\tilde{E}}{p_{\infty}/\rho_{\infty}},
    \quad
    H = \frac{\tilde{H}}{p_{\infty}/\rho_{\infty}},
    \quad
    p = \frac{\tilde{p}}{p_{\infty}},
    \quad
    t = \frac{\tilde{t}}{L_{\infty}/u_{\infty}}
\end{gathered}
\end{equation}
where tildes indicate local dimensional quantities, $\infty$ indicates reference dimensional quantities, and underlining indicates vector-valued quantities.
$E$ and $H$ are the specific total energy and enthalpy respectively.
The Euler equations in entropy variables become:
\begin{subequations} \label{eq:euler}
\begin{align}
    \label{eq:euler_pressure}
    \partial_{t} p + \underline{u}\dotp\nabla p + \gamma p\nabla\dotp\underline{u} = 0 \\
    \label{eq:euler_velocity}
    \rho\partial_{t} \underline{u} + M^{\-2}\nabla p + \rho\underline{u}\dotp\nabla\underline{u} = 0 \\
    \label{eq:euler_entropy}
    \partial_{t} s + \underline{u}\dotp\nabla s = 0
\end{align}
\end{subequations}
Where $M=\frac{\sqrt{\gamma}u_{\infty}}{a_{\infty}}$ is the reference Mach number.
The relation $\gamma p=\rho a^2$ has been used in the 3rd term in (\ref{eq:euler_pressure}), which is only valid for a perfect gas or barotropic gases which are polytropic (including isothermal or isentropic gases).
It is not valid for a general barotropic gas, although the same results can still be obtained in this case.
The entropy equation (\ref{eq:euler_entropy}) is not required for barotropic gases.

\subsection{Single timescale convective limit}
Using $M$ as a small parameter, we treat the system (\ref{eq:euler}) as a perturbation problem and expand all variables as power series of $M$:
\begin{equation} \label{eq:power_expansion}
    \psi(\underline{x},t,M) =     \ord{\psi}{0} (\underline{x},t)
                             + M  \ord{\psi}{1} (\underline{x},t)
                             + M^2\ord{\psi}{2} (\underline{x},t)
                             + \mathcalOM{3}
\end{equation}
The expansions (\ref{eq:power_expansion}) are inserted into (\ref{eq:euler}) and terms are grouped by powers of the parameter $M$.
The lowest order terms from the velocity equations are $\mathcalOM{\-2}$, $\mathcalOM{\-1}$ and $\mathcalOM{0}$ due to the $M^{\-2}$ coefficient on the pressure gradient:
\begin{subequations} \label{eq:velocity_convective_timescale}
\begin{align}
    \label{eq:velocity_convective_-2}
    \nabla\,\ord{p}{0} = 0 \\
    \label{eq:velocity_convective_-1}
    \nabla\,\ord{p}{1} = 0 \\
    \label{eq:velocity_convective_0}
    \ord{\rho}{0}\partial_{t} \ord{\underline{u}}{0} + \nabla \ord{p}{2} + \ord{\underline{\rho u}}{0}\dotp\nabla\ord{\underline{u}}{0} = 0
\end{align}
\end{subequations}
Relations (\ref{eq:velocity_convective_-2}) and (\ref{eq:velocity_convective_-1}) mean that the zeroth and first order pressure terms vary only in time.
If $\ord{p}{1}$ is zero at all points on the boundary at all times then the pressure expansion (\ref{eq:power_expansion}) can be replaced with:
\[p(\underline{x},t,M) = \ord{p}{0}(t) + M^2\ord{p}{2}(\underline{x},t) + \mathcalOM{3}\]
The spatial variations of the non-dimensional convective pressure are $\mathcalOM{2}$, so the dimensional pressure variations are $\mathcal{O}(\rho u^2)$, i.e. independent of the Mach number.
For a barotropic gas, this implies that the density also has $\mathcalOM{2}$ spatial variations.
The last relation (\ref{eq:velocity_convective_0}) is the evolution equation for the convective velocity.
Expanding the pressure equation (\ref{eq:euler_pressure}) and making use of relations (\ref{eq:velocity_convective_-2}) and (\ref{eq:velocity_convective_-1}), we find the following $\mathcalOM{0}$ and $\mathcalOM{1}$ relations:
\begin{subequations} \label{eq:pressure_convective_timescale}
\begin{align}
    \label{eq:pressure_convective_0}
    d_{t} \ord{p}{0} + \gamma\ord{p}{0}\nabla\dotp\ord{\underline{u}}{0} = 0 \\
    \label{eq:pressure_convective_1}
    d_{t} \ord{p}{1} + \gamma\ord{(p\nabla\dotp\underline{u})}{1} = 0
\end{align}
\end{subequations}
The relations (\ref{eq:pressure_convective_0}) and (\ref{eq:pressure_convective_1}) imply that the divergence of the zeroth and first order velocities are spatially uniform and react everywhere instantaneously to temporal variations in the background pressure.
Lastly, we expand the entropy equation.
Every order $n$ of the expansion is identical to the original equation, showing that the entropy is simply convected.
\begin{equation} \label{eq:entropy_convective_timescale}
    \partial_{t} \ord{s}{n} + \ord{(\underline{u}\dotp\nabla s)}{n} = 0 \\
\end{equation}
Note that there are no acoustic effects in this single-scale expansion, which should not be surprising given our choice of the convective timescale $L_{\infty}/u_{\infty}$ to non-dimensionalise the time derivatives in equation (\ref{eq:euler}).

\subsection{Two timescale convective-acoustic limit}
A two-timescale, one space scale asymptotic analysis can be used to include acoustic effects.
Defining an additional non-dimensional time $\tau$ using the acoustic speed:
\begin{equation}
    \tau = \frac{\overline{t}}{L_{\infty}/a_{\infty}} = \frac{t}{M}
\end{equation}
leads to the power series expansion:
\begin{equation} \label{eq:power_expansion_multiple}
    \psi(\underline{x},t,M) =     \ord{\psi}{0} (\underline{x},t,\tau)
                             + M  \ord{\psi}{1} (\underline{x},t,\tau)
                             + M^2\ord{\psi}{2} (\underline{x},t,\tau) + \mathcalOM{3}
\end{equation}
The time derivatives at constant $\underline{x}$ and $M$ are now:
\begin{equation} \label{eq:multiple-scale_time_derivative}
    \partial_{t} \psi\Big|_{\underline{x},M} = \big( \partial_{t} + \frac{1}{M}\partial_{\tau} \big) \psi
\end{equation}
As for the convective limit, we start with the three lowest order relations from the velocity equation:
\begin{subequations} \label{eq:velocity_multiple_timescale}
\begin{align}
    \label{eq:velocity_multiple_-2}
    \nabla \ord{p}{0} = 0 \\
    \label{eq:velocity_multiple_-1}
    \ord{\rho}{0}\partial_{\tau} \ord{\underline{u}}{0} + \nabla \ord{p}{1} = 0 \\
    \label{eq:velocity_multiple_0}
    \partial_{\tau} \ord{\underline{\rho u}}{1} + \ord{\rho}{0}\partial_{t}\ord{\underline{u}}{0} + \nabla \ord{p}{2} + \ord{\underline{\rho u}}{0}\dotp\nabla\ord{\underline{u}}{0} = 0
\end{align}
\end{subequations}
The zeroth order pressure is again constant in space, however the same is no longer true for the first order pressure term, which now varies with zeroth order velocity fluctuations on the acoustic timescale.
The second order pressure is still the relevant pressure for the zeroth order velocity fluctuations on the convective timescale (\ref{eq:velocity_multiple_0}).
Next, the pressure equation is expanded using (\ref{eq:power_expansion_multiple}) and making use of the relation (\ref{eq:velocity_multiple_-2}):
\begin{subequations} \label{eq:pressure_multiple_timescale}
\begin{align}
    \label{eq:pressure_multiple_-1}
    \partial_{\tau} \ord{p}{0} = 0 \\
    \label{eq:pressure_multiple_0}
    \partial_{\tau} \ord{p}{1} + d_{t} \ord{p}{0} + \gamma\ord{p}{0}\nabla\dotp\ord{\underline{u}}{0} = 0
\end{align}
\end{subequations}
Relations (\ref{eq:velocity_multiple_-2}) and (\ref{eq:pressure_multiple_-1}) imply that the background pressure varies only on the convective timescale.
Relation (\ref{eq:pressure_multiple_0}) shows that the leading order velocity divergence now has spatial variations on the acoustic timescale related to first order pressure variations.
In fact, relations (\ref{eq:velocity_multiple_-1}) and (\ref{eq:pressure_multiple_0}) are the equations for linear acoustics at low Mach number, with a source term $d_{t}\ord{p}{0}$ (see \cite{muller_low_1999} for more details).
The relation (\ref{eq:velocity_multiple_0}) is again the equation for the convective velocity variations.
The correct expansion for the pressure at the two-timescale limit is (\ref{eq:power_expansion_multiple}), except the zeroth order term which becomes $\ord{p}{0}(\underline{x},t,\tau)=\ord{p}{0}(t)$.
Note that a purely acoustic limit of (\ref{eq:euler}) can be found with a single-timescale expansion with the acoustic timescale $\tau$, which results in the relations (\ref{eq:velocity_multiple_-2},\ref{eq:velocity_multiple_-1},\ref{eq:pressure_multiple_-1}), and (\ref{eq:pressure_multiple_0}) without the $d_{t}\ord{p}{0}$ term.
We now expand the entropy equation using the two timescale expansion.
This time we find:
\begin{subequations} \label{eq:entropy_multiple_timescale}
\begin{align}
    \label{eq:entropy_multiple_-1}
    \partial_{\tau} \ord{s}{0} & = 0 \\
    \label{eq:entropy_multiple_n}
    \partial_{\tau} \ord{s}{n+1} + \partial_{t} \ord{s}{n} + \ord{(\underline{u}\dotp\nabla s)}{n} & = 0, \quad n\geq0
\end{align}
\end{subequations}
From which we see that the zeroth order entropy is constant on the acoustic timescale, which is consistent with a leading order approximation of isentropic sound waves.

To summarise the important points of this section, we have briefly covered two asymptotic limits of the inviscid homogeneous Euler equations at low Mach number.
Firstly, a convective limit with spatially uniform velocity divergence and $\mathcalOM{2}$ spatial pressure variations associated with convective features on the slow timescale $t$.
Secondly, a mixed convective-acoustic limit which has spatially varying divergence and $\mathcalOM{}$ spatial pressure variations associated with acoustic features on the fast timescale $\tau$ in addition to the convective features on the slow timescale.
A single-scale acoustic limit can be constructed using only the fast timescale $\tau$, which contains only the acoustic variations on the fast timescale.

\section{Previous work}\label{sec:litrev}

As highlighted in the introduction, Godfrey et al. showed in 1993 that calculating the artificial diffusion for a flux-difference-splitting scheme based on the preconditioned system produced accurate results for steady convective flows, whereas calculating the diffusion from the unmodified system produced physically inconsistent results \cite{godfrey_preconditioning_1993}.
This approach was adopted by a number of other researchers for both flux-difference-splitting methods \cite{weiss_preconditioning_1995,guillard_behaviour_1999} and central difference methods \cite{turkel_preconditioning_1999,choi_application_1993,venkateswaran_evaluation_1998}.
Turkel et al 1994 and Turkel 1999 \cite{turkel_preconditioning_1994,turkel_preconditioning_1999} used the modified equations and the known low Mach number scalings for the convective limit to show that the artificial diffusion of a standard transonic scheme is poorly balanced for this limit, having velocity diffusion which is too high and pressure diffusion which is too low.
They also showed that the preconditioned artificial diffusion rectifies both of these issues, having pressure and velocity diffusion terms which are larger and smaller by a factor of $M$ respectively compared to the unpreconditioned scheme.
We note that artificial diffusion is not the only approach to achieving stability at low Mach number. For example the kinetic energy consistent scheme of Subbareddy \& Candler 2009 \cite{subbareddy_fully_2009} is accurate and stable at low Mach number without artificial diffusion. However, artificial diffusion is the predominant method for stabilising collocated schemes in many applications.

Guillard \& Viozat 1999 \cite{guillard_behaviour_1999} used asymptotic expansions of the discrete equations to show that the unmodified Roe scheme \cite{roe_approximate_1981} produces discrete solutions which do not match the convective asymptotic solutions of the exact Euler Equations, in that they contain spurious spatial oscillations of the first order pressure $\ord{p}{1}$.
In a later paper \cite{guillard_behavior_2004} they showed that this is because the diffusion of the unmodified scheme is derived from the acoustic component of the Euler equations, so produces a residual in the acoustic pressure $\ord{p}{1}$ even for initial conditions containing purely convective variations.
Asymptotic expansion of the discrete equations of the preconditioned Roe scheme shows that this scheme produces discrete solutions which do match the exact asymptotic convective limit - the smaller velocity diffusion preventing the creation of spurious $\ord{p}{1}$ modes, and the larger pressure diffusion preventing chequerboard modes on $\ord{p}{1}$.

The preconditioned diffusion proved to be popular and successful, especially because it can be used in combination with preconditioned timestepping, resulting in a scheme which is stable under a convective CFL condition, eliminating the efficiency problem.
However, Birken \& Meister 2005 \cite{birken_stability_2005} showed that for time accurate timestepping the preconditioned diffusion has a timestep limit of $\Delta t \sim \mathcalOM{2}$ for stability of an explicit scheme, compared to $\Delta t \sim \mathcalOM{}$ for the unmodified scheme, so requires implicit timestepping to be practical for time-accurate simulations.\\

Following the finding of Birken \& Meister, a number of papers proposed a different approach to creating a diffusion scheme which is accurate at low Mach number.
Thornber et al. 2008 \cite{thornber_entropy_2008} showed that the inaccuracies of standard Godunov schemes at low Mach number are linked to spurious entropy generation by the velocity diffusion, and subsequently proposed a modification to Roe's flux which reduced the velocity diffusion by a factor of $M$, resulting in accurate solutions at low Mach number \cite{thornber_numerical_2008}.
Other methods of achieving this reduction in the velocity diffusion have been proposed, including those by Thornber et al. \cite{thornber_improved_2008}, Dellacherie \cite{dellacherie_analysis_2010}, Rieper (LMRoe) \cite{rieper_low-mach_2011}, and Miczek et al. \cite{miczek_new_2015}.
All of these methods retain the $\Delta t \sim \mathcalOM{}$ stability limit of the original scheme, but with improved low-Mach accuracy.

Dellacherie 2010 and Dellacherie et al 2016 \cite{dellacherie_analysis_2010,dellacherie_construction_2016} used the modified equations for the linear acoustic waves with artificial diffusion to rigorously show that reducing the velocity diffusion of a standard Godunov scheme by \textit{at least} a factor of $M$ will prevent the production of spurious acoustic $\ord{p}{1}$ modes and ensure accuracy for the convective low Mach number limit, provided the initial conditions are well-prepared\footnote{Well-prepared initial conditions are solutions which are \textit{close to} the solution space of the convective limit i.e. contain vanishingly small acoustic components. See Schochet \cite{schochet_fast_1994} and Dellacherie et al \cite{dellacherie_analysis_2010,dellacherie_construction_2016} for a formal definition.}.
This shows that the increased pressure diffusion in the preconditioned artificial diffusion is in fact not necessary for low-Mach accuracy.
However, it has been shown \cite{dellacherie_analysis_2010,guillard_chapter_2017} that the increased pressure diffusion introduces a Brezzi-Pitk{\"a}ranta type stabilisation \cite{brezzi_stabilization_1984}, which prevents pressure chequerboard instabilities on collocated schemes, and has previously been applied to finite-element \cite{hughes_new_1986} and finite-volume \cite{eymard_stabilized_2006} schemes for incompressible flow.\\

A third approach, the all-speed Roe scheme, introduced by Li \& Gu 2008 \cite{li_all-speed_2008,li_development_2009} reduces not only the velocity diffusion by a factor of $M$, but also the pressure diffusion compared to the standard Roe scheme, so that the diffusion is essentially a diagonal upwinding on the convective velocity scale.
The reduction in velocity diffusion means that this scheme is accurate for convective low Mach number flow, but is susceptible to severe chequerboard instabilities.
Li \& Gu resolve this by reintroducing a pressure diffusion term into the physical mass flux using a momentum interpolation method similar to that proposed by Rhie \& Chow \cite{rhie_numerical_1983}, and used by Mary and Sagaut \cite{mary_large_2002} and in AUSM schemes \cite{liou_new_1993,liou_sequel_1996,liou_sequel_2006}.
The scheme was later extended to unsteady flows by adding a timestep dependence to the pressure diffusion \cite{li_momentum_2010}.\\

The discrete forms of these three approaches are reviewed and compared by Li \& Gu 2013 \cite{li_mechanism_2013} and Guillard \& Nkonga 2017 \cite{guillard_chapter_2017}.
Li \& Gu compare the scaling of the coefficients of the pressure and velocity diffusion terms, and propose a set of guidelines on these scalings for accuracy and stability.
Guillard \& Nkonga review the origin of the accuracy problem with respect to the multiple possible low Mach number limits, and also highlight the dependence on grid type of the various low-Mach schemes.
For a first order Roe scheme on a simplex mesh (triangles in 2D or tetrahedrons in 3D), the velocity degrees of freedom are reduced such that the jumps in the normal velocity over cell interfaces vanish.
This eliminates the velocity diffusion, meaning that the unmodified Roe scheme actually provides accurate results for this cell geometry, despite its inaccuracy on other meshes.
See \cite{rieper_influence_2008,rieper_influence_2009,guillard_behavior_2009,dellacherie_influence_2010} and section 3.3 of \cite{guillard_chapter_2017} for details of this particular behaviour.\\

The literature covered so far has focused on the convective limit.
A number of papers have also covered schemes for flows with acoustic effects.
In a series of papers, Venkateswaran and Merkle showed that preconditioned dual-time schemes are ill conditioned at the acoustic limit and display poor convergence \cite{venkateswaran_dual_1995}, and that the preconditioned diffusion is inaccurate for flow with acoustics, having excessive pressure dissipation, whereas the unmodified scheme is both efficient and accurate for purely acoustic flow \cite{venkateswaran_evaluation_1998,merkle_use_1998}.
They developed an `enhanced' diffusion scheme having the same pressure diffusion as the unmodified scheme but reduced velocity diffusion, as in the preconditioned scheme \cite{venkateswaran_efficiency_2000,venkateswaran_artficial_2003}.
This diffusion scheme was shown to produce accurate results with reasonable convergence for both convective and acoustic flows.
The later low-Mach fix of Thornber/Dellacherie/Rieper \cite{thornber_numerical_2008,dellacherie_analysis_2010,rieper_low-mach_2011} is in a sense equivalent to the earlier `enhanced' scheme, although appears to have been developed independently.
Thornber and Rieper present numerical evidence that their schemes are accurate for 1D acoustics, although their analyses do not cover acoustic effects.

The enhanced/Low-Mach fix scheme is generally suitable for both convective and acoustic flows.
However, Potsdam et al. 2007 \cite{potsdam_unsteady_2007} and Sachdev et al. 2012 \cite{sachdev_improved_2012} showed numerically that under certain conditions, it is susceptible to slight instabilities on the convective pressure field and acoustic velocity field (the instability of the acoustic velocity grid-mode was also demonstrated and analysed in \cite{dellacherie_checkerboard_2009}).
By introducing timestep dependence into the diffusion formulation, they developed adaptive schemes which vary between different diffusion schemes depending on whether acoustic waves are resolved.
This adaptive methodology was shown to maintain accuracy over a range of Strouhal numbers, and is applicable to 
Roe-type \cite{potsdam_unsteady_2007,caraeni_unsteady_2017} and AUSM type \cite{sachdev_improved_2012} schemes.

Whereas the adaptive methodology aims to return to the preconditioned or unmodified scheme in situations where the enhanced/LMRoe scheme may produce oscillatory solutions, Bruel et al. 2019 \cite{bruel_low_2019} tried to modify the LMRoe scheme to eliminate the acoustic velocity instability and the related degradation in the CFL condition ($\sigma<\frac{1}{2}$) of this scheme.
Using discrete multiple-scale asymptotic analysis, they confirmed the suitability of the Roe, preconditioned Roe and LMRoe schemes for purely acoustic, purely convective, and either acoustic or convective flows respectively.
By reintroducing off-diagonal diffusion terms into LMRoe the acoustic instabilities and CFL degradation were eliminated, although the final scheme produces symmetry-breaking solutions for some convective flows which the authors attribute to the scheme not maintaining Galilean invariance.

\section{Design of artificial diffusion at low Mach number}\label{sec:design}
In this section, we find the appropriate asymptotic scaling for the artificial diffusion terms of a numerical scheme, such that the scheme provides accurate results at low Mach number.
We consider each of the three low Mach number regimes in turn.
The diffusion scaling for purely convective flow is well-known, and its form was derived in Turkel et al 1994 and Turkel 1999 \cite{turkel_preconditioning_1994,turkel_preconditioning_1999}.
We repeat this derivation here for completeness, before reapplying Turkel's method to derive the diffusion scaling for purely acoustic flow.
By examining the two schemes, we can then design a diffusion scaling suitable for mixed convective-acoustic flow.

In equation (\ref{eq:euler_modified}) we augment the continuous $x$-split 2-dimensional Euler equations with artificial diffusion terms with a form that mirrors that of the flux Jacobian, which maximises the degrees of freedom without increasing the coupling between equations.
The accuracy problem for the convective limit occurs only for multi-dimensional flow \cite{dellacherie_analysis_2010}.
However, the vast majority of schemes in the class under consideration use some form of dimension-splitting (e.g. finite-difference derivatives along grid lines, or finite-volume interface fluxes over cell faces), and the dimension-split form is simpler than the full multi-dimensional form whilst still displaying the most important low-Mach behaviours, which justifies its use over a truly multi-dimensional analysis here.
The dimension-split results can be readily extended to the $y$-split part of the 2D equations, to the 3D equations, and to the non-dimension-split equations.
For analysis of the multi-dimensional modified equations see \cite{dellacherie_analysis_2010}, and for a recent example of a truly multi-dimensional low Mach number scheme see \cite{barsukow_truly_2021}.
\begin{subequations} \label{eq:euler_modified}
    \begin{align}
        \label{eq:pressure_modified}
        \partial_t p +           u\partial_x p + \gamma p\partial_x u & = A_{11}\partial_{xx}p + A_{12}\partial_{xx}u \\
        \label{eq:velocity_modified}
        \rho\partial_t u + M^{\-2}\partial_x p +   \rho u\partial_x u & = A_{21}\partial_{xx}p + A_{22}\partial_{xx}u \\
        \label{eq:vorticity_modified}
        \partial_t v                           +        u\partial_x v & = A_{33}\partial_{xx}v \\
        \label{eq:entropy_modified}
        \partial_t s                           +        u\partial_x s & = A_{44}\partial_{xx}s
    \end{align}
\end{subequations}

We have written the artificial diffusion as linear second order terms for simplicity, but any consistent diffusive term is acceptable\footnote{A consistent diffusion term is proportional to $\Delta x^n$ for some positive n, where $\Delta x$ is the grid spacing, so vanishes as $\Delta x\to0$.}.
The results of the analysis still hold if, for example, we assume that $A\partial_{xx}\psi \sim \mathcalOM{n}$ implies both $\partial_{x}(A\partial_{x}\psi) \sim \mathcalOM{n}$ and $A\partial^4_{x}\psi \sim \mathcalOM{n}$ for non-linear or fourth order diffusion terms respectively.
This underlines the fact that we are interested in the limit of vanishing Mach number at a fixed, finite mesh spacing.
If we also took the limit of vanishing mesh spacing, then the diffusion terms on the right hand side of the system (\ref{eq:euler_modified}) would disappear, as is necessary for a consistent numerical scheme.

The aim of this section is now to derive the appropriate asymptotic scaling for the coefficients $A_{ij}$, such that the solutions to the modified system (\ref{eq:euler_modified}) have the same scaling behaviour as solutions of the original Euler equations at each low Mach number regime.
Turkel's method for deriving these scalings for purely convective flow (which we will also use for purely acoustic flow) is as follows:
\begin{enumerate}
    \item Find the limiting form of the left-hand side ($\mathcal{L}$) of equations (\ref{eq:euler_modified}) by retaining only the largest term(s) as $M\to0$.
    \item Force the artificial diffusion terms to be retained in the limit by choosing the order of the coefficients $A_{ij}$ such that all terms on the right-hand side ($\mathcal{R}$) are the same order as the largest term(s) in $\mathcal{L}$, i.e. $\mathcal{R}\sim\mathcal{O}(\mathcal{L})$
    \footnote{Strictly, this should be $\mathcal{R}\sim\Theta(\mathcal{L})$, because if $\mathcal{R}\sim o(\mathcal{L})$ then $\mathcal{R}\sim\mathcal{O}(\mathcal{L})$ still holds, but the diffusion $\mathcal{R}$ would vanish asymptotically. We will continue to use big-O instead of big-theta notation to remain consistent with the rest of the literature, with the understanding that a stricter interpretation is required for (at least the diagonal) diffusion coefficients.
    }.
\end{enumerate}
Because the mesh spacing is fixed, we can assume that $\partial_{x}\psi\sim\mathcalOM{n}\implies\partial_{xx}\psi\sim\mathcalOM{n}$ and the scalings found in section \ref{sec:theory} can be used for both $\mathcal{L}$ and $\mathcal{R}$ in step 1.
These scalings are summarised in table \ref{table:lowmach_scaling}.
We can immediately find the artificial diffusion scaling for the linear vorticity and entropy fields (\ref{eq:vorticity_modified},\ref{eq:entropy_modified}).
For these equations, $\mathcal{L}\sim\mathcalOM{0}$ in both the single- and multiple-timescale limits, so for $\mathcal{R}\sim\mathcal{O}(\mathcal{L})$ as $M\to0$, we require:
\begin{equation}\label{eq:linear_diff_scaling}
    A_{33} \sim \mathcalOM{0},
    \quad
    A_{44} \sim \mathcalOM{0}
\end{equation}
This is consistent with a standard upwind scheme with a diffusion coefficient on the order of the advecting velocity $u$, which is appropriate given that (\ref{eq:vorticity_modified},\ref{eq:entropy_modified}) have the form of the linear scalar advection equation.

The pressure and velocity equations (\ref{eq:pressure_modified},\ref{eq:velocity_modified}) form a coupled subsystem which is the distinguishing feature between the different low Mach number regimes and, by extension, between the different schemes.
The rest of this section is concerned with finding the correct scaling for the artificial diffusion coefficients for this subsystem only, assuming the scaling (\ref{eq:linear_diff_scaling}) for the linear fields.

\begin{table}
\begin{center}
\begin{tabular}{|c|c|c|} \hline
                    & Convective      & Acoustic         \\ \hline
     $p$            & \multicolumn{2}{c|}{$\ord{p}{0}   \sim\mathcalOM{0}$} \\ \hline
     $u$            & \multicolumn{2}{c|}{$\ord{u}{0}   \sim\mathcalOM{0}$} \\ \hline
     $\rho$         & \multicolumn{2}{c|}{$\ord{\rho}{0}\sim\mathcalOM{0}$} \\ \hline
     $\partial_x p$ & $\partial_x\ord{p}{2}\sim\mathcalOM{2}$ & $\partial_x\ord{p}{1}\sim\mathcalOM{1}$  \\ \hline
     $\partial_x u$ & \multicolumn{2}{c|}{$\partial_x\ord{u}{0}\sim\mathcalOM{0}$} \\ \hline
     $\partial_x v$ & \multicolumn{2}{c|}{$\partial_x\ord{v}{0}\sim\mathcalOM{0}$} \\ \hline
     $\partial_x s$ & \multicolumn{2}{c|}{$\partial_x\ord{s}{0}\sim\mathcalOM{0}$} \\ \hline
     $\partial_t p$ & $\partial_t\ord{p}{0}\sim\mathcalOM{0}$ & $\partial_{\tau}\ord{p}{1}\sim\mathcalOM{0}$  \\ \hline
     $\partial_t u$ & $\partial_t\ord{u}{0}\sim\mathcalOM{0}$ & $\partial_{\tau}\ord{u}{0}\sim\mathcalOM{-1}$ \\ \hline
     $\partial_t v$ & \multicolumn{2}{c|}{$\partial_t\ord{v}{0}\sim\mathcalOM{0}$} \\ \hline
     $\partial_t s$ & \multicolumn{2}{c|}{$\partial_t\ord{s}{0}\sim\mathcalOM{0}$} \\ \hline
\end{tabular}
\caption{Low Mach number scaling of the various terms in equations (\ref{eq:pressure_modified}, \ref{eq:velocity_modified}) for purely convective or acoustic variation.
Note that the time derivatives scaling for the acoustic variations of $p$ and $u$ make use of equation (\ref{eq:multiple-scale_time_derivative}).}
\label{table:lowmach_scaling}
\end{center}
\end{table}

\subsection{Artificial diffusion for purely convective flows}
First, we find the required scaling of the coefficients $A^c_{ij}$, $i,j=1,2$ for purely convective flow.
Using the scalings in table \ref{table:lowmach_scaling}, the largest terms in $\mathcal{L}$ for (\ref{eq:pressure_modified}, \ref{eq:velocity_modified}) are $\mathcalOM{0}$.
The diagonal coefficients $A_{11}$ and $A_{22}$ should be $\mathcalOM{\-2}$ and $\mathcalOM{0}$ respectively for the diagonal artificial diffusion terms to remain in the limit equations.
By the same argument, the off-diagonal coefficients $A_{12}$ and $A_{21}$ should be $\mathcalOM{0}$ and $\mathcalOM{\-2}$ respectively\footnote{
    It is acceptable for the off-diagonal terms to be $o(\mathcal{L})$ and vanish in the limit, however the diagonal terms should be $\Theta(\mathcal{L})$ for stability.}.
Collecting these results, the convective scaling of the diffusion matrix is:
\begin{equation} \label{eq:convective_diffusion_scaling}
    \uuline{A}^c \sim \mathcal{O}
    \begin{matrix}
    \begin{pmatrix}
    M^{\-2} & M^0 \\
    M^{\-2} & M^0 \\
    \end{pmatrix}
    \end{matrix}
\end{equation}
Retaining only the leading order $\mathcalOM{0}$ terms from the modified equations (\ref{eq:pressure_modified},\ref{eq:velocity_modified}), the limit equations for convective flow variations and convective diffusion scaling are\footnote{
Turkel \cite{turkel_preconditioning_1999} uses $p\sim\mathcalOM{2}$ for both the background pressure and the pressure gradient, so the term $\partial_t p$ in (\ref{eq:pressure_cvar_cdiff}) disappears from his limit equation (11a in \cite{turkel_preconditioning_1999}). For steady flows these are equivalent, but the time-derivative must be retained for flows where the background pressure varies in time, for example due to heat addition or non-zero net mass flow over the boundaries \cite{muller_low_1999}.}:
\begin{subequations} \label{eq:cvar_cdiff_limit}
    \begin{align}
        \label{eq:pressure_cvar_cdiff}
        \partial_t \ord{p}{0} + \gamma\ord{p}{0}\partial_x \ord{u}{0} & = \ord{A}{\-2}_{11}\partial_{xx}\ord{p}{2} + \ord{A}{0}_{12}\partial_{xx}\ord{u}{0} \\
        \label{eq:velocity_cvar_cdiff}
        \ord{\rho}{0}\partial_t \ord{u}{0} + \partial_x \ord{p}{2} + \ord{\rho u}{0}\partial_x \ord{u}{0} & = \ord{A}{\-2}_{21}\partial_{xx}\ord{p}{2} + \ord{A}{0}_{22}\partial_{xx}\ord{u}{0}
    \end{align}
\end{subequations}
where $\ord{A}{n}_{ij}\implies A_{ij}\sim\Theta(M^{n})$.
The limit equations (\ref{eq:cvar_cdiff_limit}) are identical to the $\mathcalOM{0}$ pressure and velocity relations (\ref{eq:pressure_convective_0},\ref{eq:velocity_convective_0}) from the single-scale asymptotic expansion, with the addition of artificial diffusion terms.
They also closely resemble the governing equations of Chorin's Artificial Compressibility Method \cite{chorin_numerical_1967}, which was influential in much of the early work on low Mach number preconditioning \cite{turkel_review_1993,turkel_preconditioning_1994,weiss_preconditioning_1995}.
As is appropriate for close-to-incompressible flow, the magnitude of both limit equations becomes independent of the Mach number, and at steady-state the velocity divergence will approach zero (up to the value of the diffusion terms) according to the pressure equation (\ref{eq:pressure_cvar_cdiff}).

\subsection{Artificial diffusion for purely acoustic flows}
Next we consider flow with purely acoustic variation, repeating the process above to obtain the correct scaling of the artificial diffusion $\uuline{A}^a$ in this regime.
Using the acoustic flow variations from table \ref{table:lowmach_scaling}, the largest terms in $\mathcal{L}$ for (\ref{eq:pressure_modified},\ref{eq:velocity_modified}) are $\mathcalOM{0}$ and $\mathcalOM{\-1}$ respectively.
For all artificial diffusion terms to be retained in the limit, we require:
\begin{equation} \label{eq:acoustic_diffusion_scaling}
    \uuline{A}^a \sim \mathcal{O}
    \begin{matrix}
    \begin{pmatrix}
    M^{\-1} & M^0 \\
    M^{\-2} & M^{\-1} \\
    \end{pmatrix}
    \end{matrix}
\end{equation}
The limit equations using this diffusion scaling and acoustic flow variations are:
\begin{subequations} \label{eq:avar_adiff_limit}
    \begin{align}
        \label{eq:pressure_avar_adiff}
        \partial_{\tau}\ord{p}{1} + \gamma\ord{p}{0}\partial_x \ord{u}{0} & = \ord{A}{\-1}_{11}\partial_{xx}\ord{p}{1} + \ord{A}{0}_{12}\partial_{xx}\ord{u}{0} \\
        \label{eq:velocity_avar_adiff}
        \ord{\rho}{0}\partial_{\tau}\ord{u}{0} + \phantom{\ord{p}{0}}\partial_x \ord{p}{1} & = \ord{A}{\-2}_{21}\partial_{xx}\ord{p}{1} + \ord{A}{\-1}_{22}\partial_{xx}\ord{u}{0}
    \end{align}
\end{subequations}
which are the equations governing linear acoustics at low Mach number, as we found in section \ref{sec:theory} with the relations (\ref{eq:pressure_multiple_0}) and (\ref{eq:velocity_multiple_-1}), plus the artificial diffusion terms.
Most schemes designed for transonic flow approach a diagonal approximation to this scheme, which resembles upwinding of the two acoustic waves (see appendix \ref{app:transonic_scaling}).
Note that if the off-diagonal terms vanish ($A_{12}\sim o(M^0)$ and $A_{21}\sim o(M^{\-2})$), then the limit equations (\ref{eq:avar_adiff_limit}) are equivalent to Dellacherie's first order modified equations of a standard Godunov scheme for the linear acoustic equations (equations (61) and (62) in \cite{dellacherie_analysis_2010} with $\nu_r=\nu_{u_k}=a_*\Delta x/2M$), which are shown to destroy convective low Mach number accuracy.
In practice, this scaling is not used for low Mach number simulations because it will create spurious acoustic waves from any non-trivial convective variations (as we will see later), and if the flow has zero (or enforced) convective variations then specialist acoustics solvers are much more suitable.

\subsection{Artificial diffusion for mixed convective-acoustic flows}
Acoustic variations are often the result of, or coexist with, convective phenomena, making it desirable to have a scheme with acceptable limit equations for both convective and acoustic variations.
By examining the limit equations (\ref{eq:cvar_cdiff_limit}) and (\ref{eq:avar_adiff_limit}), we can identify two requirements for an acceptable form of the limit equations.
The first is that all terms in $\mathcal{L}$ from the relevant physical relations are retained, i.e. the momentum and divergence relations (\ref{eq:pressure_convective_0},\ref{eq:velocity_convective_0}) for convective flow, and the linear acoustics relations (\ref{eq:pressure_multiple_0},\ref{eq:velocity_multiple_-1}) for acoustic flow, which is necessary for the solutions to the modified equations to have the same scaling as solutions to the original equations.
The second property is that some diffusion terms are retained in $\mathcal{R}$, which will be necessary once we attempt to construct stable discrete schemes.

Next we will see if the limit equations for convective diffusion with acoustic variations, and for acoustic diffusion with convective variations satisfy these requirements.
The limit equations for convective diffusion scaling $\uuline{A}^c$, equation (\ref{eq:convective_diffusion_scaling}), and acoustic flow variations are:
\begin{subequations} \label{eq:avar_cdiff_limit}
    \begin{align}
        \label{eq:pressure_avar_cdiff}
        0                                                & = \ord{A}{\-2}_{11}\partial_{xx}\ord{p}{1} \\
        \label{eq:velocity_avar_cdiff}
        \ord{\rho}{0}\partial_{\tau}\ord{u}{0} + \partial_x\ord{p}{1} & = \ord{A}{\-2}_{21}\partial_{xx}\ord{p}{1}
    \end{align}
\end{subequations}
The diagonal diffusion term in the pressure equation (\ref{eq:pressure_avar_cdiff}) is too large $\ord{A}{\-2}_{11}\partial_{xx}\ord{p}{1}\sim\mathcalOM{\-1}$, so $\mathcal{R}\sim\omega(\mathcal{L})$ resulting in a parabolic equation for the acoustic pressure $\ord{p}{1}$.
This means that the convective scheme $\uuline{A}^c$ effectively filters acoustic variations from the solution, just as the single-timescale expansion of the original Euler equations does, making this scheme unsuitable for purely acoustic or mixed convective-acoustic flow.
Conversely, the velocity equation (\ref{eq:velocity_avar_cdiff}) retains all the required terms in $\mathcal{L}$ from the acoustic velocity relation (\ref{eq:velocity_multiple_-1}), although the diagonal diffusion term vanishes asymptotically.
The limit equations for the acoustic diffusion scaling $\uuline{A}^a$, equation (\ref{eq:acoustic_diffusion_scaling}), and convective flow variations are:
\begin{subequations} \label{eq:cvar_adiff_limit}
\begin{align}
    \label{eq:pressure_cvar_adiff}
    \partial_t \ord{p}{0} + \gamma\ord{p}{0}\partial_x \ord{u}{0} & = \ord{A}{0}_{12}\partial_{xx}\ord{u}{0} \\
    \label{eq:velocity_cvar_adiff}
                                                          0 & = \ord{A}{\-1}_{22}\partial_{xx}\ord{u}{0}
\end{align}
\end{subequations}
The diagonal diffusion term in the velocity equation (\ref{eq:velocity_cvar_adiff}) is too large $\ord{A}{\-1}_{22}\partial_{xx}\ord{u}{0}\sim\mathcalOM{\-1}$, so $\mathcal{R}\sim\omega(\mathcal{L})$ resulting in a parabolic equation for the convective velocity $\ord{u}{0}$.
This will smooth out any convective variations of the velocity, making this scheme unsuitable for purely convective or mixed convective-acoustic flow.
On the other hand, the pressure equation (\ref{eq:pressure_cvar_adiff}) retains all the required terms in $\mathcal{L}$ from the convective pressure relation (\ref{eq:pressure_convective_0}), but the diagonal diffusion term is asymptotically vanishing.
This over-damping of the velocity field and under-damping of the pressure field is a well-known problem with conventional (transonic) compressible schemes at the single-scale low Mach number limit \cite{guillard_behaviour_1999,guillard_behavior_2004,guillard_chapter_2017}.

We have shown that, unsurprisingly, neither the convective scheme $\uuline{A}^c$ nor the acoustic scheme $\uuline{A}^a$ completely fulfil the requirements for a mixed convective-acoustic scheme.
However, the pressure limit equation (\ref{eq:pressure_cvar_adiff}) with $\uuline{A}^a$ and the velocity limit equation (\ref{eq:velocity_avar_cdiff}) with $\uuline{A}^c$ both retain the relevant terms in $\mathcal{L}$, even if some diffusion terms vanish.
This suggests we can form a scheme for mixed flow by combining the acoustic pressure diffusion $A^a_{11}$, $A^a_{12}$ and the convective velocity diffusion $A^c_{21}$, $A^c_{22}$:
\begin{equation} \label{eq:mixed_diffusion_scaling}
    \uuline{A}^m \sim \mathcal{O}
    \begin{matrix}
    \begin{pmatrix}
    M^{\-1} & M^{0} \\
    M^{\-2} & M^{0} \\
    \end{pmatrix}
    \end{matrix}
\end{equation}
which we call the mixed scheme.
We shall see during the discrete analysis in section \ref{sec:discrete} that most modern schemes use this scaling, although to the authors' knowledge the only studies which explicitly identify the use of acoustic scaling on the pressure equation and convective scaling on the velocity equation are those of Venkateswaran and co-workers \cite{venkateswaran_efficiency_2000,venkateswaran_artficial_2003,potsdam_unsteady_2007,sachdev_improved_2012}.
The limit equations for the mixed diffusion scaling with convective flow variations are:
\begin{subequations} \label{eq:cvar_mdiff_limit}
\begin{align}
    \label{eq:pressure_cvar_mdiff}
    \partial_t \ord{p}{0}                         + \gamma\ord{p}{0}\partial_x \ord{u}{0} & = \phantom{\ord{A}{\-2}_{21}\partial_{xx}\ord{p}{2} + } \ord{A}{0}_{12}\partial_{xx}\ord{u}{0} \\
    \label{eq:velocity_cvar_mdiff}
    \ord{\rho}{0}\partial_t \ord{u}{0} + \partial_x \ord{p}{2} + \ord{\rho u}{0}\partial_x \ord{u}{0} & =          \ord{A}{\-2}_{21}\partial_{xx}\ord{p}{2} +    \ord{A}{0}_{22}\partial_{xx}\ord{u}{0}
\end{align}
\end{subequations}
and with acoustic flow variations are:
\begin{subequations} \label{eq:avar_mdiff_limit}
\begin{align}
    \label{eq:pressure_avar_mdiff}
    \partial_{\tau} \ord{p}{1} + \gamma\ord{p}{0}\partial_x \ord{u}{0} & = \ord{A}{\-1}_{11}\partial_{xx}\ord{p}{1} + \ord{A}{0}_{12}\partial_{xx}\ord{u}{0} \\
    \label{eq:velocity_avar_mdiff}
    \ord{\rho}{0}\partial_{\tau} \ord{u}{0} + \phantom{\ord{p}{0}}\partial_x \ord{p}{1} & = \ord{A}{\-2}_{21}\partial_{xx}\ord{p}{1}
\end{align}
\end{subequations}
from which we see that for both convective and acoustic flow variations, all terms in $\mathcal{L}$ are retained compared to the desired form of the limit equations (\ref{eq:cvar_cdiff_limit},\ref{eq:avar_adiff_limit}).
The limit equations for the convective velocity and the acoustic pressure (\ref{eq:velocity_cvar_mdiff},\ref{eq:pressure_avar_mdiff}) retain both artificial diffusion terms, however the convective pressure and acoustic velocity limit equations (\ref{eq:pressure_cvar_mdiff}, \ref{eq:velocity_avar_mdiff}) retain only the off-diagonal diffusion terms.
This may degrade the stability of the corresponding discrete scheme, especially if the off-diagonal diffusion terms vanish.
The issues of degraded stability on the convective pressure and acoustic velocity were numerically demonstrated in \cite{sachdev_improved_2012}, and the acoustic velocity instability was analysed in \cite{dellacherie_checkerboard_2009,bruel_low_2019}; here we show that the cause of both can be identified in the continuous setting.
Note that for vanishing off-diagonal terms, the limit equations (\ref{eq:avar_mdiff_limit}) are equivalent to Dellacherie's first order modified equations for a low Mach Godunov scheme applied to the linear acoustic equations (equations (61) and (62), or (75), in \cite{dellacherie_analysis_2010} with $\nu_r=a_*\Delta x/2M$ and $\nu_{u_k}=0$), which are shown to be accurate for convective low Mach number flow.

\subsection{Adaptive schemes}
We have identified three scalings of the diffusion matrix - convective, acoustic, and mixed - each suited to different flow regimes.
However, it would be useful to have a single numerical method that could select the most appropriate scaling, either so the same method could be used for multiple problems, or because different flow regimes exist in a single problem \cite{potsdam_unsteady_2007}.
The major difference between the schemes is whether they allow acoustic features, so a sensible metric for which scheme to use would be whether acoustic waves can be resolved.
This can be estimated by the ratio of the acoustic timescale $\tau$ and the simulation timestep $\Delta t$:
\begin{equation} \label{eq:unsteady_mach}
    \frac{\tau}{\Delta t} = \frac{L_{\infty}/a}{\Delta t} = \frac{L_{\infty}/\Delta t}{a} = M_u
\end{equation}
When $\Delta t > \tau$ acoustic waves cannot be temporally resolved and $M_u$ is small.
When $\Delta t < \tau$ acoustic waves can be resolved and $M_u$ is large.
The parameter $\tau/\Delta t$ is normally presented as the unsteady Mach number $M_u$.
This parameter was first introduced by Venkateswaran \& Merkle 1995 \cite{venkateswaran_dual_1995} to improve the conditioning of a low Mach preconditioned dual-time scheme, and has since been used to control the diffusion scaling \cite{venkateswaran_artficial_2003,potsdam_unsteady_2007,sachdev_improved_2012,caraeni_unsteady_2017}.
In practice, $M_u$ is bounded between $M$ and $1$ so $M_u\to1$ as $\Delta t\to0$ and $M_u\to M$ as $\Delta t\to\infty$.
The coefficients $A_{11}$ and $A_{22}$ can be made to vary between the convective scaling when $M_u \approx M$ and the acoustic scaling when $M_u \approx 1$ using:
\begin{equation}\label{eq:adaptive_diff_scaling}
    A_{11}\sim\mathcal{O}(M^{\-1}M_u^{\-1}),
    \quad
    A_{22}\sim\mathcal{O}(M^{\-1}M_u)
\end{equation}

The effect of adaptive diffusion on accuracy and efficiency is discussed more in \cite{potsdam_unsteady_2007} and \cite{sachdev_improved_2012} but will not be covered in any more detail here.

In this section we have shown how artificial diffusion schemes can be derived for purely convective or purely acoustic low Mach number flow using Turkel's method \cite{turkel_preconditioning_1999} of balancing the artificial diffusion terms $\mathcal{R}$ with the physical terms $\mathcal{L}$ in the limit as $M\to0$.
These schemes were then combined to create a mixed scheme suitable for both convective and acoustic flow, although this scheme has asymptotically vanishing artificial diffusion on the convective pressure and acoustic velocity equations.

We emphasise that these three diffusion scalings are not novel to the present study - all are in use today for low Mach number simulations, with the exception of the purely acoustic scheme, for the reasons discussed above.
However, we have shown that many of the low-Mach behaviours of this class of schemes can be demonstrated independently of any specific discretisation.

In deriving and analysing these schemes so far, we have enforced the variations to obey either the convective or acoustic scalings from table \ref{table:lowmach_scaling}.
While this is useful for designing the diffusion schemes, we do not yet know how the schemes will perform if the variations are not enforced.

\section{Analysis of the continuous Euler equations with artificial diffusion}\label{sec:continuous}
In this section we will further analyse the three artificial diffusion schemes described in the previous section to better understand their behaviour.
The bulk of this section will be six asymptotic expansions: a single-timescale and a multiple-timescale expansion of the modified equations (\ref{eq:euler_modified}) for each of the three diffusion schemes $\uuline{A}^c$, $\uuline{A}^a$ and $\uuline{A}^m$.
Because all variables will be expanded according to either equation (\ref{eq:power_expansion}) or (\ref{eq:power_expansion_multiple}), the scaling of the variations will not be enforced, overcoming the main limitation of the previous section.
At the end of this section we estimate the asymptotic scaling for the spectral radius of each scheme, which will affect their stability.

Before carrying out the asymptotic expansions, we consider how to judge whether a relation obtained from the expansion of the modified equations (\ref{eq:euler_modified}) ``matches'' - in some sense - the corresponding relation from the expansion of the physical equations (\ref{eq:euler}).
All terms in $\mathcal{L}$ will match exactly by construction.
However, we identify three types of relation with differing requirements on the terms in $\mathcal{R}$:
\begin{enumerate}
    \item Pressure variation relations.
        Relations (\ref{eq:velocity_convective_-2},\ref{eq:velocity_convective_-1}) must be enforced exactly, else lower order pressure variations will swamp the higher order terms of interest, leading to catastrophic loss of accuracy.
    \item Transport equation relations.
        The convective momentum relation (\ref{eq:velocity_convective_0}), and the acoustic relations (\ref{eq:velocity_multiple_-1},\ref{eq:pressure_multiple_0}) should retain some artificial diffusion terms for the scheme to remain stable.
    \item Divergence constraint relations.
        It can be argued that relations (\ref{eq:pressure_convective_0},\ref{eq:pressure_convective_1}) should be matched exactly, so that the $\mathcalOM{0}$ velocity field becomes divergence-free \cite{li_mechanism_2013}.
        However, the \textit{inf-sup} condition means that some discrete schemes are susceptible to pressure-velocity decoupling (chequer-board modes) in the incompressible limit.
        Certain collocated schemes \cite{hughes_new_1986,eymard_stabilized_2006} can avoid this issue by using Brezzi-Pitk{\"a}ranta stabilisation \cite{brezzi_stabilization_1984}, which introduces a pressure diffusion term into the continuity equation.
        As noted in \cite{guillard_chapter_2017,dellacherie_analysis_2010} this technique can be applied to density-based schemes, so it can also be argued that the velocity divergence should be zero only up to the value of some pressure diffusion term.
        Both choices are valid, and the impact of this choice will be discussed later.
\end{enumerate}

\subsection{Expansion of the Euler equations with convective diffusion}
First, we carry out a single timescale expansion of (\ref{eq:pressure_modified},\ref{eq:velocity_modified}) with convective diffusion scaling (\ref{eq:convective_diffusion_scaling}).
The $\mathcalOM{\-2}$, $\mathcalOM{\-1}$ and $\mathcalOM{0}$ velocity relations are:
\begin{subequations} \label{eq:velocity_1time_cdiff}
\begin{align}
    \label{eq:velocity_1time_cdiff_-2}
    \partial_x \ord{p}{0} & = \ord{A}{\-2}_{21}\partial_{xx}\ord{p}{0} \\
    \label{eq:velocity_1time_cdiff_-1}
    \partial_x \ord{p}{1} & = \ord{A}{\-2}_{21}\partial_{xx}\ord{p}{1} \\
    \label{eq:velocity_1time_cdiff_0}
    \ord{\rho}{0}\partial_{t} \ord{u}{0} + \partial_x \ord{p}{2} + \ord{\rho u}{0}\partial_x \ord{u}{0} & = \ord{A}{\-2}_{21}\partial_{xx}\ord{p}{2} + \ord{A}{0}_{22}\partial_{xx}\ord{u}{0} 
\end{align}
\end{subequations}
The relation (\ref{eq:velocity_1time_cdiff_0}) is exactly the physical momentum relation (\ref{eq:velocity_convective_0}) plus all the artificial diffusion terms, as we required, and as found in the limit equation (\ref{eq:velocity_cvar_cdiff}).
Upon first inspection, the pressure variation relations (\ref{eq:velocity_1time_cdiff_-2},\ref{eq:velocity_1time_cdiff_-1}) do not appear to satisfy the physical relations (\ref{eq:velocity_convective_-2},\ref{eq:velocity_convective_-1}) because of the presence of the artificial diffusion terms.
Expanding the pressure equation gives the $\mathcalOM{\-2}$, $\mathcalOM{\-1}$ and $\mathcalOM{0}$ relations:
\begin{subequations} \label{eq:pressure_1time_cdiff}
\begin{align}
    \label{eq:pressure_1time_cdiff_-2}
    0 & = \ord{A}{\-2}_{11}\partial_{xx}\ord{p}{0} \\
    \label{eq:pressure_1time_cdiff_-1}
    0 & = \ord{A}{\-2}_{11}\partial_{xx}\ord{p}{1} \\
    \label{eq:pressure_1time_cdiff_0}
    d_{t} \ord{p}{0} + \gamma\ord{p}{0}\partial_x \ord{u}{0} & = \ord{A}{\-2}_{11}\partial_{xx}\ord{p}{2} + \ord{A}{0}_{12}\partial_{xx}\ord{u}{0} 
\end{align}
\end{subequations}
Now we see that the spatial variations of $\ord{p}{0,1}$ are constrained by the two systems (\ref{eq:velocity_1time_cdiff_-2},\ref{eq:pressure_1time_cdiff_-2}) and (\ref{eq:velocity_1time_cdiff_-1},\ref{eq:pressure_1time_cdiff_-1}) respectively.
Both systems admit the desired constant solutions, but we cannot yet tell whether they also admit non-constant solutions\footnote{
We do not make use of $A\partial_{xx}p = 0 \implies \partial_{xx}p = 0$ as this is invalid for nonlinear artificial diffusion terms of the form $\partial_x (A \partial_x p)$.}.
We will assume $\ord{p}{0,1}$ have constant initial conditions, and boundary conditions which are either periodic, or constant in both time and space. As for the physical relations (\ref{eq:velocity_convective_timescale}), $\ord{p}{1}$ is assumed to have initial and boundary conditions equal to zero.
These assumptions lead to constant solutions, satisfying the pressure variation relations (\ref{eq:velocity_convective_-2},\ref{eq:velocity_convective_-1}).
See Guillard \& Viozat \cite{guillard_behaviour_1999} for a more in-depth discussion of the conditions under which the discrete equivalents of these relations (which we will see later) lead to constant $\ord{p}{0,1}$ solutions.
Note that it is useful to have restraints on both the first and second derivatives, despite the more involved reasoning about the constancy of $\ord{p}{0,1}$.
Assuming that central gradient approximations ($\partial_x p |_i \approx \frac{p_{i+1}-p_{i-1}}{2\Delta x}$) will be used in the final discrete schemes, the discrete $\partial_x p = 0$ constraint alone could lead to odd-even decoupling and allow non-constant grid modes.
These modes are suppressed by the additional constraint $\partial_{xx} p = 0$.

Assuming constant $\ord{p}{0,1}$ then $\partial_t \ord{p}{0} \to d_t\ord{p}{0}$ and relation (\ref{eq:pressure_1time_cdiff_0}) is exactly the physical divergence relation (\ref{eq:pressure_convective_0}), plus all the artificial diffusion terms.
The retention of the artificial pressure diffusion term allows for Brezzi-Pitk{\"a}ranta stabilisation as discussed above.
Therefore, all relevant relations from the single-timescale expansion are reproduced by the convective diffusion scheme.

Next, we carry out a multiple-timescale expansion of (\ref{eq:pressure_modified},\ref{eq:velocity_modified}) with convective diffusion scaling (\ref{eq:convective_diffusion_scaling}).
The $\mathcalOM{\-2}$, $\mathcalOM{\-1}$ and $\mathcalOM{0}$ velocity relations are:
\begin{subequations} \label{eq:velocity_2time_cdiff}
\begin{align}
    \label{eq:velocity_2time_cdiff_-2}
    \partial_x \ord{p}{0} & = \ord{A}{\-2}_{21}\partial_{xx}\ord{p}{0} \\
    \label{eq:velocity_2time_cdiff_-1}
    \ord{\rho}{0}\partial_{\tau}\ord{u}{0} + \partial_x \ord{p}{1} & = \ord{A}{\-2}_{21}\partial_{xx}\ord{p}{1} \\
    \label{eq:velocity_2time_cdiff_0}
    \partial_{\tau}\ord{\rho u}{1} + \ord{\rho}{0}\partial_{t}\ord{u}{0} + \partial_x\ord{p}{2} + \ord{\rho u}{0}\partial_x\ord{u}{0} & = \ord{A}{\-2}_{21}\partial_{xx}\ord{p}{2} + \ord{A}{0}_{22}\partial_{xx}\ord{u}{0} 
\end{align}
\end{subequations}
The $\mathcalOM{\-2}$, $\mathcalOM{\-1}$ and $\mathcalOM{0}$ pressure relations are:
\begin{subequations} \label{eq:pressure_2time_cdiff}
\begin{align}
    \label{eq:pressure_2time_cdiff_-2}
    0 & = \ord{A}{\-2}_{11}\partial_{xx}\ord{p}{0} \\
    \label{eq:pressure_2time_cdiff_-1}
    \partial_{\tau}\ord{p}{0} & = \ord{A}{\-2}_{11}\partial_{xx}\ord{p}{1} \\
    \label{eq:pressure_2time_cdiff_0}
    \partial_{\tau}\ord{p}{1} + d_{t}\ord{p}{0} + \gamma\ord{p}{0}\partial_x \ord{u}{0} & = \ord{A}{\-2}_{11}\partial_{xx}\ord{p}{2} + \ord{A}{0}_{12}\partial_{xx}\ord{u}{0} 
\end{align}
\end{subequations}
The $\ord{p}{0}$ relations (\ref{eq:velocity_2time_cdiff_-2},\ref{eq:pressure_2time_cdiff_-2}) are identical to the single-timescale relations (\ref{eq:velocity_1time_cdiff_-2},\ref{eq:pressure_1time_cdiff_-2}), so $\ord{p}{0}$ is constant under the same assumptions as before.
If we assume no forcing of $\ord{p}{0}$ on the acoustic timescale - which seems reasonable - then $\partial_{\tau}\ord{p}{0}=0$ and $\partial_t \ord{p}{0} \to d_t\ord{p}{0}$, and relation (\ref{eq:pressure_2time_cdiff_-1}) becomes a parabolic equation for $\ord{p}{1}$ equivalent to relation (\ref{eq:pressure_avar_cdiff}).
This will quickly damp any acoustic variations in the pressure, confirming that the convective diffusion scaling $\uuline{A}^c$ is unsuitable for simulations with acoustic variations.

\subsection{Expansion of the Euler equations with acoustic diffusion}
The single-timescale expansion of the modified equations (\ref{eq:pressure_modified},\ref{eq:velocity_modified}) with $\uuline{A}^a$ leads to the following $\mathcalOM{\-2}$ and $\mathcalOM{\-1}$ velocity relations:
\begin{subequations} \label{eq:velocity_1time_adiff}
\begin{align}
    \label{eq:velocity_1time_adiff_-2}
    \partial_x \ord{p}{0} & = \ord{A}{\-2}_{21}\partial_{xx}\ord{p}{0} \\
    \label{eq:velocity_1time_adiff_-1}
    \partial_x \ord{p}{1} & = \ord{A}{\-2}_{21}\partial_{xx}\ord{p}{1} +  \ord{A}{\-1}_{22}\partial_{xx}\ord{u}{0}
\end{align}
\end{subequations}
and the $\mathcalOM{\-1}$ and $\mathcalOM{0}$ pressure relations:
\begin{subequations} \label{eq:pressure_1time_adiff}
\begin{align}
    \label{eq:pressure_1time_adiff_-1}
    0 & = \ord{A}{\-1}_{11}\partial_{xx}\ord{p}{0} \\
    \label{eq:pressure_1time_adiff_0}
    d_t \ord{p}{0} + \gamma\ord{p}{0}\partial_x \ord{u}{0} & = \ord{A}{\-1}_{11}\partial_{xx}\ord{p}{1} +  \ord{A}{0}_{12}\partial_{xx}\ord{u}{0}
\end{align}
\end{subequations}
The $\ord{p}{0}$ relations (\ref{eq:velocity_1time_adiff_-2},\ref{eq:pressure_1time_adiff_-1}) lead to constant $\ord{p}{0}$ under the assumptions described above.
On the other hand, the relation (\ref{eq:velocity_1time_adiff_-1}) means that $\partial_x \ord{p}{1} \neq 0$ unless the velocity field is trivial and the term $\ord{A}{\-1}_{22}\partial_{xx}\ord{u}{0}=0$, making the acoustic diffusion scheme unsuitable for convective low Mach number flows.
For a diagonal scheme at steady-state, relations (\ref{eq:velocity_1time_adiff_-1},\ref{eq:pressure_1time_adiff_0}) resemble the Stokes flow equations, indicating that the solution will be dominated by the diffusive effects, balanced by variations in $\ord{p}{1}$.
The production of unphysical acoustic modes by $\ord{A}{\-1}_{22}$, even from a well-prepared initial field, has been shown numerous times for specific discrete schemes \cite{guillard_behaviour_1999,thornber_numerical_2008,li_mechanism_2013,rieper_low-mach_2011}, and in the continuous setting in \cite{guillard_behavior_2004,dellacherie_analysis_2010}.

A two timescale expansion of equations (\ref{eq:pressure_modified},\ref{eq:velocity_modified}) with acoustic diffusion gives the $\mathcalOM{\-2}$ and $\mathcalOM{\-1}$ velocity relations:
\begin{subequations} \label{eq:velocity_2time_adiff}
\begin{align}
    \label{eq:velocity_2time_adiff_-2}
    \partial_x \ord{p}{0} & = \ord{A}{\-2}_{21}\partial_{xx}\ord{p}{0} \\
    \label{eq:velocity_2time_adiff_-1}
    \ord{\rho}{0}\partial_{\tau}\ord{u}{0} + \partial_x \ord{p}{1} & = \ord{A}{\-2}_{21}\partial_{xx}\ord{p}{1} + \ord{A}{\-1}_{22}\partial_{xx}\ord{u}{0}
\end{align}
\end{subequations}
and the $\mathcalOM{\-1}$ and $\mathcalOM{0}$ pressure relations:
\begin{subequations} \label{eq:pressure_2time_adiff}
\begin{align}
    \label{eq:pressure_2time_adiff_-1}
    \partial_{\tau} \ord{p}{0} & = \ord{A}{\-1}_{11}\partial_{xx}\ord{p}{0} \\
    \label{eq:pressure_2time_adiff_0}
    \partial_{\tau} \ord{p}{1} + d_{t}\ord{p}{0} + \gamma\ord{p}{0}\partial_x \ord{u}{0} & = \ord{A}{\-1}_{11}\partial_{xx}\ord{p}{1} + \ord{A}{0}_{12} \partial_{xx} \ord{u}{0}
\end{align}
\end{subequations}
The $\ord{p}{0}$ relations (\ref{eq:velocity_2time_adiff_-2},\ref{eq:pressure_2time_adiff_-1}) lead to constant $\ord{p}{0}$ under the assumptions described above.
The relations (\ref{eq:velocity_2time_adiff_-1},\ref{eq:pressure_2time_adiff_0}) are exactly the physical acoustic relations (\ref{eq:velocity_multiple_-1},\ref{eq:pressure_multiple_0}) plus all artificial diffusion terms, so we expect this scheme to be able to simulate at least purely acoustic flow.
However, this scheme is still unsuitable for mixed convective-acoustic flow, as it will be unable to properly resolve the convective component as shown in the single timescale expansion.

\subsection{Expansion of the Euler equations with mixed diffusion}
Finally, we carry out the asymptotic expansions with the mixed diffusion scheme $\uuline{A}^m$ (\ref{eq:mixed_diffusion_scaling}).
The single timescale expansion results in the $\mathcalOM{\-2}$, $\mathcalOM{\-1}$ and $\mathcalOM{0}$ velocity relations:
\begin{subequations} \label{eq:velocity_1time_mdiff}
\begin{align}
    \label{eq:velocity_1time_mdiff_-2}
    \partial_x\ord{p}{0} & = \ord{A}{\-2}_{21}\partial_{xx}\ord{p}{0} \\
    \label{eq:velocity_1time_mdiff_-1}
    \partial_x\ord{p}{1} & = \ord{A}{\-2}_{21}\partial_{xx}\ord{p}{1} \\
    \label{eq:velocity_1time_mdiff_0}
    \ord{\rho}{0}\partial_{t}\ord{u}{0} + \partial_x\ord{p}{2} + \ord{\rho u}{0}\partial_x\ord{u}{0} & = \ord{A}{\-2}_{21}\partial_{xx}\ord{p}{2} + \ord{A}{0}_{22}\partial_{xx}\ord{u}{0} 
\end{align}
\end{subequations}
and the $\mathcalOM{\-1}$ and $\mathcalOM{0}$ pressure relations:
\begin{subequations} \label{eq:pressure_1time_mdiff}
\begin{align}
    \label{eq:pressure_1time_mdiff_-1}
    0 & = \ord{A}{\-1}_{11}\partial_{xx}\ord{p}{0} \\
    \label{eq:pressure_1time_mdiff_0}
    d_t \ord{p}{0} + \gamma\ord{p}{0}\partial_x \ord{u}{0} & = \ord{A}{\-1}_{11}\partial_{xx}\ord{p}{1} +  \ord{A}{0}_{12}\partial_{xx}\ord{u}{0}
\end{align}
\end{subequations}
The $\ord{p}{0}$ relations (\ref{eq:velocity_1time_mdiff_-2},\ref{eq:pressure_1time_mdiff_-1}) lead to constant $\ord{p}{0}$ under the assumptions described above.
On the other hand $\ord{p}{1}$ is constrained by relation (\ref{eq:velocity_1time_mdiff_-1}) and the continuity relation (\ref{eq:pressure_1time_mdiff_0}).
While the constant vector is clearly a solution to (\ref{eq:velocity_1time_mdiff_-1}) under the previous assumptions, it is now less clear whether it is the only solution.
As we shall see later, almost all schemes which use the mixed diffusion scaling have asymptotically diagonal diffusion.
In this case, (\ref{eq:velocity_1time_mdiff_-1}) becomes $\partial_{x}\ord{p}{1}=0$, so the physical relation (\ref{eq:velocity_convective_-1}) is matched exactly, although as discussed earlier, the discrete scheme will be susceptible to odd-even decoupling on $\ord{p}{1}$ without the constraint on $\partial_{xx}\ord{p}{1}$.
At steady-state, the continuity relation (\ref{eq:pressure_1time_mdiff_0}) for a diagonal scheme becomes $\ord{p}{0}\partial_x\ord{u}{0} = \ord{A}{\-1}_{11}\partial_{xx}\ord{p}{1}$.
This means that if $\ord{p}{1}$ is constant then $\ord{u}{0}$ is divergence-free, just as the incompressibility condition requires, but cannot be stabilised against chequer-board modes on $\ord{p}{\geq2}$ in the manner of the Brezzi-Pitk{\"a}ranta.

The last relation, velocity relation (\ref{eq:velocity_1time_mdiff_0}), exactly matches (\ref{eq:velocity_1time_cdiff_0}), which matches the physical relation (\ref{eq:velocity_convective_0}) with both diffusion terms retained.
As every relevant physical relation is matched by the mixed diffusion scheme, we can expect that discrete schemes with matching modified equations will be suitable for simulations of purely convective low Mach number flow, although potentially at the risk of chequer-board modes on $\ord{p}{\geq2}$.

Two timescale expansion of (\ref{eq:pressure_modified},\ref{eq:velocity_modified}) with mixed diffusion scaling (\ref{eq:mixed_diffusion_scaling}) gives the $\mathcalOM{\-2}$, $\mathcalOM{\-1}$ and $\mathcalOM{0}$ velocity relations:
\begin{subequations} \label{eq:velocity_2time_mdiff}
\begin{align}
    \label{eq:velocity_2time_mdiff_-2}
    \partial_x\ord{p}{0} & = \ord{A}{\-2}_{21}\partial_{xx}\ord{p}{0} \\
    \label{eq:velocity_2time_mdiff_-1}
    \ord{\rho}{0}\partial_{\tau}\ord{u}{0} + \partial_x\ord{p}{1} & = \ord{A}{\-2}_{21}\partial_{xx}\ord{p}{1} \\
    \label{eq:velocity_2time_mdiff_0}
    \partial_{\tau}\ord{\rho u}{1} + \ord{\rho}{0}\partial_{t}\ord{u}{0} + \partial_x\ord{p}{2} + \ord{\rho u}{0}\partial_x\ord{u}{0} & = \ord{A}{\-2}_{21}\partial_{xx}\ord{p}{2} + \ord{A}{0}_{22}\partial_{xx}\ord{u}{0} 
\end{align}
\end{subequations}
and the $\mathcalOM{\-1}$ and $\mathcalOM{0}$ pressure relations:
\begin{subequations} \label{eq:pressure_2time_mdiff}
\begin{align}
    \label{eq:pressure_2time_mdiff_-1}
    \partial_{\tau} \ord{p}{0} & = \ord{A}{\-1}_{11}\partial_{xx}\ord{p}{0} \\
    \label{eq:pressure_2time_mdiff_0}
    \partial_{\tau}\ord{p}{1} + d_{t}\ord{p}{0} + \gamma\ord{p}{0}\partial_x \ord{u}{0} & = \ord{A}{\-1}_{11}\partial_{xx}\ord{p}{1} + \ord{A}{0}_{12} \partial_{xx} \ord{u}{0}
\end{align}
\end{subequations}
The $\ord{p}{0}$ relations (\ref{eq:velocity_2time_mdiff_-2},\ref{eq:pressure_2time_mdiff_-1}) again lead to constant $\ord{p}{0}$ under the assumptions described above.
Relations (\ref{eq:velocity_2time_mdiff_-1},\ref{eq:pressure_2time_mdiff_0}) match the physical acoustic relations (\ref{eq:velocity_multiple_-1},\ref{eq:pressure_multiple_0}), plus some artificial diffusion terms.
The presence of $\partial_{xx}\ord{p}{1}$ in relation (\ref{eq:pressure_2time_mdiff_0}) now provides the appropriate diffusion on the acoustic pressure variations.
On the other hand, the relation (\ref{eq:velocity_2time_mdiff_-1}) lacks a diagonal diffusion term, just as in the limit equation (\ref{eq:velocity_avar_mdiff}), so a diagonal scheme will have no diffusion on the acoustic velocity variations.

Relations (\ref{eq:velocity_2time_mdiff_-1},\ref{eq:pressure_2time_mdiff_0}) are almost equivalent to the modified equations of the scheme proposed by Bruel et al. (section 3.1.5.1 in \cite{bruel_low_2019}).
This scheme differs from the current scheme in the precise form of the diffusion in the off-diagonal terms, but importantly the Mach number scaling of these terms matches ours.
These differences will be discussed further in section \ref{sec:discrete}.
The mixed diffusion scheme $\uuline{A}^m$ is suitable for both convective and mixed convective-acoustic low Mach number flows.
however, the scheme is not without disadvantages, having potentially degraded stability on the convective pressure and acoustic velocity variations.

\subsection{Spectral radius estimates}
We now estimate the scaling of the spectral radii $\Lambda(\uuline{A})$ of $\uuline{A}^c$, $\uuline{A}^a$ and  $\uuline{A}^m$ as $M\to0$.
The spectral radius of the combined flux components determines the CFL bound for stability of explicit timestepping schemes and affects the convergence of implicit schemes.
The flux Jacobian of the Euler equations has spectral radius $u+a \sim \mathcalOM{\-1}$, so the artificial diffusion will induce a more stringent stability bound if $\Lambda(\uuline{A})\sim\omega(M^{\-1})$.
Because diffusion matrices are positive (semi-)definite, we can estimate the scaling of the spectral radius with the scaling of the trace\footnote{
    The trace cannot grow faster than the spectral radius for any fixed rank matrix.
    The trace of a matrix can grow more slowly than the spectral radius due to cancellation.
    For example, the trace of the flux Jacobian of the Euler equations in $N$ dimensions is $(N+2)u$, which is $\mathcalOM{0}$, even though the spectral radius $u+a$ is $\mathcalOM{-1}$.
    However, cancellation is impossible for a positive (semi-)definite matrix by definition.
    As the trace of a positive (semi-)definite matrix cannot grow asymptotically faster or slower than the spectral radius, it is an appropriate estimate for the scaling of the spectral radius.}.
By inspection, the traces of the coefficient matrices (\ref{eq:convective_diffusion_scaling}), (\ref{eq:acoustic_diffusion_scaling}) and (\ref{eq:mixed_diffusion_scaling}) are:
\begin{equation} \label{eq:trace_growths}
    \text{tr}(\underline{\underline{A}}^c) \sim \mathcalOM{\-2},
    \quad
    \text{tr}(\underline{\underline{A}}^a) \sim \text{tr}(\underline{\underline{A}}^m) \sim \mathcalOM{\-1}
\end{equation}
$tr(\uuline{A}^a)$ and $tr(\uuline{A}^m)$ are the same order as the physical spectral radius, so should not change the stability requirements - up to a constant factor.
On the other hand, $tr(\uuline{A}^c)$ is larger than the physical spectral radius by an order of $M^{\-1}$, so will require a CFL bound which decreases one order faster than the physical CFL bound in order to remain stable as $M\to0$.
This scaling was first proven by Birken \& Meister 2005 \cite{birken_stability_2005} for the specific case of preconditioned matrix-Rusanov artificial diffusion.
The estimate (\ref{eq:trace_growths}) also agrees with \cite{dellacherie_analysis_2010} (equation (80) with $\kappa_r=M^{\-1}$) that this restriction will hold for any scheme whose pressure diffusion converges to the convective limit as $M\to0$.
For example, in the companion paper we will see that this is true for existing AUSM schemes that converge to this limit, which explains the problems encountered in \cite{matsuyama_performance_2014,kitamura_reduced_2016,chen_improved_2018}.\\

In this section, we have used single and multiple-scale asymptotic analysis to investigate each artificial diffusion scheme at the convective and mixed convective-acoustic limits.
We have demonstrated several known properties of discrete low Mach number schemes.
Including: the spurious forcing of the acoustic pressure $\ord{p}{1}$ by the acoustic scheme $\uuline{A}^a$; the over-damping of the acoustic pressure by the convective scheme $\uuline{A}^c$; the susceptibility to chequer-board modes on the convective pressure $\ord{p}{2}$ and instabilities on the acoustic velocity variations of the mixed scheme $\uuline{A}^m$; and the CFL limit of the convective scheme.
So far we have worked in the entropy variables and the continuous setting to simplify the presentation, but in the next section we transfer the artificial diffusion to the conservative variables.

\section{Analysis of the discrete Euler equations with artificial diffusion}\label{sec:discrete}

\begin{table}
    \centering
    \begin{tabular}{|c|c|}
         \hline
         Symbol & Definition \\\hline
         $\psi_{i}$           & Quantity $\psi$ in cell $i$ \\\hline
         $\psi$               & Edge/face average of quantity $\psi$ (only in edge/face summations) \\\hline
         $\Omega_i$           & Area/volume of cell $i$ \\\hline
         $S_{il}$             & Length/area of edge/face between cells $i$ and $l$ \\\hline
         $\underline{n}_{il}$ & Outgoing normal vector of edge/face between cells $i$ and $l$ \\\hline
         $\mathcal{V}(i)$     & Set of all cells neighbouring cell $i$ \\\hline
         $U$                  & Edge/face normal velocity $\underline{u}\cdot\underline{n}_{il}$ \\\hline
         $\Delta_{il} \psi$   & Interface jump $\psi_l - \psi_i$ \\\hline
         $\Delta_{il} U$      & Interface jump in normal velocity $(\underline{u}_l - \underline{u}_i)\cdot\underline{n}_{il}$ \\\hline
    \end{tabular}
    \caption{Nomenclature used for the discrete asymptotic expansions.}
    \label{tab:discrete_expansion_nomenclature}
\end{table}

The continuous analysis of the previous section is general to most schemes with the modified equations (\ref{eq:euler_modified}).
In this section we demonstrate how the findings of the continuous analysis transfer to the discrete setting in the particular case of a first order cell-centred finite volume Roe-type scheme.
First we identify the form of the interface flux in the conserved variables that matches the diffusion in the entropy variables (\ref{eq:euler_modified}), which we can then compare against previous Roe-type schemes from the literature.
We then carry out the same six asymptotic expansions as in the previous section, but this time on the discrete equations.
We finish the section by comparing the von Neumann symbols for each of the three schemes.

\subsection{A general form for low Mach number finite-volume schemes}

We consider the semi-discrete equations for a first order cell-centred finite-volume scheme in conservative variables $\underline{q}=(\rho,\underline{\rho u},\rho E)^T$, where the time derivatives are left continuous using a method of lines approach.
Using the nomenclature of Guillard \& Viozat \cite{guillard_behaviour_1999} (table \ref{tab:discrete_expansion_nomenclature}), the evolution in time of the solution at each cell $i$ is described by the ODE:
\begin{equation} \label{eq:fv_discrete_equations}
    \Omega_i \frac{d\underline{q}_i}{dt} + \facesum S_{il}\{\underline{f}\}_{il} = \facesum S_{il}\underline{f}_{il}^d
\end{equation}
Where $\{\underline{f}\}_{il}=\frac{1}{2}(\underline{f}(\underline{q}_i)+\underline{f}(\underline{q}_l))$ is the central approximation of the exact physical flux $\underline{f}(\underline{q})$ and $\underline{f}_{il}^d$ is the artificial diffusion flux between cells $i$ and $l$.
This nomenclature is general to any cell-centred scheme, but we will consider only quadrilaterals in 2D and hexahedra in 3D because, as mentioned in section \ref{sec:litrev}, the accuracy problem for the convective limit disappears on simplex meshes \cite{rieper_influence_2008,rieper_influence_2009,guillard_behavior_2009,dellacherie_influence_2010,guillard_chapter_2017}.
In the non-dimensional entropy variables and a 2D face-aligned coordinate frame, we choose the elements $A_{ij}$ of the Jacobian of the diffusive flux to have the form:
\begin{equation} \label{eq:fv-entropyvar-diffusion}
\renewcommand{\arraystretch}{1.25}
    |\uuline{A}| =
    \mu_u\frac{|U|}{2}\uuline{\mathcal{I}} +
    \frac{1}{2}
    \begin{matrix} \begin{pmatrix}
        M^{\-2}\dfrac{\gamma p}{\rho|v|}\mu_{11} & \pm\gamma p\mu_{12} & 0 & 0 \\
       \pm M^{\-2}\mu_{21} & \rho|v|\mu_{22}     & 0 & 0 \\
        0                  & 0                   & 0 & 0 \\
        0                  & 0                   & 0 & 0
    \end{pmatrix} \end{matrix}
\end{equation}
The first term is the convective upwinding.
The second term is the diffusion on the pressure (first row) and normal velocity (second row) which was discussed in the previous sections.
$|v|$ is some $\mathcalOM{0}$ velocity scale which is included to ensure correct dimensionality.
$\mu_{\alpha}$ ($\alpha\in\{u,11,12,21,22\}$) are positive expressions whose form and Mach number scaling are specific to each discrete scheme.
The precise forms of the elements of this Jacobian are somewhat arbitrary, so long as their scaling can be chosen to match one of the three diffusion scalings $\uuline{A}^{c,a,m}$ - achieved here by varying the scaling of the coefficients $\mu_{\alpha}$.
This particular form is chosen because it simplifies the diffusion coefficients in the conserved variables.
The off-diagonal terms may be negative so long as positive (semi-)definiteness of the Jacobian is maintained - this will be discussed in more detail below.
Transforming (\ref{eq:fv-entropyvar-diffusion}) to the dimensional conserved variables, the diffusive flux between cells $i$ and $l$ is:
\begin{equation} \label{eq:WeissSmith_diffusion}
\renewcommand{\arraystretch}{1.25}
    \underline{\tilde{f}}_{il}^d = \frac{1}{2}
    \Bigg[
    \mu_u|\tilde{U}|
    \begin{matrix} \begin{pmatrix}
        \Delta_{il} \tilde{\rho}  \\
        \Delta_{il} \tilde{\rho}\underline{\tilde{u}} \\
        \Delta_{il} \tilde{\rho}\tilde{E}
    \end{pmatrix} \end{matrix}
    +
    \delta U
    \begin{matrix} \begin{pmatrix}
        \tilde{\rho}   \\
        \tilde{\rho}\underline{\tilde{u}} \\
        \tilde{\rho}\tilde{H}
    \end{pmatrix} \end{matrix}
    +
    \delta p
    \begin{matrix} \begin{pmatrix}
        0 \\
        \underline{n} \\
        \tilde{U}
    \end{pmatrix} \end{matrix}
    \Bigg]
\end{equation}
where the interface velocity and pressure perturbations $\delta U$ and $\delta p$ are defined as:
\begin{equation} \label{eq:interface_delta_up}
    \begin{split}
        \delta U & = \dfrac{\mu_{11}}{\tilde{\rho}|\tilde{v}|}\Delta_{il}\tilde{p} +  {\dfrac{\tilde{U}_{il}}{|\tilde{U}_{il}|}}\mu_{12}\Delta_{il}\tilde{U} \\
        \delta p & = {\dfrac{\tilde{U}_{il}}{|\tilde{U}_{il}|}}\mu_{21}\Delta_{il}\tilde{p} + \tilde{\rho}|\tilde{v}|\mu_{22}\Delta_{il}\tilde{U} 
    \end{split}
\end{equation}
This is precisely the Liu \& Vinokur form \cite{liu_upwind_1989}, which was also used by Weiss \& Smith \cite{weiss_preconditioning_1995} and Li \& Gu \cite{li_mechanism_2013} for low Mach number Roe-type fluxes.
The first term is the natural upwinding for the convective system.
From the definitions of $\delta U$ and $\delta p$ we can see that the second and third terms are, respectively, the diffusion on the pressure and velocity equations in the entropy variables.
We expand the coefficients $\mu_{\alpha}$ as:
\begin{equation}\label{eq:mu_coefficient_form}
    \mu_{\alpha} = \epsilon_{\alpha}\nu_{\alpha}M^{n}
\end{equation}
$\epsilon_{\alpha}$ is a constant positive real valued diffusion coefficient, for example $\epsilon=1$ gives the standard first order upwind diffusion, or $1/64 < \epsilon < 1/32$ can be used for fourth derivative diffusion \cite{blazek_computational_2015}.
Usually $\epsilon_{\alpha}$ are all equal for a particular scheme, although they do not have to be.
The most common example of this is $\epsilon_{11}$ for the convective scheme, which should instead be chosen in the recommended range for the Brezzi-Pitk{\"a}ranta stabilisation coefficient.
For the rest of this section we will use $\epsilon_{\alpha}=1$ for simplicity.
$\nu_{\alpha}\sim\mathcalOM{0}$ is some (non-dimensionalised) scheme specific expression and is the distinguishing feature between different schemes with the same diffusion scaling.
The exponent $n$ determines whether the diffusion has the convective, acoustic or mixed scaling, with the required exponents listed in table \ref{table:delta_up_parameter_scaling}.
The $M^{\-2}$ coefficients on $\mu_{11}$ and $\mu_{21}$ in (\ref{eq:fv-entropyvar-diffusion}) mean that the convective scaling exponents are all zero (i.e. $M$ independent).
Clearly, the convective upwinding coefficient should be $\mu_u=\epsilon_u$ for all regimes.\footnote{$\mu_u=\epsilon_u\nu_u$ could be used where $\nu_u$ is some non-linear function used to control the dissipation level, such as in \cite{kitamura_reduced_2016}.}
Although expanding out $\mu_{\alpha}$ in this manner adds some complexity, it allows the separation of: the magnitude of the diffusion ($\epsilon_{\alpha}$); the type of diffusion scaling ($M^{n}$); and the specific discrete form ($\nu_{\alpha}$) of each scheme.
The full expressions for the diffusive fluxes can be found in appendix \ref{app:fv-diffusive-fluxes}.

\subsection{Comparison to previous work}

\subsubsection{Previous guidelines}

We can compare the scalings in table \ref{table:delta_up_parameter_scaling} with guidelines put forward by previous authors.
Dellacherie proved in \cite{dellacherie_analysis_2010} that for Godunov type upwind schemes, the velocity diffusion must be $\mu_{22}\sim\mathcalOM{0}$ for accuracy at the convective limit, which was also shown in \cite{guillard_behaviour_1999,turkel_preconditioning_1999}.
It is clear that our findings for the convective and mixed scalings agree with this scaling.
Dellacherie primarily studies the mixed scaling, but notes that the convective diffusion scaling (specifically the preconditioned Roe-Turkel scheme) also satisfies $\mu_{22}\sim\mathcalOM{0}$, although with a larger pressure diffusion equivalent to the Brezzi-Pitk{\"a}ranta stabilisation.
These two scalings are both accurate for convective flow, but by considering both the convective and acoustic limits we have shown why this choice exists, and that they are distinct schemes with significantly different properties for acoustic and mixed flows.
It is also shown in \cite{dellacherie_analysis_2010} that $\mu_{22}\sim o(M^{0})$ is accurate for convective flow.
We can see from (\ref{eq:fv-entropyvar-diffusion}) that if $\mu_{22}\sim o(M^{0})$ then there will still be $\mathcalOM{0}$ velocity diffusion in this position from the convective upwinding term, albeit with a slightly different form, so the overall diffusion scaling is comparable to the $\mu_{22}\sim\mathcalOM{0}$ scheme.

In their survey of low Mach number Roe-type schemes for the convective limit, Gu \& Li \cite{li_mechanism_2013} offered three guidelines for the design of low Mach number schemes.
The first guideline states that $\mu_{22}\sim\mathcalOM{0}$ or smaller is necessary for accuracy at the convective limit, as previously discussed for \cite{dellacherie_analysis_2010}.
The second guideline states that $\mu_{11}$ should be between $\mathcalOM{}$ and $\mathcalOM{0}$ inclusive, where $\mu_{11}\sim\mathcalOM{}$ may allow some small pressure chequerboards, but $\mu_{11}\sim\mathcalOM{0}$ will suppress all pressure chequerboards.
This guideline matches the analysis for the convective and mixed scalings in the previous section, but again by including acoustic effects in the analysis we can see why this choice exists, and which flows each scaling is suitable for.
Gu \& Li also state that if $\mu_{11}\sim\mathcalOM{0}$ only in the continuity equation, very little improvement in the control of pressure chequerboards is seen compared to $\mu_{11}\sim\mathcalOM{}$.
Equations (\ref{eq:fv-entropyvar-diffusion}) and (\ref{eq:WeissSmith_diffusion},\ref{eq:interface_delta_up}) show that if $\mu_{11}$ in $\delta U$ is different in each equation then the equivalent diffusion in the entropy variables will not exactly match (\ref{eq:fv-entropyvar-diffusion}), which could explain the degraded performance.
The third guideline states that a cut-off Mach number should only be used in denominators, i.e. when it decreases the diffusion.
The issue of cut-off Mach numbers has not been covered here, but broadly speaking this guideline means, firstly, that $\mu_{22}$ will not be increased beyond $\mathcalOM{0}$ by a cut-off Mach number, which would compromise the accuracy for convective flow features, and secondly, $\mu_{11}$ can still be decreased in the convective scaling, which alleviates the stability issue related to the $\mathcalOM{\-2}$ spectral radius.
See \cite{li_mechanism_2013} for details of this implementation.

\begin{table}
\begin{center}
\begin{tabular}{|c|c|c|c|} \hline
     $\alpha$   & Convective      & Acoustic         & Mixed            \\ \hline
     $u$        & $\mathcalOM{0}$ & $\mathcalOM{0}$  & $\mathcalOM{0}$  \\ \hline
     $11$       & $\mathcalOM{0}$ & $\mathcalOM{}$   & $\mathcalOM{}$   \\ \hline
     $12$       & $\mathcalOM{0}$ & $\mathcalOM{0}$  & $\mathcalOM{0}$  \\ \hline
     $21$       & $\mathcalOM{0}$ & $\mathcalOM{0}$  & $\mathcalOM{0}$  \\ \hline
     $22$       & $\mathcalOM{0}$ & $\mathcalOM{-1}$ & $\mathcalOM{0}$  \\ \hline
\end{tabular}
\caption{Required scaling of the coefficients in $\delta U$ and $\delta p$.}
\label{table:delta_up_parameter_scaling}
\end{center}
\end{table}

\subsubsection{Classification of existing Roe-type schemes}
Now that we have an expression for the artificial diffusion in the conserved variables (\ref{eq:WeissSmith_diffusion},\ref{eq:interface_delta_up}), we can classify existing schemes as having either convective, mixed or acoustic diffusion scaling at low Mach number.
Table \ref{tab:lm-fv-schemes} details many low-Mach Roe-type schemes from the literature.
Some are adaptive schemes, and have different scalings for large timesteps using the convective CFL, $\sigma_u = u\Delta t/\Delta x \approx 1$, and for small timesteps using the acoustic CFL, $\sigma_a = a\Delta t/\Delta x \approx 1$.
We have also noted whether each diffusion scheme is asymptotically diagonal ($\mu_{12},\mu_{21}\sim o(M^{0})$) or upper/lower triangular in the entropy variables.
The discrete form of $\mu_{ij}$ for a number of schemes in table \ref{tab:lm-fv-schemes} can be found in \cite{li_mechanism_2013}.
Several trends are apparent in this table.

Almost all of the earlier schemes use the convective scaling \cite{godfrey_preconditioning_1993,turkel_review_1993,weiss_preconditioning_1995,guillard_behaviour_1999,guillard_behavior_2004,mary_large_2002}.
These early schemes use a diffusive Jacobian derived from the preconditioned physical Jacobian (except \cite{mary_large_2002}), which leads naturally to the convective diffusion scaling (see \cite{turkel_preconditioning_1999} and \cite{guillard_behaviour_1999} for more detail).
On the other hand, many of the more recent schemes use the mixed diffusion \cite{venkateswaran_efficiency_2000,venkateswaran_artficial_2003,potsdam_unsteady_2007,thornber_numerical_2008,thornber_improved_2008,dellacherie_analysis_2010,rieper_low-mach_2011,sachdev_improved_2012,oswald_l2roe_2016,bruel_low_2019} even when the target regime is the convective limit.
These schemes are derived primarily by identifying that for the original Roe scheme, which has the acoustic scaling, the accuracy problem is caused by $\ord{A_{22}}{\-1}$.
These schemes selectively reduce $A_{22}$ by a factor of $M$, thus reaching the mixed scheme.
Most of these studies do not discuss the acoustic low Mach capability of the scheme - exceptions being Venkateswaran and coauthors \cite{venkateswaran_efficiency_2000,venkateswaran_artficial_2003,potsdam_unsteady_2007,sachdev_improved_2012} and Bruel et al. \cite{bruel_low_2019}\footnote{Dellacherie \cite{dellacherie_analysis_2010} analyses the acoustic equations in depth, but makes only a minor distinction between the convective and mixed diffusion scalings, instead focusing on the necessity of reducing $\mu_{22}$ compared to the acoustic scaling.}.
All non-adaptive mixed schemes are diagonal except for the scheme of Bruel et al. \cite{bruel_low_2019}.

The main development of adaptive schemes is due to Venkateswaran and co-authors \cite{venkateswaran_dual_1995,venkateswaran_efficiency_2000,venkateswaran_artficial_2003,potsdam_unsteady_2007,sachdev_improved_2012}, to allow accurate simulation of acoustic phenomena with the mixed scheme, but to return to the more favourable convergence and stability properties of the convective scaling when $\Delta t$ is too large to resolve the acoustic waves.
To the authors' knowledge, \cite{venkateswaran_efficiency_2000} is the earliest use of the mixed diffusion scaling in the literature.
The scalar diffusion schemes in \cite{venkateswaran_efficiency_2000,venkateswaran_artficial_2003,potsdam_unsteady_2007} are diagonal for both large and small timesteps.
The matrix diffusion schemes in \cite{sachdev_improved_2012} return to the preconditioned Roe scheme for large timesteps, where all four diffusion coefficients are well balanced, but for small timesteps they approach mixed schemes which are upper/lower triangular in the entropy variables (i.e. either $\mu_{12}$ or $\mu_{21}$ is $o(M^0)$).\footnote{
This can be seen by identifying that $C\phi^T$ and $C'_p$ in equations (12,13) in \cite{sachdev_improved_2012} correspond to the $\delta U$ and $\delta p$ terms respectively in (\ref{eq:WeissSmith_diffusion},\ref{eq:interface_delta_up}).}
The first scheme (equation (14) in \cite{sachdev_improved_2012}) has adaptive scaling on the $\delta U$ diffusion terms $\mu_{11}$ and $\mu_{12}$, meaning they both return to the values in the original Roe scheme for small timestep - $\mu_{11}$ becomes $\mathcalOM{}$ and $\mu_{12}$ becomes $o(M^0)$, so $\uuline{A}$ is lower triangular.
Conversely, the second scheme (equation (15) in \cite{sachdev_improved_2012}) has adaptive scaling on the pressure diffusion terms $\mu_{11}$ and $\mu_{21}$, so becomes upper triangular.
The authors state that the matrix diffusion scheme of Potsdam et al. \cite{potsdam_unsteady_2007} is almost equivalent to the second scheme in \cite{sachdev_improved_2012}.
Despite these differences, Sachdev et al. report no visible differences between results with these three schemes.

Interestingly, the unsteady momentum interpolation method of Li \& Gu \cite{li_momentum_2010} also appears to be an adaptive scheme.
The earlier All-speed scheme \cite{li_all-speed_2008,li_development_2009} approaches the convective scheme of Mary \& Sagaut \cite{mary_large_2002} at low Mach number.
However, in \cite{li_momentum_2010} the pressure diffusion $\mu_{11}$ is scaled by the timestep so will be a factor of $M$ smaller with $\sigma_a\approx1$ than with $\sigma_u\approx1$, hence approaching the mixed scaling for small timesteps.
This can be seen from equations (19,20) and figures 3-4 of \cite{li_momentum_2010} where slight chequerboards are seen for small timesteps but are controlled for larger timesteps.

The final point to note from table \ref{tab:lm-fv-schemes} is that a few schemes reduce at least some components of the convective upwinding term by a factor of $M$ \cite{thornber_improved_2008,dellacherie_analysis_2010,oswald_l2roe_2016}.
We cannot assess the effects of this change using our analysis so far, but we will return to consider it at the end of the section when we examine the von Neumann symbols of the first order scheme (\ref{eq:fv_discrete_equations}).

\begin{table}
    \centering
    \begin{tabularx}{\textwidth}{|p{0.25\textwidth}|c|c|c|X|} \hline
         \multirow{2}{*}{Scheme} & \multicolumn{2}{c|}{Scaling}                                 & \multirow{2}{*}{Diagonal} & \multirow{2}{*}{Comments} \\ \cline{2-3}
                                 & $\sigma_u\sim\mathcal{O}(1)$  & $\sigma_a\sim\mathcal{O}(1)$ &                           &                           \\ \hline
         Roe 1981 \cite{roe_approximate_1981} & \multicolumn{2}{c|}{Acoustic}                                  & Yes & Original transonic scheme \\ \hline
         Preconditioned Roe 1993 \cite{godfrey_preconditioning_1993} & \multicolumn{2}{c|}{Convective}      & No  & Preconditioned Roe scheme \\ \hline
         Venkateswaran \& Merkle 1995 \cite{venkateswaran_dual_1995} & Convective & Acoustic & Yes & Central difference dissipation \\ \hline
         Weiss \& Smith 1995 \cite{weiss_preconditioning_1995}  & \multicolumn{2}{c|}{Convective} & No  & Preconditioned Roe scheme \\ \hline
         Guillard \& Viozat 1999 \cite{guillard_behaviour_1999} & \multicolumn{2}{c|}{Convective} & No  & First asymptotic expansion of discrete preconditioned Roe scheme \\ \hline
         Venkateswaran \& Merkle 20(00/03) \cite{venkateswaran_efficiency_2000,venkateswaran_artficial_2003} & Convective & Mixed & Yes & Adaptive scalar diffusion. $\mu_{11}$ only applied to the continuity equation \\ \hline
         Mary \& Sagaut 2002 \cite{mary_large_2002} & \multicolumn{2}{c|}{Convective} & Yes & No $\delta p$, and introduces only $\mu_{11}$ into the mass flux in $\mathcal{L}$, similar to \cite{rhie_numerical_1983,liou_new_1993} \\ \hline
         Guillard \& Murrone 2004 \cite{guillard_behavior_2004} & \multicolumn{2}{c|}{Convective} & No & Godunov scheme based on the preconditioned system \\ \hline
         Postdam et al. 2007 \cite{potsdam_unsteady_2007} & Convective & Mixed & (Yes/No)/(Yes/UT) & Two adaptive diffusion schemes (scalar/matrix) \\ \hline
         Thornber et al. 2008 \cite{thornber_numerical_2008,thornber_improved_2008} & \multicolumn{2}{c|}{Mixed} & Yes & \cite{thornber_improved_2008} also modifies inviscid flux and multiplies momentum upwinding by $M$ \\ \hline
         Li \& Gu 2008/9 All-speed Roe \cite{li_all-speed_2008,li_development_2009} & \multicolumn{2}{c|}{Convective} & Yes & \multirow{2}{*}{\parbox{\linewidth}{No $\delta p$, and introduces only $\mu_{11}$ as in \cite{mary_large_2002}. \cite{li_momentum_2010} extends \cite{li_all-speed_2008,li_development_2009} with $\Delta t$ dependence in $\mu_{11}$}} \\ \cline{1-4}
         Li \& Gu 2010 time-marching MIM \cite{li_momentum_2010} & Convective & Mixed & Yes & \\ \hline
         Dellacherie 2010 \cite{dellacherie_analysis_2010} and Dellacherie et al 2016 \cite{dellacherie_construction_2016} & \multicolumn{2}{c|}{\parbox{0.2\linewidth}{\centering Mixed/Convective}} & Yes & Recommends mixed scaling with $\delta p$ removed \cite{dellacherie_analysis_2010} or reduced by $M$ \cite{dellacherie_construction_2016}. Analysis of $\delta p$ applies also to convective scaling \\ \hline
         Rieper 2011 (LMRoe) \cite{rieper_low-mach_2011} & \multicolumn{2}{c|}{Mixed} & Yes & Reduces $\mu_{22}$ in Roe scheme by factor of $M$ \\ \hline
         Sachdev et al. 2012 \cite{sachdev_improved_2012} & Convective & Mixed & No/(UT/LT) & Two adaptive schemes with different forms for small timesteps \\ \hline
         O{\ss}wald et al. 2016 (L2Roe) \cite{oswald_l2roe_2016} & \multicolumn{2}{c|}{Mixed} & Yes & Modifies LMRoe to reduce vorticity upwinding by factor of $M$ \\ \hline
         Caraeni \& Weiss 2017 \cite{caraeni_unsteady_2017} & Convective & Mixed & No &  Modified $\mu_{11}$ of \cite{weiss_preconditioning_1995} to approach the mixed scheme as $\Delta t\to0$ \\ \hline
         Bruel et al. 2019 \cite{bruel_low_2019} & \multicolumn{2}{c|}{Mixed} & No & Restores off-diagonal terms to mixed scheme for acoustic flows. \\ \hline
    \end{tabularx}
    \caption{Diffusion scaling of a number of existing low Mach number Roe-type or central difference schemes. Some schemes have different scalings when the timestep is calculated using the convective CFL ($\sigma_u$) and $M_u\sim\mathcal{O}(M)$, or using the acoustic CFL ($\sigma_a$) and $M_u\sim\mathcal{O}(1)$.
    A diagonal scheme has vanishing $\mu_{12},\mu_{21}\sim o(M^{0})$. UT stands for upper triangular, with vanishing $\mu_{21}\sim o(M^0)$ and $\mu_{12}\sim\mathcalOM{0}$. LT stands for lower triangular with vanishing $\mu_{12}\sim o(M^0)$ and $\mu_{21}\sim\mathcalOM{0}$.}
    \label{tab:lm-fv-schemes}
\end{table}

\subsubsection{Behaviour of the off-diagonal diffusion}\label{sec:off-diagonal}

The off-diagonal diffusion terms in (\ref{eq:interface_delta_up}) contain the switch $U_{il}/|U_{il}|=\pm1$.
This ensures that these terms have a diffusive character but also gives the possibility of negative diffusion coefficients.\footnote{The authors would like to thank the anonymous reviewer who demonstrated that these terms did not have a diffusive form in an earlier draft. Correcting this led to the discussion in section \ref{sec:off-diagonal}.}
Previous schemes often use just $U_{il}$, but we prefer a non-dimensional $\mathcal{O}(1)$ expression.
In 2D the off-diagonal terms result in anisotropic diffusion of the form $(\pm\partial_{xx}u\pm\partial_{yy}v)$ on all equations through $\delta U$, and $\pm\partial_{xx}p$ on the $x$-momentum equation and $\pm\partial_{yy}p$ on the $y$-momentum equation through $\delta p$.
The sign on $\partial_{xx}u$ and $\partial_{xx}p$ will be the same, as will the sign on $\partial_{yy}v$ and $\partial_{yy}p$, but the sign may differ between the $x$ and $y$ derivatives.
As stated earlier, off-diagonal anti-diffusion is tolerable so long as positive (semi-)definiteness is maintained.
This requires $\mu_{11}\mu_{22}-\mu_{12}\mu_{21}\geq0$, where equality implies semi-definiteness.
For asymptotically diagonal or triangular schemes this holds trivially.
This covers all acoustic schemes, almost all mixed schemes and some convective schemes in table \ref{tab:lm-fv-schemes}.
The full convective schemes are due to the preconditioned diffusion, and the inequality holds for these schemes.\footnote{As $M\to0$ the coefficients of this scheme become $\mu_{11}\approx2|\tilde{v}|/(U\sqrt{5})$, $\mu_{22}\approx U(3-\sqrt{5})/(|\tilde{v}|\sqrt{5})$, and $\mu_{12}=\mu_{21}\approx1/\sqrt{5}$.}
For the mixed scheme $\mu_{11}\mu_{22}$ vanishes asymptotically for both convective and acoustic variations, therefore $\mu_{12}\mu_{21}$ must also vanish asymptotically else the diffusion will be negative definite and the scheme unstable.\footnote{When iterative methods that require diagonal dominance are used for implicit timestepping, the semi-definiteness of the mixed scheme also necessitates the use of inconsistent diffusion on the left-hand-side \cite{potsdam_unsteady_2007,sachdev_improved_2012,shima_new_2013}.}
The one exception to this is the full scheme of Bruel et al \cite{bruel_low_2019}, where the off-diagonal terms are designed to remove the acoustic instability and to obtain a CFL limit $\sigma_{a}<1$ compared to $\sigma_{a}<0.5$ for the diagonal mixed scheme.
This scheme modifies the off-diagonal diffusion to be isotropic with the form $\nabla^{2}u + \nabla^{2}v$ in $\delta U$ and $\nabla^{2}p$ on all momentum equations through $\delta p$.
Positive semi-definiteness is ensured by enforcing that the off-diagonal terms have opposite sign, but which term is positive and which negative is indeterminate.
However, the scheme produces asymmetric solutions for flow around a circular cylinder, which Bruel et al attribute to this free choice of sign and the resulting loss of Galilean invariance.

\subsection{Expansion of the discrete Euler equations with artificial diffusion}

In this section we carry out single- and multiple-scale asymptotic expansions of the discrete equations (\ref{eq:fv_discrete_equations}) with diffusive flux (\ref{eq:WeissSmith_diffusion},\ref{eq:interface_delta_up}) for each of the three scalings in table (\ref{table:delta_up_parameter_scaling}).
For simplicity we have taken $\epsilon_{\alpha}=1$ throughout.
We will keep the discussion of these expansions brief, because most points have already been discussed in the previous section, and many of the expansions have been presented in the literature (although usually for one specific flux scheme).
The discrete single-timescale expansion with the convective diffusion is investigated in detail in the seminal papers from Guillard \& coauthors \cite{guillard_behaviour_1999,guillard_behavior_2004}, and the most important features are also reported in the reviews by Guillard \& Nkonga \cite{guillard_chapter_2017} and Li \& Gu \cite{li_mechanism_2013}.
The discrete single-timescale expansion with the acoustic diffusion can be found in a several papers, as it shows why classical transonic schemes fail at low Mach number, see \cite{guillard_behaviour_1999,guillard_behavior_2004,guillard_chapter_2017,rieper_low-mach_2011}.
Several of these papers also present the discrete single-timescale expansion with the mixed diffusion \cite{dellacherie_analysis_2010,rieper_low-mach_2011,li_all-speed_2008,lin_density_2018,li_mechanism_2013,guillard_chapter_2017}.
To the authors' knowledge, the only previous study in the literature to carry out multiple-timescale expansions of the discrete equations is Bruel et al. \cite{bruel_low_2019}, who present multiple-timescale expansions of the baratropic Euler equations with all three schemes.

\subsubsection{Expansion with convective diffusion}

First, we carry out the single timescale expansion of the discrete equations (\ref{eq:fv_discrete_equations}) with convective diffusion scaling.
The relations are exactly equivalent to those found by Guillard \& Viozat \cite{guillard_behaviour_1999}, except for the more general notation for the diffusion coefficients, so will only be discussed briefly - consult \cite{guillard_behaviour_1999} for more in-depth discussion on these discrete equations, they are included here mainly for completeness.
The $\mathcalOM{\-2}$ and $\mathcalOM{\-1}$ relations are identical, and are:
\begin{subequations} \label{eq:discrete_cdiff_1scale_-2}
\begin{align}
    \label{eq:discrete_cdiff_1scale_-2_density}
    0 & = \facesum S_{il}\ord{\rho}{0}\frac{\nu_{11}}{\rho|v|} \Delta_{il}\ord{p}{0,1} \\
    \label{eq:discrete_cdiff_1scale_-2_momentum}
    \facesum S_{il} \ord{p_l}{0,1} \underline{n}_{il} & = \facesum S_{il}( \ord{\underline{\rho u}}{0}\frac{\nu_{11}}{\rho|v|} + \frac{U_{il}}{|U_{il}|}\nu_{21}\underline{n}_{il} ) \Delta_{il}\ord{p}{0,1} \\
    \label{eq:discrete_cdiff_1scale_-2_energy}
    0 & = \facesum S_{il}\ord{\rho H}{0}\frac{\nu_{11}}{\rho|v|} \Delta_{il}\ord{p}{0,1}
\end{align}
\end{subequations}
The right-hand sides of (\ref{eq:discrete_cdiff_1scale_-2_density},\ref{eq:discrete_cdiff_1scale_-2_energy}) are in the form of the discrete Laplacian.
In $N$ dimensions, equations (\ref{eq:discrete_cdiff_1scale_-2}) are two elliptic systems each of $N+2$ equations for $\ord{p}{0,1}$, which are the discrete equivalents of the continuous relations (\ref{eq:velocity_1time_cdiff_-2},\ref{eq:pressure_1time_cdiff_-2}) and (\ref{eq:velocity_1time_cdiff_-1},\ref{eq:pressure_1time_cdiff_-1}).
Under a small number of reasonable assumptions, they enforce constant $\ord{p}{0,1}$ over the entire domain \cite{guillard_behaviour_1999}.
Using the non-dimensional $\mathcalOM{0}$ equation of state $\ord{p}{0}=(\gamma-1)\ord{\rho E}{0} = \ord{\rho}{0}\ord{T}{0}$ \cite{muller_low_1999}, the $\mathcalOM{0}$ relations are:
\begin{subequations} \label{eq:discrete_cdiff_1scale_0}
\begin{alignat}{2}
    \label{eq:discrete_cdiff_1scale_0_density}
    &\Omega_i \frac{d\ord{\rho_i}{0}}{dt}
       && + \frac{1}{2}\facesum S_{il} \ord{\rho_l}{0}\ord{U_l}{0} \\
    &  && = \frac{1}{2}\facesum S_{il}\big(|\ord{U}{0}|\Delta_{il}\ord{\rho}{0} +
            \ord{\rho}{0}(\frac{\nu_{11}}{\rho|v|} \Delta_{il}\ord{p}{2}
          + \frac{U_{il}}{|U_{il}|}\nu_{12} \Delta_{il}\ord{U}{0}) \big) \notag \\
    \label{eq:discrete_cdiff_1scale_0_momentum}
    &\Omega_i \frac{d\ord{\underline{\rho u}_i}{0}}{dt}
       && + \frac{1}{2}\facesum S_{il}(\ord{\underline{\rho u}_l}{0}\ord{U_l}{0} + \ord{p_l}{2}\underline{n}_{il}) \\
    &  && = \frac{1}{2} \facesum S_{il}\big(|\ord{U}{0}|\Delta_{il}\ord{\underline{\rho u}}{0}
          + \ord{\underline{\rho u}}{0}(\frac{\nu_{11}}{\rho|v|}\Delta_{il}\ord{p}{2} + \frac{U_{il}}{|U_{il}|}\nu_{12}\Delta_{il}\ord{U}{0})
          + \underline{n}_{il}(\frac{U_{il}}{|U_{il}|}\nu_{21}\Delta_{il}\ord{p}{2} + \rho|v|\nu_{22}\Delta_{il}\ord{U}{0}) \big) \notag \\
    \label{eq:discrete_cdiff_1scale_0_energy}
    &\Omega_i \frac{d\ord{\rho E_i}{0}}{dt}
       && + \frac{1}{2}\facesum S_{il} \ord{\rho H_l}{0}\ord{U_l}{0}
          = \frac{\Omega_i}{\gamma-1} \frac{d\ord{p}{0}}{dt}
          + \frac{\ord{\rho H}{0}}{2}\facesum S_{il} \ord{U_l}{0} \\
    &  && = \frac{\ord{\rho H}{0}}{2}\facesum S_{il}\big(
            \frac{\nu_{11}}{\rho|v|} \Delta_{il}\ord{p}{2}
          + \frac{U_{il}}{|U_{il}|}\nu_{12} \Delta_{il}\ord{U}{0} \big) \notag
\end{alignat}
\end{subequations}
Where we have used constant $\ord{p}{0}$ to imply constant $\ord{\rho E}{0}$ and $\ord{\rho H}{0}$ to simplify relation (\ref{eq:discrete_cdiff_1scale_0_energy}).
If the zeroth order entropy $\ord{s}{0}$ is constant, then the zeroth order temperature and density $\ord{T}{0}$ and $\ord{\rho}{0}$ also become constant.
The second term in (\ref{eq:discrete_cdiff_1scale_0_density}) is the discrete divergence, and relations (\ref{eq:discrete_cdiff_1scale_0_density},\ref{eq:discrete_cdiff_1scale_0_energy}) are the discrete equivalents of the continuity relation (\ref{eq:pressure_1time_cdiff_0}).
At steady state on a regular grid, the pressure diffusion takes exactly the form of the Brezzi-Pitk\"aranta stabilisation \cite{brezzi_stabilization_1984} used with a finite-volume scheme by Eymard et al. \cite{eymard_stabilized_2006}, so this scheme is expected to be free of chequerboard instabilities.
The momentum relation (\ref{eq:discrete_cdiff_1scale_0_momentum}) is the discrete equivalent of relation (\ref{eq:velocity_1time_cdiff_0})\footnote{Properly, it is a discrete combination of both (\ref{eq:velocity_1time_cdiff_0}) and (\ref{eq:pressure_1time_cdiff_0}), as can be seen from the diffusion terms.}, having all the necessary terms in $\mathcal{L}$, and properly balanced diffusion in $\mathcal{R}$.
The discrete single scale expansion with the convective diffusion scaling therefore reproduces all the necessary relations for the convective low Mach number limit, so is a consistent scheme for this limit.

Next, we carry out the two timescale expansion with the convective diffusion scaling.
The $\mathcalOM{\-2}$ relations are identical to those from the single-timescale expansion, so we assume $\ord{p}{0}$ and related zeroth order thermodynamic quantities are constant again.
The $\mathcalOM{\-1}$ relations are:
\begin{subequations} \label{eq:discrete_cdiff_2scale_-1}
\begin{align}
    \label{eq:discrete_cdiff_2scale_-1_density}
    \Omega_i \frac{d\ord{\rho_i}{0}}{d\tau}
        & = \frac{1}{2}\facesum S_{il}\ord{\rho}{0}\frac{\nu_{11}}{\rho|v|} \Delta_{il}\ord{p}{1} \\
    \label{eq:discrete_cdiff_2scale_-1_momentum}
    \Omega_i \frac{d\ord{\underline{\rho u}_i}{0}}{d\tau} + \frac{1}{2}\facesum S_{il} \ord{p_l}{1} \underline{n}_{il}
        & = \frac{1}{2}\facesum S_{il}( \ord{\underline{\rho u}}{0}\frac{\nu_{11}}{\rho|v|} + \frac{U_{il}}{|U_{il}|}\nu_{21} \underline{n}_{il} ) \Delta_{il}\ord{p}{1} \\
    \label{eq:discrete_cdiff_2scale_-1_energy}
    \Omega_i \frac{d\ord{\rho E_i}{0}}{d\tau}
        & = \frac{\ord{\rho H}{0}}{2}\facesum S_{il}\frac{\nu_{11}}{\rho|v|} \Delta_{il}\ord{p}{1}
\end{align}
\end{subequations}
Using the $\mathcalOM{0}$ equation of state, we can see that relations (\ref{eq:discrete_cdiff_2scale_-1_density},\ref{eq:discrete_cdiff_2scale_-1_energy}) are the discrete equivalents to (\ref{eq:pressure_2time_cdiff_-1}), which is is a parabolic equation for $\ord{p}{1}$, assuming $\partial_{\tau}\ord{p}{0}\to0$.
This means $\ord{p}{1}$ will become constant on the acoustic timescale $\tau$ unless $\ord{p}{0}$ is forced on the acoustic timescale (which would violate the physical relation (\ref{eq:pressure_multiple_-1})), so the convective diffusion scaling is indeed unsuitable for acoustic or mixed convective-acoustic simulations.

\subsubsection{Expansion with acoustic diffusion}
The single timescale expansion of the discrete equations (\ref{eq:fv_discrete_equations}) with acoustic diffusion scaling leads to the $\mathcalOM{\-2}$ momentum relation:
\begin{equation} \label{eq:discrete_adiff_1scale_-2_momentum}
    \facesum S_{il} \ord{p_l}{0} \underline{n}_{il}
        = \facesum S_{il}\frac{U_{il}}{|U_{il}|}\nu_{21} \underline{n}_{il} \Delta_{il}\ord{p}{0}
\end{equation}
and the $\mathcalOM{\-1}$ relations:
\begin{subequations} \label{eq:discrete_adiff_1scale_-1}
\begin{align}
    \label{eq:discrete_adiff_1scale_-1_density}
    0 & = \facesum S_{il}\ord{\rho}{0}\frac{\nu_{11}}{\rho|v|} \Delta_{il}\ord{p}{0} \\
    \label{eq:discrete_adiff_1scale_-1_momentum}
    \facesum S_{il} \ord{p_l}{1} \underline{n}_{il}
        & = \facesum S_{il} \underline{n}_{il} \big( \frac{U_{il}}{|U_{il}|}\nu_{21} \Delta_{il}\ord{p}{1} + \rho|v|\nu_{22} \Delta_{il}\ord{U}{0} \big) \\
    \label{eq:discrete_adiff_1scale_-1_energy}
    0 & = \facesum S_{il}\ord{\rho H}{0}\frac{\nu_{11}}{\rho|v|} \Delta_{il}\ord{p}{0}
\end{align}
\end{subequations}
The relations (\ref{eq:discrete_adiff_1scale_-2_momentum},\ref{eq:discrete_adiff_1scale_-1_density},\ref{eq:discrete_adiff_1scale_-1_energy}) are the same elliptic system for $\ord{p}{0}$ as we have seen previously, so we assume $\ord{p}{0}$ is constant.
We have used a constant $\ord{p}{0}$ to simplify relation (\ref{eq:discrete_adiff_1scale_-1_momentum}), which is a discretisation of the continuous relation (\ref{eq:velocity_1time_adiff_-1}).
The first order pressure $\ord{p}{1}$ cannot be constant unless the jumps in normal velocity $\Delta_{il}\ord{U}{0}$ are zero at every cell face - this is even clearer if the scheme is diagonal, which the acoustic scheme usually is.
As previously stated, this condition is actually enforced for simplex meshes, leading to a divergence free solution with the correct pressure scaling \cite{rieper_influence_2008,rieper_influence_2009,guillard_behavior_2009,dellacherie_influence_2010}.
This is why we restricted our analysis to quadrilaterals and hexahedra at the beginning of this section.
For these cell types, (\ref{eq:discrete_adiff_1scale_-1_momentum}) forces either trivial velocity fields or non-constant $\ord{p}{1}$, so this scheme is unsuitable for convective low Mach number flows.

The two timescale expansion with acoustic diffusion scaling leads to the same $\mathcalOM{\-2}$ momentum relation (\ref{eq:discrete_adiff_1scale_-2_momentum}), and the following $\mathcalOM{\-1}$ relations:
\begin{subequations} \label{eq:discrete_adiff_2scale_-1}
\begin{align}
    \label{eq:discrete_adiff_2scale_-1_density}
    \Omega_i \frac{d\ord{\rho_i}{0}}{d\tau}
        & = \frac{1}{2}\facesum S_{il}\ord{\rho}{0}\frac{\nu_{11}}{\rho|v|} \Delta_{il}\ord{p}{0} \\
    \label{eq:discrete_adiff_2scale_-1_momentum}
    \Omega_i \frac{d\ord{\underline{\rho u}_i}{0}}{d\tau} + \frac{1}{2}\facesum S_{il} \ord{p_l}{1} \underline{n}_{il}
        & = \frac{1}{2}\facesum S_{il} \underline{n}_{il} \big( \frac{U_{il}}{|U_{il}|}\nu_{21} \Delta_{il}\ord{p}{1}
          + \rho|v|\nu_{22} \Delta_{il}\ord{U}{0} \big) \\
    \label{eq:discrete_adiff_2scale_-1_energy}
    \Omega_i \frac{d\ord{\rho E_i}{0}}{d\tau}
        & = \frac{1}{2}\facesum S_{il}\ord{\rho H}{0}\frac{\nu_{11}}{\rho|v|} \Delta_{il}\ord{p}{0}
\end{align}
\end{subequations}
The $\mathcalOM{0}$ density relation is:
\begin{equation}\label{eq:discrete_adiff_2scale_0_density}
    \Omega_i \big( \frac{d\ord{\rho_i}{1}}{d\tau} + \frac{d\ord{\rho_i}{0}}{dt} \big)
      + \frac{1}{2}\facesum S_{il}\ord{\rho}{0}\ord{U_l}{0}
      = \frac{1}{2}\facesum S_{il}\big(|\ord{U}{0}|\Delta_{il}\ord{\rho}{0} +
            \ord{\rho}{0}(\frac{\nu_{11}}{\rho|v|} \Delta_{il}\ord{p}{1}
          + \frac{U_{il}}{|U_{il}|}\nu_{12} \Delta_{il}\ord{U}{0}) \big)
\end{equation}
Relations (\ref{eq:discrete_adiff_1scale_-2_momentum},\ref{eq:discrete_adiff_2scale_-1_density},\ref{eq:discrete_adiff_2scale_-1_energy}) are now an unsteady elliptic system on the acoustic timescale for $\ord{p}{0}$, leading variations in $\ord{p}{0}$ to be dissipated on the acoustic timescale.
Constant $\ord{p}{0}$ has been used to simplify relations (\ref{eq:discrete_adiff_2scale_-1_momentum},\ref{eq:discrete_adiff_2scale_0_density}), which for constant entropy (i.e. constant $\ord{\rho}{0}$) are the discrete equivalents of the linear acoustics relations (\ref{eq:velocity_2time_adiff_-1},\ref{eq:pressure_2time_adiff_0}).
This shows that the scheme is suitable for simulating purely acoustic low Mach number flows (for example 1D shocktube flows), but not mixed convective-acoustic flows due to the issues highlighted in the single-scale expansion.

\subsubsection{Expansion with mixed diffusion}
Lastly, we carry out the expansions with the mixed diffusion scaling.
For much of the discussion we assume that the scheme is diagonal for simplicity, but we have retained both off-diagonal diffusion terms in the expansions on the understanding that at least one of $\mu_{12}$ and $\mu_{21}$ must vanish according to the discussion in section \ref{sec:off-diagonal}.
The single timescale expansion leads to the  $\mathcalOM{\-2}$ momentum relation:
\begin{equation}\label{eq:discrete_mdiff_1scale_-2_momentum}
    \facesum S_{il} \ord{p_l}{0} \underline{n}_{il}
        = \facesum S_{il}\frac{U_{il}}{|U_{il}|}\nu_{21} \underline{n}_{il} \Delta_{il}\ord{p}{0}
\end{equation}
and the $\mathcalOM{\-1}$ relations:
\begin{subequations} \label{eq:discrete_mdiff_1scale_-1}
\begin{align}
    \label{eq:discrete_mdiff_1scale_-1_density}
    0 & = \facesum S_{il}\ord{\rho}{0}\frac{\nu_{11}}{\rho|v|} \Delta_{il}\ord{p}{0} \\
    \label{eq:discrete_mdiff_1scale_-1_momentum}
    \facesum S_{il} \ord{p_l}{1} \underline{n}_{il}
        & = \facesum S_{il} \big( \ord{\underline{\rho u}}{0}\frac{\nu_{11}}{\rho|v|} \Delta_{il}\ord{p}{0} + \frac{U_{il}}{|U_{il}|}\nu_{21} \underline{n}_{il} \Delta_{il}\ord{p}{1} \big) \\
    \label{eq:discrete_mdiff_1scale_-1_energy}
    0 & = \facesum S_{il}\ord{\rho H}{0}\frac{\nu_{11}}{\rho|v|} \Delta_{il}\ord{p}{0}
\end{align}
\end{subequations}
Relations (\ref{eq:discrete_mdiff_1scale_-2_momentum},\ref{eq:discrete_mdiff_1scale_-1_density},\ref{eq:discrete_mdiff_1scale_-1_energy}) are the elliptic system for $\ord{p}{0}$.
Assuming constant $\ord{p}{0}$, (\ref{eq:discrete_mdiff_1scale_-1_momentum}) becomes an elliptic equation for $\ord{p}{1}$, for which constant $\ord{p}{1}$ may not be the only solution.
For diagonal or upper triangular schemes the right hand side of relation (\ref{eq:discrete_mdiff_1scale_-1_momentum}) vanishes.
On a regular 2D Cartesian grid, the momentum relations then become $\ord{p}{1}_{i+1,j}-\ord{p}{1}_{i-1,j}=0$ and $\ord{p}{1}_{i,j+1}-\ord{p}{1}_{i,j-1}=0$, which will, on their own, admit chequer-board modes.
The $\mathcalOM{0}$ relations are:
\begin{subequations} \label{eq:discrete_mdiff_1scale_0}
\begin{alignat}{2}
    \label{eq:discrete_mdiff_1scale_0_density}
    &\Omega_i \frac{d\ord{\rho_i}{0}}{dt}
       && + \frac{1}{2}\facesum S_{il}\ord{\rho}{0}\ord{U_l}{0} \\
    &  && = \frac{1}{2}\facesum S_{il}\big(|\ord{U}{0}|\Delta_{il}\ord{\rho}{0} +
            \ord{\rho}{0}(\frac{\nu_{11}}{\rho|v|} \Delta_{il}\ord{p}{1}
          + \frac{U_{il}}{|U_{il}|}\nu_{12} \Delta_{il}\ord{U}{0}) \big) \notag \\
    \label{eq:discrete_mdiff_1scale_0_momentum}
    &\Omega_i \frac{d\ord{\underline{\rho u}_i}{0}}{dt}
       && + \frac{1}{2}\facesum S_{il}(\ord{\underline{\rho u}_l}{0}\ord{U_l}{0} + \ord{p_l}{2}\underline{n}_{il}) \\
    &  && = \frac{1}{2} \facesum S_{il}\big(|\ord{U}{0}|\Delta_{il}\ord{\underline{\rho u}}{0}
          + \ord{\underline{\rho u}}{0}(\frac{\nu_{11}}{\rho|v|}\Delta_{il}\ord{p}{1} + \frac{U_{il}}{|U_{il}|}\nu_{12}\Delta_{il}\ord{U}{0})
          + \underline{n}_{il}(\frac{U_{il}}{|U_{il}|}\nu_{21}\Delta_{il}\ord{p}{2} + \rho|v|\nu_{22}\Delta_{il}\ord{U}{0}) \big) \notag \\
    \label{eq:discrete_mdiff_1scale_0_energy}
    &\Omega_i \frac{d\ord{\rho E_i}{0}}{dt}
       && + \frac{\ord{\rho H}{0}}{2}\facesum S_{il} \ord{U_l}{0} \\
    &  && = \frac{\ord{\rho H}{0}}{2}\facesum S_{il}\big(
            \frac{\nu_{11}}{\rho|v|} \Delta_{il}\ord{p}{1}
          + \frac{U_{il}}{|U_{il}|}\nu_{12} \Delta_{il}\ord{U}{0} \big) \notag
\end{alignat}
\end{subequations}
Relations (\ref{eq:discrete_mdiff_1scale_0_density},\ref{eq:discrete_mdiff_1scale_0_energy}) are discrete equivalents of the continuity relation (\ref{eq:pressure_1time_mdiff_0}).
Assuming $\mu_{12}\sim o(M^{0})$, at steady state these relations mean that $\ord{p}{1}$ is free of chequer-boards \textit{iff} the discrete divergence of the zeroth order velocity is zero.
In practice, we know of no case - either in the literature or in our own studies - where this scheme has produced steady solutions with non-constant $\ord{p}{1}$, and Rieper \cite{rieper_low-mach_2011} provides an argument that this scheme will damp the divergence on the convective timescale on Cartesian grids.
Although approaching a divergence free solution as $M\to0$ is appealing, it means that this scheme will be susceptible to chequerboard-modes on $\ord{p}{2}$, as we shall see in numerical examples later.
Lastly, the momentum relation (\ref{eq:discrete_mdiff_1scale_0_momentum}) is consistent with the continuous relation (\ref{eq:velocity_1time_mdiff_0}). 
The scheme therefore reproduces all the relevant asymptotic relations, and is suitable for simulating convective low Mach number flow, but with the caveat of potential  chequer-board modes on $\ord{p}{2}$.

The two timescale expansion with mixed diffusion scaling gives the same $\mathcalOM{\-2}$ momentum relation and $\mathcalOM{\-1}$ density and energy relations as for the two timescale expansion with acoustic diffusion (\ref{eq:discrete_adiff_1scale_-2_momentum},\ref{eq:discrete_adiff_2scale_-1_density},\ref{eq:discrete_adiff_2scale_-1_energy}), which lead to constant $\ord{p}{0}$.
Subsequently, the $\mathcalOM{\-1}$ momentum relation is:
\begin{equation}
    \label{eq:discrete_mdiff_2scale_-1_momentum}
    \Omega_i \frac{d\ord{\underline{\rho u}_i}{0}}{d\tau} + \facesum S_{il} \ord{p_l}{1} \underline{n}_{il}
          = \facesum S_{il}
            \frac{U_{il}}{|U_{il}|}\nu_{21} \underline{n}_{il} \Delta_{il}\ord{p}{1}
\end{equation}
and the $\mathcalOM{0}$ relations are:
\begin{subequations} \label{eq:discrete_mdiff_2scale_0}
\begin{alignat}{2}
    \label{eq:discrete_mdiff_2scale_0_density}
    &\Omega_i \big( \frac{d\ord{\rho_i}{1}}{d\tau} && + \frac{d\ord{\rho_i}{0}}{dt} \big)
       + \frac{1}{2}\facesum S_{il}\ord{\rho}{0}\ord{U_l}{0} \\
    &  && = \frac{1}{2}\facesum S_{il}\big(|\ord{U}{0}|\Delta_{il}\ord{\rho}{0} +
            \ord{\rho}{0}(\frac{\nu_{11}}{\rho|v|} \Delta_{il}\ord{p}{1}
          + \frac{U_{il}}{|U_{il}|}\nu_{12} \Delta_{il}\ord{U}{0}) \big) \notag \\
    \label{eq:discrete_mdiff_2scale_0_momentum}
    &\Omega_i \big( \frac{d\ord{\underline{\rho u}_i}{1}}{d\tau} && + \frac{d\ord{\underline{\rho u}_i}{0}}{dt} \big)
       + \frac{1}{2}\facesum S_{il}(\ord{\underline{\rho u}_l}{0}\ord{U_l}{0} + \ord{p_l}{2}\underline{n}_{il}) \\
    &  && = \frac{1}{2} \facesum S_{il}\big(|\ord{U}{0}|\Delta_{il}\ord{\underline{\rho u}}{0}
          + \ord{\underline{\rho u}}{0}(\frac{\nu_{11}}{\rho|v|}\Delta_{il}\ord{p}{1} + \frac{U_{il}}{|U_{il}|}\nu_{12}\Delta_{il}\ord{U}{0})
          + \underline{n}_{il}(\frac{U_{il}}{|U_{il}|}\nu_{21}\Delta_{il}\ord{p}{2} + \rho|v|\nu_{22}\Delta_{il}\ord{U}{0}) \big) \notag \\
    \label{eq:discrete_mdiff_2scale_0_energy}
    &\Omega_i \big( \frac{d\ord{\rho E_i}{1}}{d\tau} && + \frac{d\ord{\rho E_i}{0}}{dt} \big)
       + \frac{\ord{\rho H}{0}}{2}\facesum S_{il} \ord{U_l}{0} \\
    &  && = \frac{\ord{\rho H}{0}}{2}\facesum S_{il}\big(
            \frac{\nu_{11}}{\rho|v|} \Delta_{il}\ord{p}{1}
          + \frac{U_{il}}{|U_{il}|}\nu_{12} \Delta_{il}\ord{U}{0} \big) \notag
\end{alignat}
\end{subequations}
The relations (\ref{eq:discrete_mdiff_2scale_-1_momentum}) and (\ref{eq:discrete_mdiff_2scale_0_density},\ref{eq:discrete_mdiff_2scale_0_energy}) are the discrete equivalents of the linear acoustics relations (\ref{eq:velocity_2time_mdiff_-1},\ref{eq:pressure_2time_mdiff_0}).
It can be seen that, as in the continuous analysis, the velocity diffusion vanishes from the acoustic momentum relation (\ref{eq:discrete_mdiff_2scale_-1_momentum}), while the acoustic relations (\ref{eq:discrete_mdiff_2scale_0_density},\ref{eq:discrete_mdiff_2scale_0_energy}) retain the pressure diffusion term.
Therefore this scheme is consistent for acoustic flows, however potentially has instabilities in the acoustic velocity variations.
The momentum relation (\ref{eq:discrete_mdiff_2scale_0_momentum}) is the discrete equivalent of (\ref{eq:velocity_2time_mdiff_0}), confirming that this scheme is suitable for mixed convective-acoustic flows.\\

The discrete expansions of the first order finite-volume scheme have confirmed that all findings of the continuous analysis in sections \ref{sec:design} and \ref{sec:continuous} carry over to the discrete setting for this particular discrete form, with each diffusion scaling behaving as expected for both the single- and multiple-scale limits.

\subsection{Stability of the discrete scheme}
We now consider the stability of the discrete schemes by finding expressions for the eigenvalues of the artificial diffusion Jacobian and the von Neumann symbols of the fully discrete first order scheme.

\subsubsection{Eigenvalues of the artificial diffusion Jacobian}
The eigenvalues of the diffusive flux Jacobian (\ref{eq:fv-entropyvar-diffusion}) are:
\begin{subequations}\label{eq:fv_diffusion_eigenvalues}
\begin{alignat}{1}
    \lambda_{1,2} & = \frac{M^{\-2}|u|\mu_{11} + |u|\mu_{22} + 2|U| \pm \sqrt{|u|^2(M^{\-2}\mu_{11} - \mu_{22})^2 - M^{-2}p\mu_{12}\mu_{21}}}{2} \\
    \lambda_{3,4} & = |U|
\end{alignat}
\end{subequations}
If $\mu_{12}\mu_{21}=0$ then:
\begin{equation}
    \lambda_{1,2,3,4} = \{ M^{\-2}|u|\mu_{11} + |U|,\, |u|\mu_{22} + |U|,\, |U|,\, |U| \}
\end{equation}
and the spectral radius is:
\begin{equation}\label{eq:simple_fv_diffusion_spectral_radius}
    \Lambda(\uuline{A}) = \max(\lambda_i) = |u|\max(M^{-2}\mu_{11}, \mu_{22}) + |U|
\end{equation}
Using the scalings in table \ref{table:delta_up_parameter_scaling}, we see that the scaling of $\Lambda(\uuline{A})$, either from (\ref{eq:simple_fv_diffusion_spectral_radius}) or from a leading order approximation of (\ref{eq:fv_diffusion_eigenvalues}), agrees with the trace estimates (\ref{eq:trace_growths}) for all three diffusion scalings.

\subsubsection{von Neumann symbols of the first order scheme}
We now find the stability limits for the linearised first order scheme with each diffusion scaling using von Neumann analysis.
Discretising equation (\ref{eq:fv_discrete_equations}) using central differences in space and forward Euler in time, we have the linearised update for the solution vector $\underline{q}^{n}_{i}$ in cell $i$ at time level $n$:
\begin{equation}\label{eq:linearised_discrete_equations}
    \frac{\underline{q}^{n+1}_{i} - \underline{q}^{n}_{i}}{\Delta t}
    + \uuline{J}\bigg(\frac{\underline{q}^{n}_{i+1} - \underline{q}^{n}_{i-1}}{2\Delta x}\bigg)
    =
    \uuline{A}\bigg(\frac{\underline{q}^{n}_{i+1} - 2\underline{q}^{n}_{i} + \underline{q}^{n}_{i-1}}{2\Delta x}\bigg)
\end{equation}
where $\uuline{J}$ is the physical flux Jacobian and $\uuline{A}$ is the diffusive flux Jacobian.
In this section we assume all diffusion matrices are diagonal in the entropy variables, as it simplifies the presentation.
For stability analysis with the off-diagonal terms included, see \cite{birken_stability_2005} for the convective scheme and \cite{bruel_low_2019} for the mixed scheme.
Substituting $\underline{q}^{n}_{j} = \underline{\hat{q}}^{n}e^{ikj\Delta x}$, where $k=n\pi\Delta x/l$ is the discrete wavenumber, we find the amplification matrix in Fourier space is:
\begin{equation}\label{eq:fv_amplification_matrix}
    \uuline{\hat{G}}(k) =
    \uuline{\mathcal{I}}
    -\frac{\Delta t}{\Delta x}\bigg(
        2\skt^2\uuline{A} + i\sk\uuline{J}
    \bigg)
\end{equation}
where $\uuline{\mathcal{I}}$ is the identity matrix, $\skt=\sin(k/2)$ and $\sk=\sin(k)$.
For stability, the eigenvalues $\lambda_i$ of $(\uuline{\hat{G}}(k))$ must be $|\lambda_i|\leq1\;\forall\;k$.
We consider the 1D equations, as stability of the 1D scheme is a necessary condition for stability of the multidimensional schemes.
In the entropy variables the entropy equation decouples and the associated eigenvalue is identical for all three diffusion schemes, and is exactly that of the first order upwind scheme for scalar advection:
\begin{equation}\label{eq:vn_symbols_3}
    \lambda_3 = 1 - (\sigma_a M)(2\skt^2 + i\sk)
\end{equation}
where $\sigma_a=a\Delta t / \Delta x$ is the acoustic CFL number.
This eigenvalue depends on the convective CFL number $\sigma_a M$, as we would expect, and is stable for $0\leq(\sigma_a M)\leq1$.
The eigenvalues for the pressure-velocity subsystem using the acoustic diffusion scheme are:
\begin{equation}\label{eq:vn_symbols_adiff_12}
    \lambda^a_{1,2} = 1 - (\sigma_a M)(2\skt^2 + i\sk) - \sigma_a(2\skt^2 \pm i\sk)
\end{equation}
Where we can identify the upwind discretisations of the convective system in the second term, and the forward/backward travelling acoustic waves in the third term.
The stability is dominated by the acoustic terms so, to leading order, the scheme is stable for $0\leq\sigma_a\leq1$.
\begin{figure}
    \centering
    \includegraphics[width=0.95\linewidth]{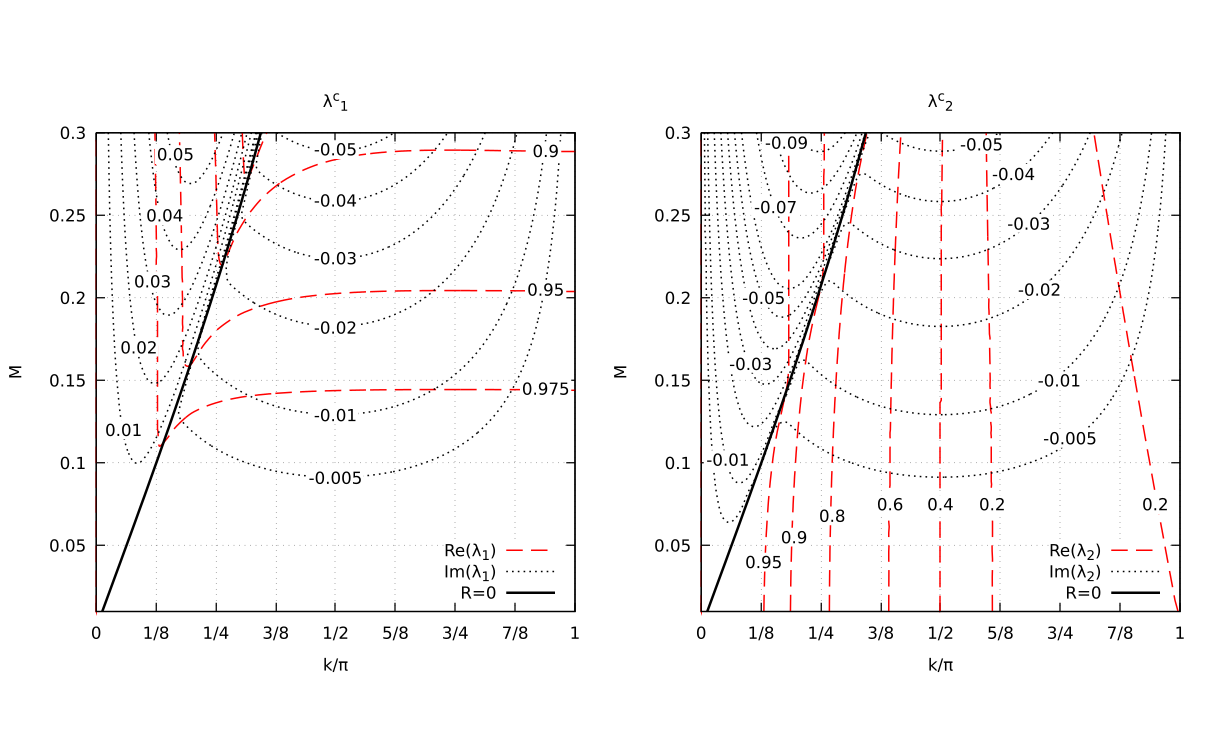}
    \caption{Real and imaginary components for the von Neumann symbols for the convective scheme  (\ref{eq:vn_symbols_cdiff_full}) with $(\sigma_a M^{-1}) = 0.6$. $\lambda^c_1$ (left) and $\lambda^c_2$ (right).}
    \label{fig:convective_vonneumann}
\end{figure}
The full expression for the eigenvalues for the convective scheme is:
\begin{equation}\label{eq:vn_symbols_cdiff_full}
    \lambda^c_{1,2} =
        1 - (\sigma_a M)(2\skt^2 + i\sk) - \sigma_a\Big( M^{\-1}\skt^2
        \pm i\sqrt{\sk^2 - M^{\-2}\skt^4 }\Big)
\end{equation}
The second term is the convective subsystem, the third term is due to the large $(\sigma_a M^{\-1})$ pressure diffusion, and the radicand has contributions from both the pressure diffusion and from the acoustic waves.
The leading order term is the $(\sigma_a M^{\-1})$ pressure diffusion, so the stability limit will scale a factor of $M$ worse than the usual stability limit (which depends on $\sigma_a$), as originally shown for preconditioned diffusion by Birken \& Meister \cite{birken_stability_2005} and as predicted for all convective schemes by the spectral radius estimates (\ref{eq:trace_growths}) in section \ref{sec:continuous}.
Contours of the real and imaginary parts of $\lambda^c_1$ and $\lambda^c_2$ for varying $k$ and $M$ are plotted in figure \ref{fig:convective_vonneumann}.
The eigenvalues have two distinct regions, either side of the locus $M=\tan(k/2)/2$ where the radicand goes to zero (labelled $R=0$ in figure \ref{fig:convective_vonneumann}).
In the low wavenumber region to the left of this locus the radicand is positive so this term is imaginary, and to the right in the moderate/high wavenumber region the radicand is negative and this term is real.
We can simplify the square root in each region using binomial expansions for $k<<M$ and $k>>M$.
In the low wavenumber region this leads to the following expression:
\begin{equation}\label{eq:vn_symbols_cdiff_lowk}
    \lambda^c_{1,2} =
        1 - (\sigma_a M)(2\skt^2 + i\sk) - (\sigma_a M^{\-1})\skt^2
        \pm i\sigma_a\sk
\end{equation}
and in the high wavenumber region:
\begin{subequations}\label{eq:vn_symbols_cdiff_highk}
\begin{align}
    \lambda^c_1 & = 1 - (\sigma_a M)(2 + i\sk) \\
    \lambda^c_2 & = 1 - (\sigma_a M)((4\skt^2-2) + i\sk) - (\sigma_a M^{\-1})(2\skt^2)
\end{align}
\end{subequations}
These expressions can be interpreted from the perspective of a multiple-space scale approach, where acoustic and convective phenomena occur on large and small spatial scales respectively, instead of the multiple-time scale approach used in the rest of the paper.
In the low wavenumber `acoustic' region, the leading order imaginary term for both eigenvalues is due to the $\sigma_a$ forward/backward travelling acoustic  waves.
However, these terms are dominated by the $(\sigma_a M^{\-1})$ diffusion, so to leading order both eigenvalues resemble those of a diffusion equation, corresponding to the rapid damping of acoustic waves by this scheme.
Long wavelength variations in both pressure and (normal) velocity are due to acoustic effects and must be damped.

On the other hand, in the high wavenumber `convective' region the leading order imaginary terms in both eigenvalues are on the $(\sigma_a M)$ convective scale, and only $\lambda^c_2$ is dominated by the $(\sigma_a M^{\-1})$ diffusion.
At short wavelengths, (normal) velocity is no longer an acoustic quantity, but joins entropy (and vorticity in 2/3D) as a convected quantity.
The pressure is still dominated by the large diffusion, which means that the Brezzi-Pitk{\"a}ranta type stabilisation is retained across all wavenumbers.

The boundary between the two regions reduces approximately linearly with $M$ because of the scale separation between acoustic and convective phenomena as $M\to0$.
In real applications implicit timestepping is used for this scheme to overcome the stringent $(\sigma_a M^{\-1})$ stability limit, but the von Neumann analysis of the explicit first order scheme provides an interesting insight into the behaviour of the scheme.

\begin{figure}
    \centering
\begin{subfigure}[t]{0.49\textwidth}
    \centering
    \includegraphics[width=0.99\linewidth]{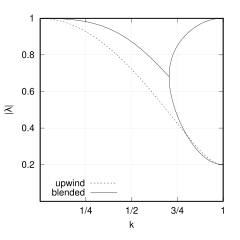}
    \caption{}
    \label{fig:mixed_amplification}
\end{subfigure}
\begin{subfigure}[t]{0.49\textwidth}
    \centering
    \includegraphics[width=0.99\linewidth]{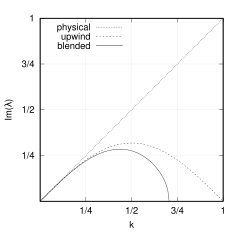}
    \caption{}
    \label{fig:mixed_dispersion}
\end{subfigure}
    \caption{von Neumann symbols for the 1D mixed diffusion scheme with forward Euler time integration. (a) Magnitude of the von Neumann symbols (amplification). (b) Imaginary component of the von Neumann symbols (dispersion).}
    \label{fig:mixed_amplification_dispersion}
\end{figure}

The first two eigenvalues of the mixed scheme are:
\begin{equation}\label{eq:vn_symbols_mdiff_12}
    \lambda^m_{1,2} = 1 - (\sigma_a M)(2\skt^2 + i\sk) - \sigma_a\Big(\skt^2 \pm i\sqrt{\sk^2 - \skt^4}\Big)
\end{equation}
We can identify the convective second term proportional to $(\sigma_a M)$.
The acoustic third term is proportional to $(\sigma_a)$ but the diffusive term $\skt^2$ is half that of the standard upwind scheme, and the advective term has an additional contribution from the second term in the radicand.
The magnitudes and imaginary component of the eigenvalues $\lambda^m_{1,2}$, ignoring the convective term, are plotted in figure \ref{fig:mixed_amplification_dispersion} alongside the equivalent values for the standard upwind scheme.
For the low wavenumber range $0 \leq k < 2\arctan(2)\approx2.2$, the radicand is real and the scheme behaves similarly to the upwind scheme except with lower diffusion, evident in figure \ref{fig:mixed_amplification}, and an additional dispersion error, evident in figure \ref{fig:mixed_dispersion}.
For the high wavenumber range $2\arctan(2) \leq k \leq \pi$, the radicand is imaginary and the scheme is entirely diffusive.
Figure \ref{fig:mixed_amplification} shows that $|\lambda_{i}|$ bifurcates in the high wavenumber range.
As $\pi-k=k'\to0$, the eigenvalues $\lambda^m_{1,2}$ can be expressed as:
\begin{subequations}
\begin{align}
    \lambda^m_1 & \approx 1 - (\sigma_a)(2\skt^2) + \mathcal{O}(k'^2\sigma_a) \\
    \lambda^m_2 & \approx 1 - (\sigma_a)(2\ckt^2) + \mathcal{O}(k'^4\sigma_a) \approx 1 - \mathcal{O}(k'^2\sigma_a)
\end{align}
\end{subequations}
where $\ckt=\cos(k/2)$.
$\lambda^m_1$ approaches the pure diffusion scheme as $k\to\pi$, as does the upwind scheme.
To leading order, $\lambda^m_2$ approaches 1 independently of $\sigma_a$, indicating that there is a grid mode which is undamped by the mixed scheme.
This was shown by Dellacherie \cite{dellacherie_checkerboard_2009} by considering the time evolution of the energy of the grid mode using the mixed scheme.
The low wavenumber range is stable under the reduced CFL condition $0<\sigma_a<0.5$, which is the same as found by Dellacherie \cite{dellacherie_analysis_2010}, and Bruel et al. \cite{bruel_low_2019}.
The high wavenumber range is (linearly) stable under the CFL condition $0<\sigma_a<1$, although nonlinear interactions in the full scheme will excite the undamped grid mode in many circumstances.

Lastly, we return to the question of reducing the convective upwinding by a factor of M, as done by Thornber \cite{thornber_improved_2008} and O{\ss}wald et al. \cite{oswald_l2roe_2016} in the velocity equations.
The eigenvalue of the upwind scheme for the convective terms (\ref{eq:vn_symbols_3}) using a diffusion coefficient of $\alpha|U|/2$ instead of $|U|/2$ is:
\begin{equation}\label{eq:vn_symbols_reduced_upwind}
    \lambda^{\alpha} = 1 - (\sigma_a M)(2\alpha\skt^2 - i\sk)
\end{equation}
which is linearly stable under the CFL condition $0\leq (\sigma_a M) \leq \alpha$.
If $\alpha=M$ and $\sigma_a \approx 1$, then the upwind scheme with reduced diffusion is linearly stable.
This argument also justifies reducing the convective upwinding for every component, not just the velocity.
However, because we rely on $\sigma_{a}\approx1$ this argument does not hold for preconditioned schemes where $\sigma_u\approx1$ and $\sigma_a\approx1/M$.

\section{Numerical examples}\label{sec:examples}
We now show numerical examples which demonstrate the behaviours of the three low Mach number schemes.
The discrete equations for a cell-centred finite volume scheme (\ref{eq:fv_discrete_equations}) are solved using a first order scheme in both time and space.
The first order explicit Euler scheme is used for time integration.
Diagonal diffusion is used for all examples.
Following the discussion of section \ref{sec:off-diagonal}, the mixed scheme with full diffusion is unstable with the diffusion form (\ref{eq:WeissSmith_diffusion},\ref{eq:interface_delta_up}), and upper/lower triangular mixed diffusion was found by Potsdam et al \cite{potsdam_unsteady_2007} and Sachdev et al \cite{sachdev_improved_2012} to have the same behaviour as the diagonal scheme.
For the acoustic scheme diagonal diffusion is equivalent to standard upwinding (see appendix \ref{app:transonic_scaling}).
For the convective scheme almost no difference was seen between full or diagonal diffusion, so the diagonal diffusion results are shown for consistency with the other schemes.
If higher order reconstructions and time integration are used, higher resolution results are obtained, but the qualitative behaviour remains the same, although smaller differences between the fluxes are observed due to the reduction in the  diffusion.
All simulations are carried out with an ideal gas with ratio of specific heats $\gamma=1.4$ and specific gas constant $R=287.058$.

\subsection{One dimensional examples}
In one dimension, the only solutions which are compatible with the convective limit (\ref{eq:velocity_convective_timescale},\ref{eq:pressure_convective_timescale}) are trivial, with both constant pressure and velocity \cite{dellacherie_analysis_2010}.
As such, all flow variations are either acoustic or entropy waves, and one dimensional examples can be used to isolate the behaviour of the numerical fluxes for purely acoustic flow.

\subsubsection{Isolated soundwave}
\begin{figure}
    \centering
\begin{subfigure}[t]{0.49\textwidth}
    \centering
    \includegraphics[width=0.99\linewidth]{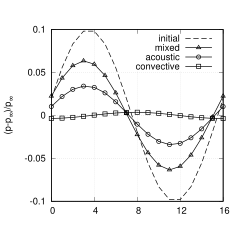}
    \caption{Isolated soundwave after travelling for a single period.}
    \label{fig:soundwave1D}
\end{subfigure}
\begin{subfigure}[t]{0.49\textwidth}
    \centering
    \includegraphics[width=0.99\linewidth]{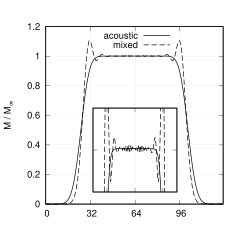}
    \caption{Low Mach number shocktube. Inset figure is a close-up of $16\leq x\leq 112$ and $0.96\leq y\leq 1.04$.}
    \label{fig:shockwave1D}
\end{subfigure}
    \caption{1D acoustic examples}
    \label{fig:acoustic_examples_1D}
\end{figure}

The first test case is an isolated soundwave in one dimension.
This test case will show the differences between the fluxes for a smooth purely acoustic flow.
The pressure $p(x,t)$ is initialised with a sinusoidal profile, the density $\rho(x,t)$ is initialised assuming isentropic flow, and the velocity $u(x,t)$ is initialised using the Riemann invariant for a forward travelling sound wave:
\begin{equation} \label{eq:acoustic_profile}
\begin{aligned}
    p(x,0) & = p_{\infty}( 1 + M_{\infty} g(x) ) \\
    \rho(x,0) & = \rho_{\infty}\bigg(\frac{p}{p_{\infty}}\bigg)^{\frac{1}{\gamma}} \\
    u(x,0) & = u_{\infty} + \frac{2(a-a_{\infty})}{\gamma-1}
\end{aligned}
\end{equation}
using the following conditions:
\begin{equation*}
    p_{\infty} = 1,
    \quad
    \rho_{\infty} = 1,
    \quad
    M_{\infty} = 0.01,
    \quad
    g(x) = c_0 sin(2\pi x/l),
    \quad
    c_0 = 0.1
\end{equation*}
Figure \ref{fig:soundwave1D} shows the non-dimensional gauge pressure distributions $(p(x,t)-p_{\infty})/p_{\infty}$ after one period using 16 points per wavelength and a CFL number $\sigma=0.125$, requiring $128$ timesteps for the acoustic and mixed fluxes, and $128/M=12,800$ timesteps for the convective flux.
The results for the acoustic and mixed schemes show the expected behaviour: the wave has travelled almost one wavelength and has been slightly diffused.
The slight dispersion dispersion error of $2-3\%$ for the acoustic and mixed schemes is consistent with what would be expected from the von Neumann analysis in equations (\ref{eq:vn_symbols_adiff_12}) and (\ref{eq:vn_symbols_mdiff_12}).
The acoustic scheme has diffused the pressure peaks roughly twice as much as the mixed scheme.
The convective scheme however has almost entirely smoothed out the wave, which has travelled only a small portion of a wavelength.
These results are entirely consistent with our earlier analysis which found that the acoustic and mixed schemes are both suitable for smooth acoustic flow, whereas the convective scheme is overly diffusive of acoustic variations.

\subsubsection{Low Mach number shocktube}
The second one dimensional case is a low Mach number shock tube.
This case demonstrates the performance of the different fluxes for discontinuous purely acoustic flow, and was first used by Sachdev et al. \cite{sachdev_improved_2012} for testing adaptive schemes.
The left and right initial states are:
\begin{equation*}
    p_L = 100028.04\text{Pa}, \quad p_R = 100000\text{Pa}, \quad u_{L/R} = 0, \quad T_{L/R} = 300\text{K}
\end{equation*}
The very small pressure difference produces a contact wave moving to the right at $M_{\infty}=0.0001$ between two receding weak shockwaves.
Figure \ref{fig:shockwave1D} shows a close-up of the Mach number distributions between the two shock waves obtained with the acoustic and the mixed flux schemes after $96$ timesteps with a CFL of $\sigma=0.4$.
The acoustic flux gives a monotone solution, as expected for a first order, upwind scheme.
However, the solution found with the mixed flux has significant oscillations originating at the shockwaves - an undamped grid mode as predicted by the von Neumann analysis.\\

From the one-dimensional examples we can verify that: the convective scheme is completely unsuitable for flow with acoustic variations; the acoustic scheme is suitable for flows with both smooth and discontinuous acoustic variations; and the mixed scheme is suitable for flows with smooth acoustic variations, but has too little diffusion to properly handle acoustic variations close to the grid scale.

\subsection{Two dimensional examples}
Non-trivial solutions to the convective limit exist in two and higher dimensions.
We present two numerical examples in 2D, one which tests the schemes' capabilities for steady purely convective flow, and one which tests the capabilities for unsteady mixed convective-acoustic flow.

\subsubsection{Circular cylinder}
This classic test case will demonstrate the performance of each flux scheme for steady purely convective flow.
The correct solutions should tend to the incompressible solution as $M\to0$.
The farfield state is:
\begin{equation*}
    \rho_{\infty} = 1, \quad u_{\infty} = 1, \quad M_{\infty}=0.01
\end{equation*}
The background pressure is calculated from $p_{\infty}=\rho_{\infty}/(\gamma M^2_{\infty})$.
The cylinder is centred at the origin with radius $r=1$, and is meshed using an O-mesh which extends out to 50 radii from the origin, with 64 and 48 cells in the circumferential and radial directions.
The first row of cells next to the cylinder has a radial height of $0.036r$. The cell height grows at a rate of 1.11, with the final row having a radial height of $4.9r$.
Curvature-corrected boundary conditions are used at the inviscid cylinder wall \cite{dadone_surface_1994}, and ghost cells at freestream boundaries are set using the upstream velocity and entropy and the downstream static pressure.
Convergence is reached by pseudo-timestepping at a CFL of $\sigma=0.4$, which is accelerated by local timestepping and preconditioning using Weiss \& Smith's preconditioning matrix \cite{weiss_preconditioning_1995} (preconditioning is only used for the convective and mixed schemes, as preconditioning does not reduce the spectral radius of the acoustic scheme \cite{godfrey_preconditioning_1993}).

\begin{figure}
    \centering
    \begin{subfigure}{0.49\textwidth}
        \centering
        \includegraphics[width=\textwidth]{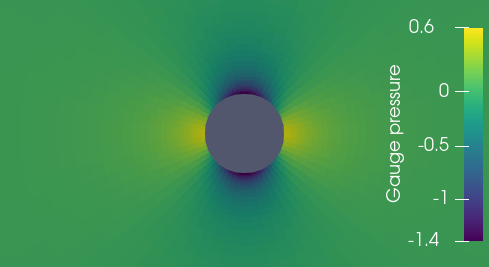}
        \caption{Exact solution}
        \label{fig:cylinder_exact}
    \end{subfigure}
    \hfill
    \begin{subfigure}{0.49\textwidth}
        \centering
        \includegraphics[width=\textwidth]{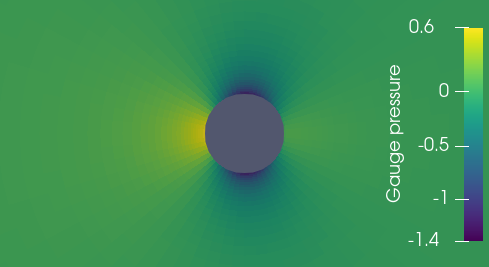}
        \caption{Convective scheme}
        \label{fig:cylinder_convective}
    \end{subfigure}
    \vskip\baselineskip
    \begin{subfigure}{0.49\textwidth}
        \centering
        \includegraphics[width=\textwidth]{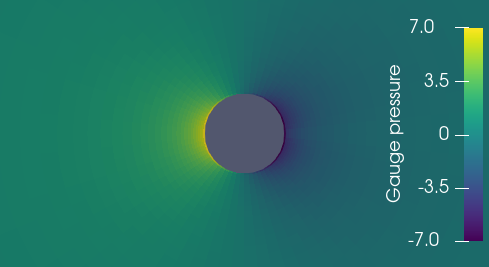}
        \caption{Acoustic scheme}
        \label{fig:cylinder_acoustic}
    \end{subfigure}
    \hfill
    \begin{subfigure}{0.49\textwidth}
        \centering
        \includegraphics[width=\textwidth]{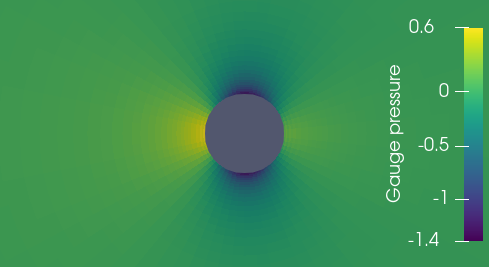}
        \caption{Mixed scheme}
        \label{fig:cylinder_mixed}
    \end{subfigure}
    \caption{Gauge pressure around a 2D cylinder at $M=0.01$ using the various flux scalings.}
    \label{fig:cylinder}
\end{figure}

\begin{figure}
    \centering
\begin{subfigure}[t]{0.45\textwidth}
    \centering
    \includegraphics[width=0.99\linewidth]{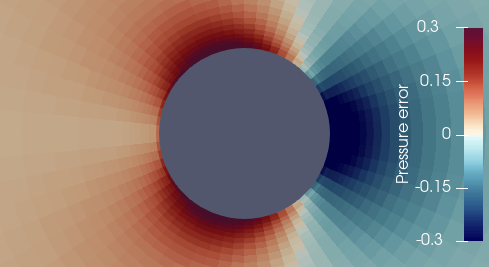}
    \caption{Convective scheme}
    \label{fig:cylinder_convective_error}
\end{subfigure}
\begin{subfigure}[t]{0.45\textwidth}
    \centering
    \includegraphics[width=0.99\linewidth]{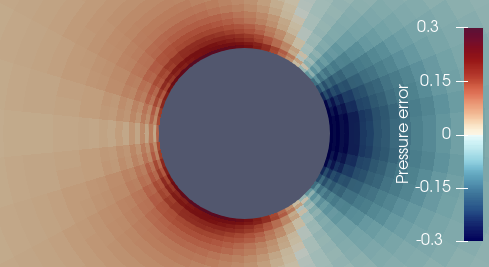}
    \caption{Mixed scheme}
    \label{fig:cylinder_mixed_error}
\end{subfigure}
    \caption{Gauge pressure error around a 2D cylinder for the convective and mixed flux scalings.}
    \label{fig:cylinder_error}
\end{figure}

The results for the convective flux is shown in figure \ref{fig:cylinder_convective}, and closely resemble the exact pressure distribution from the classical incompressible potential solution in figure \ref{fig:cylinder_exact}.
The solution found with the acoustic flux is shown in figure \ref{fig:cylinder_acoustic}, and is visibly very different from the incompressible solution.
The artificial diffusion dominates over the other terms, and the flow more closely resembles a Stokes flow than the inviscid solution, with significantly larger pressure variations.
Lastly, the solution found with the mixed flux is shown in figure \ref{fig:cylinder_mixed} and is very similar to the convective scheme and the incompressible solution.
However, on close inspection there is a small chequer-board mode in the radial direction.
This is more easily seen in figure \ref{fig:cylinder_error} which shows the difference between the exact pressure and the pressure found with the mixed and convective flux schemes.
The error for the convective flux is smooth with no chequer-board modes, however the error for the mixed flux clearly shows chequer-board modes in the radial direction.

\subsubsection{Soundwave-vortex interaction}
The last numerical example is the interaction of a one-dimensional soundwave passing through a low Mach number Gresho vortex \cite{gresho_theory_1990,liska_comparison_2003} with stationary background conditions.
This provides a very simple test of a numerical scheme's ability to not only simulate both convective and acoustic features, but also their interaction, and has been used previously by \cite{bruel_low_2019,miczek_new_2015}.

\begin{figure}
    \centering
    \begin{subfigure}{0.49\textwidth}
        \centering
        \includegraphics[width=\textwidth]{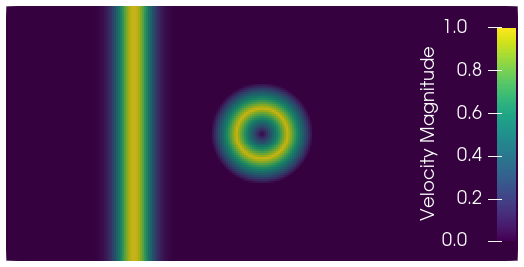}
        \caption{Initial condition}
        \label{fig:gresho_dtau_initial}
    \end{subfigure}
    \hfill
    \begin{subfigure}{0.49\textwidth}
        \centering
        \includegraphics[width=\textwidth]{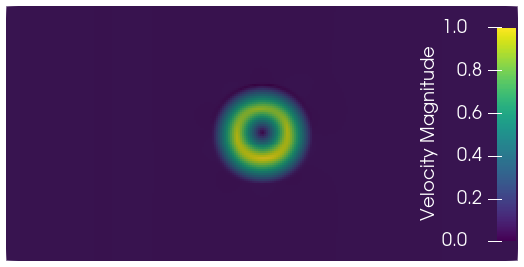}
        \caption{Convective scheme}
        \label{fig:gresho_dtau_convective}
    \end{subfigure}
    \vskip\baselineskip
    \begin{subfigure}{0.49\textwidth}
        \centering
        \includegraphics[width=\textwidth]{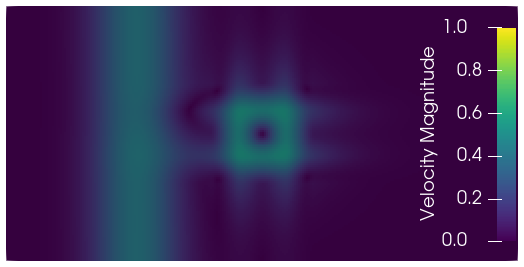}
        \caption{Acoustic scheme}
        \label{fig:gresho_dtau_acoustic}
    \end{subfigure}
    \hfill
    \begin{subfigure}{0.49\textwidth}
        \centering
        \includegraphics[width=\textwidth]{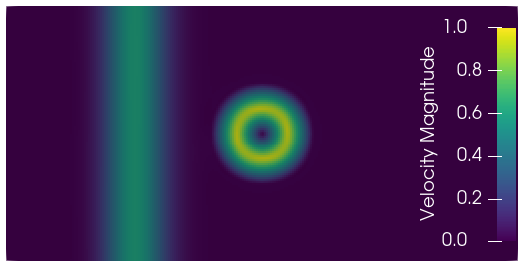}
        \caption{Mixed scheme}
        \label{fig:gresho_dtau_mixed}
    \end{subfigure}
    \caption{Velocity profiles for an acoustic wave and Gresho vortex after 1 acoustic period $\tau=l/a$ with various flux schemes.}
    \label{fig:gresho_dtau}
\end{figure}

The domain is $(x,y) \in [-2h,2h)\times[-h,h)$ with periodic boundary conditions in both dimensions, and is discretised with $256\times128$ square cells.
The initial conditions are shown in figure \ref{fig:gresho_dtau_initial}, composed of the superposition of a Gresho vortex centred at $(0,0)$ with a diameter $0.4h$, and a right-travelling soundwave with a Gaussian profile centred along the line $(-h,y)$.
The soundwave profile is calculated according to equations (\ref{eq:acoustic_profile}), with:
\begin{equation*}
    \rho_{\infty} = 1,
    \quad
    M_{\infty} = 0.01,
    \quad
    u_{\infty} = 0,
    \quad
    g(x) = c_1 G(x+h)
\end{equation*}
Where $G(x)$ is the Gaussian function with standard deviation of $0.1h$, $c_1$ is chosen such that the maximum acoustic velocity is $1$, and $p_{\infty} = \rho_{\infty}/(\gamma M^2_{\infty})$ again.
The Gresho vortex has circumferential velocity $u_{\theta}(r)$ and pressure $p(r)$ according to \cite{miczek_new_2015}:
\begin{equation}
    u_{\theta}(r) =
    \begin{cases}
        5r,   & 0   \leq r < 0.2 \\
        2-5r, & 0.2 \leq r < 0.4 \\
        0,    & 0.4 \leq r
    \end{cases}
\end{equation}
\begin{equation}
    p(r) =
    \begin{cases}
        p_0 + \frac{25}{2}r^2,                                     & 0   \leq r < 0.2 \\
        p_0 + \frac{25}{2}r^2 + 4(1-5r-\text{ln}0.2 + \text{ln}r), & 0.2 \leq r < 0.4 \\
        p_0 - 2 + 4\text{ln}2,                                     & 0.4 \leq r
    \end{cases}
\end{equation}
with $p_0 = p_{\infty} + 2 - 4\text{ln}2$.

\begin{figure}
    \centering
    \begin{subfigure}{0.49\textwidth}
        \centering
        \includegraphics[width=\textwidth]{roe2D/soundwave_vortex/initial}
        \caption{Initial condition}
        \label{fig:gresho_dt_initial}
    \end{subfigure}
    \hfill
    \begin{subfigure}{0.49\textwidth}
        \centering
        \includegraphics[width=\textwidth]{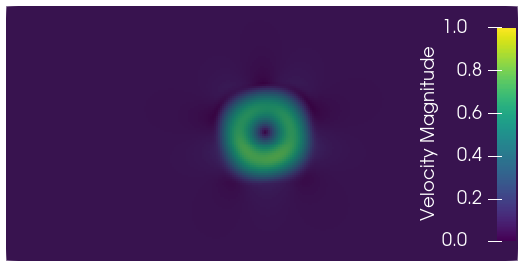}
        \caption{Convective scheme}
        \label{fig:gresho_dt_convective}
    \end{subfigure}
    \vskip\baselineskip
    \begin{subfigure}{0.49\textwidth}
        \centering
        \includegraphics[width=\textwidth]{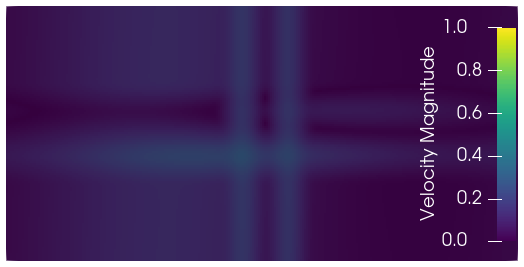}
        \caption{Acoustic scheme}
        \label{fig:gresho_dt_acoustic}
    \end{subfigure}
    \hfill
    \begin{subfigure}{0.49\textwidth}
        \centering
        \includegraphics[width=\textwidth]{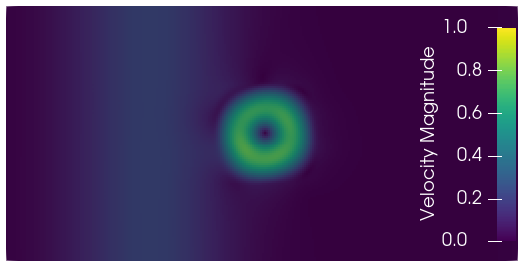}
        \caption{Mixed scheme}
        \label{fig:gresho_dt_mixed}
    \end{subfigure}
    \caption{Velocity profiles for an acoustic wave and Gresho vortex after 10 acoustic periods (0.1 convective time periods $l/u$) with various flux schemes.}
    \label{fig:gresho_dt}
\end{figure}

The results for each flux after a single acoustic period ($4h/a_{\infty}$) and after ten acoustic periods (equivalent to 1/10th of a convective period $4h/(a_{\infty}M_{\infty})$) are shown in figures \ref{fig:gresho_dtau} and \ref{fig:gresho_dt} respectively, using a CFL of $\sigma=0.4$.

Figure \ref{fig:gresho_dtau_convective} shows the solution found with the convective scheme after one acoustic time.
The vortex shape is very well preserved, with barely any reduction in peak velocity.
The soundwave on the other hand has been completely dissipated and is no longer visible.
The only indication of the soundwave is the slight asymmetry in the vortex velocity over the $y$ axis, which happens as the soundwave is smeared out over the domain.

The solution from the acoustic scheme is shown in figure \ref{fig:gresho_dtau_acoustic}.
The vortex has a drastically reduced peak velocity, and is misshapen, having become aligned with the grid due to the anisotropic artificial diffusion on the face-normal velocity.
The soundwave is still visible, having travelled once around the domain.
It is also damped, although not to an unsurprising degree for a first order upwind scheme, given there were fewer than $20$ cells across the initial width of the soundwave.

As predicted by the asymptotic analysis, the results for the mixed scheme (figure \ref{fig:gresho_dtau_mixed}) combine the favourable behaviour of both the convective and acoustic schemes.
Both the vortex and the soundwave are well preserved.
The vortex shows comparable diffusion to the convective scheme (although without the asymmetry), and the soundwave is less damped than that found with the acoustic scheme, in agreement with the von Neumann analysis in the previous section.\\

The results for the convective scheme after 10 acoustic periods (0.1 convective periods) are shown in \ref{fig:gresho_dt_convective}.
The vortex shape is mostly preserved with only minimal distortion around the corners, where the flow is most misaligned with the grid.
The peak vortex velocity is around $75\%$ of the initial peak value.
The acoustic scheme results are shown in figure \ref{fig:gresho_dt_acoustic}.
The vortex is even more distorted than in figure \ref{fig:gresho_dtau_acoustic}, and has polluted a large part of the domain.
After integration over 10 periods, the soundwave is barely visible on the left side of the figure behind this distortion.
The result for the mixed scheme again retains the favourable properties of the other two schemes (figure \ref{fig:gresho_dt_mixed}).
The velocity magnitude of the vortex is similar to that found with the convective scheme, but with marginally less distortion to the shape.
The soundwave is almost completely diffused, although less so than with the acoustic scheme.\\

\begin{figure}
    \centering
    \begin{subfigure}{0.49\textwidth}
        \centering
        \includegraphics[width=\textwidth]{roe2D/soundwave_vortex/mixed_dt}
        \caption{Original mixed scheme}
        \label{fig:gresho_reduced_mixed}
    \end{subfigure}
    \hfill
    \begin{subfigure}{0.49\textwidth}
        \centering
        \includegraphics[width=\textwidth]{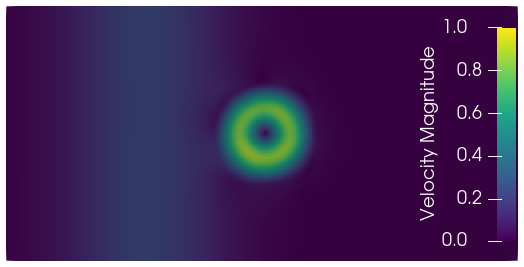}
        \caption{Reduced upwinding}
        \label{fig:gresho_reduced_mU}
    \end{subfigure}
    \vskip\baselineskip
    \begin{subfigure}{0.49\textwidth}
        \centering
        \includegraphics[width=\textwidth]{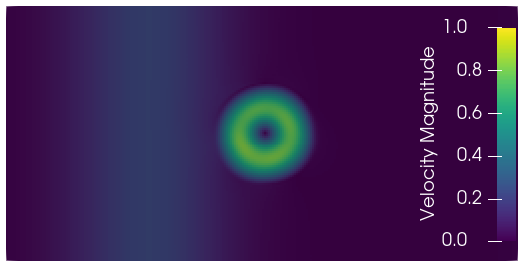}
        \caption{Reduced $\nu_{22}$}
        \label{fig:gresho_reduced_mnu22}
    \end{subfigure}
    \hfill
    \begin{subfigure}{0.49\textwidth}
        \centering
        \includegraphics[width=\textwidth]{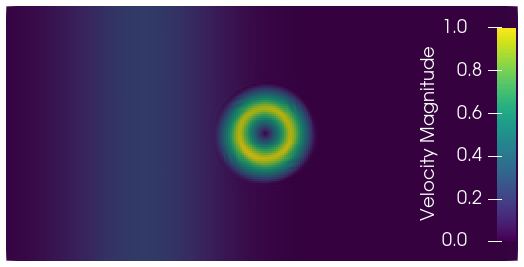}
        \caption{Reduced upwinding and $\nu_{22}$}
        \label{fig:gresho_reduced_mUnu22}
    \end{subfigure}
    \caption{Velocity profiles for an acoustic wave and Gresho vortex after 10 acoustic times $\tau$ with the mixed flux scheme, with various diffusion terms reduced by a factor of $M$.}
    \label{fig:gresho_reduced}
\end{figure}

We now investigate the effect of reduced convective upwinding, as proposed in \cite{thornber_improved_2008,rieper_low-mach_2011,oswald_l2roe_2016}.
Figure \ref{fig:gresho_reduced_mU} shows the solution after 10 acoustic times found with the mixed scheme where the convective upwinding - the first term in (\ref{eq:fv-entropyvar-diffusion}) - has been reduced by a factor of $M$.
The peak vortex velocity is 87\% of the initial value compared to 75\% with the original mixed scheme (figure \ref{fig:gresho_reduced_mixed}), a moderate improvement.
Inspection of the diffusion matrix (\ref{eq:fv-entropyvar-diffusion}) shows that for the mixed scheme the $\mu_{22}$ term is of the same magnitude as the convective upwind term.
Figure \ref{fig:gresho_reduced_mnu22} shows the solution found by reducing only the $\mu_{22}$ term by a factor of $M$.
The peak velocity is 82\% of the initial value, a comparable improvement to reducing the convective upwind term.
Figure \ref{fig:gresho_reduced_mUnu22} shows the results found by reducing both the convective upwinding and $\mu_{22}$ terms by a factor of $M$ compared to the original mixed scheme.
The improvement is striking, with a peak vortex velocity of 99\% of the initial value.
Some small instabilities can be seen around the corners of the vortex, where the flow is most misaligned with the grid and the 1D von Neumann analysis of section \ref{sec:discrete} is least valid.
Although figure \ref{fig:gresho_reduced_mUnu22} shows a significant reduction in the diffusion without significant loss of stability, recall from the von Neumann analysis that the stability of this scheme relies on $\sigma_{a}\sim\mathcal{O}(1)$, making it unsuitable for preconditioned schemes where $\sigma_{u}\sim\mathcal{O}(1)$.

The convective upwinding reductions in \cite{rieper_low-mach_2011,thornber_improved_2008,oswald_l2roe_2016} do not exactly correspond to those discussed here.
Rieper \cite{rieper_low-mach_2011} reduces the jump in the face-normal velocity component by a factor of $M$, corresponding to reducing the $\mu_{22}$ term from the acoustic scheme scaling to the mixed scheme scaling, and reducing by a factor of $M$ the convective upwinding on the face-normal component of velocity only.
Thornber et al. \cite{thornber_improved_2008} and O{\ss}wald et al. \cite{oswald_l2roe_2016} reduce the jumps in all velocity components by a factor of $M$, so for vortical flows are equivalent to the reduced convective upwinding scheme in figure \ref{fig:gresho_reduced_mU}.
However, reducing the entire convective upwinding term also reduces the diffusion on the entropy convection, so can be expected to give better results for cases with significant entropy variations such as heat transfer simulations.\\

In this section we have used numerical examples to verify the findings of sections \ref{sec:design}, \ref{sec:continuous} and \ref{sec:discrete}.
The one dimensional examples demonstrated that the acoustic scheme is well suited to purely acoustic flow, while the convective scheme is unable to resolve any acoustic phenomena.
The mixed scheme is suitable for smooth acoustic flow - resolving these phenomena with less diffusion than the acoustic scheme - although is unstable for discontinuous acoustic flow, possessing an undamped grid mode.
Using steady flow around a two dimensional cylinder, we verified that the acoustic scheme is unsuitable for purely convective flow, whilst the convective scheme approaches the incompressible limit as $M\to0$.
The mixed scheme also approaches the incompressible limit, avoiding the catastrophic failure of the acoustic scheme, however we saw it is susceptible to chequerboard modes on the second order convective pressure.
Finally we saw that neither the acoustic nor convective scheme is suitable for mixed convective-acoustic flow.
The mixed scheme on the other hand can resolve both convective and acoustic features in the same flow, and can also achieve this with dramatically reduced convective upwind diffusion.

\section{Conclusions}\label{sec:conclusions}
Low Mach number flows include a range of applications of scientific and engineering interest.
Collocated, density based solvers for compressible flow are one method of simulating low Mach number flow, but care must be taken to ensure that the artificial diffusion does not compromise their accuracy in this regime.
In this paper we have reviewed the behaviour of the artificial diffusion in this class of schemes at low Mach number using the modified equations.
By considering both the convective and acoustic low Mach number limits, we have shown how three artificial diffusion scalings naturally arise in the entropy variables - one suitable for purely convective flow, one for purely acoustic flow, and one for flow with both convective and acoustic features.
Single- and multiple-scale asymptotic expansions of the modified equations established the behaviours of each scheme for different flows.
The convective scaling is compatible with convective flows, but damps out acoustic variations on a very fast timescale.
The acoustic scaling is compatible with acoustic flows, but will produce spurious pressure waves for convective flow.
The mixed scaling is compatible with both convective, acoustic, and mixed flow, but has vanishing diffusion on the convective pressure and the acoustic velocity, which may lead to pressure chequerboard modes and acoustic grid-mode instabilities respectively.

Transforming the artificial diffusion to a Roe-type finite volume scheme in the conserved variables enabled us to compare to a number of existing low Mach number methods.
Each of these methods matched one of the three scalings, and our analysis agrees with previous theoretical analyses and well-known behaviours of these existing methods.
The convective and mixed scalings conform with previous guidelines for accurate schemes for convective low Mach number flow \cite{dellacherie_analysis_2010,li_mechanism_2013}, but by considering accuracy requirements for both convective and acoustic effects, we were able to explain why there are two possible scalings suitable for the convective limit, and what their relative advantages and disadvantages are.
The price of the mixed scheme's flexibility is its compromised stability and the additional constraints on the off-diagonal diffusion.
It is the authors' belief that remedying the stability of this scheme, and in particular whether the off-diagonal terms can be leveraged to achieve this, is the most pertinent open question in this research area.
Asymptotic expansion of the discrete equations showed that all findings of the continuous analysis apply for the first order Roe-type finite volume scheme, and von Neumann analysis confirmed the stability estimates obtained from the continuous analysis.
Finally, four numerical examples demonstrated the performance of each diffusion scaling for acoustic, convective and mixed convective-acoustic flow.

There is a significant body of literature investigating this class of schemes in the limit of vanishing Mach number.
We have shown that the most important behaviours of this class of schemes at low Mach number can be found and explained in a simple manner using the continuous modified equations in the entropy variables.
This form can be used to compare schemes and predict their capabilities independently of the specific discretisation, as well as to provide guidelines for the development of novel low-Mach schemes.

\section*{Competing interest}
The authors declare that they have no known competing financial interests or personal relationships that could have appeared to influence the work reported in this paper.

\section*{Acknowledgements}
The authors gratefully acknowledge support from the EPSRC Center for Doctoral Training in Gas Turbine Aerodynamics and Rolls-Royce plc.
We also thank the anonymous reviewers whose feedback significantly improved the manuscript, particularly in the treatment of the off-diagonal diffusion terms.

\printbibliography

@article{turkel_preconditioning_1999,
	title = {Preconditioning {Techniques} in {Computational} {Fluid} {Dynamics}},
	volume = {31},
	url = {https://doi.org/10.1146/annurev.fluid.31.1.385},
	doi = {10.1146/annurev.fluid.31.1.385},
	number = {1},
	urldate = {2021-04-08},
	journal = {Annual Review of Fluid Mechanics},
	author = {Turkel, E.},
	year = {1999},
	note = {\_eprint: https://doi.org/10.1146/annurev.fluid.31.1.385},
	pages = {385--416},
}

@article{birken_stability_2005,
	title = {Stability of {Preconditioned} {Finite} {Volume} {Schemes} at {Low} {Mach} {Numbers}},
	volume = {45},
	issn = {1572-9125},
	url = {https://doi.org/10.1007/s10543-005-0009-0},
	doi = {10.1007/s10543-005-0009-0},
	language = {en},
	number = {3},
	urldate = {2021-04-08},
	journal = {BIT Numerical Mathematics},
	author = {Birken, P. and Meister, A.},
	month = sep,
	year = {2005},
	pages = {463--480},
}

@incollection{caraeni_unsteady_2017,
	title = {Unsteady {Low}-{Mach} {Preconditioning} for {Roe} {Flux}-{Differencing} {Scheme}},
	url = {https://arc.aiaa.org/doi/abs/10.2514/6.2017-4402},
	urldate = {2021-04-08},
	booktitle = {23rd {AIAA} {Computational} {Fluid} {Dynamics} {Conference}},
	publisher = {American Institute of Aeronautics and Astronautics},
	author = {Caraeni, Doru A. and Weiss, Jonathan M.},
	year = {2017},
	doi = {10.2514/6.2017-4402},
	note = {\_eprint: https://arc.aiaa.org/doi/pdf/10.2514/6.2017-4402},
}

@article{dadone_surface_1994,
	title = {Surface boundary conditions for the numerical solution of the {Euler} equations},
	volume = {32},
	issn = {0001-1452},
	url = {https://doi.org/10.2514/3.11983},
	doi = {10.2514/3.11983},
	number = {2},
	urldate = {2021-04-08},
	journal = {AIAA Journal},
	author = {Dadone, A. and Grossman, B.},
	year = {1994},
	note = {Publisher: American Institute of Aeronautics and Astronautics
\_eprint: https://doi.org/10.2514/3.11983},
	pages = {285--293},
}

@article{guillard_behaviour_1999,
	title = {On the behaviour of upwind schemes in the low {Mach} number limit},
	volume = {28},
	issn = {0045-7930},
	url = {https://www.sciencedirect.com/science/article/pii/S0045793098000176},
	doi = {10.1016/S0045-7930(98)00017-6},
	language = {en},
	number = {1},
	urldate = {2021-04-08},
	journal = {Computers \& Fluids},
	author = {Guillard, Hervé and Viozat, Cécile},
	month = jan,
	year = {1999},
	pages = {63--86},
}

@incollection{guillard_chapter_2017,
	series = {Handbook of {Numerical} {Methods} for {Hyperbolic} {Problems}},
	title = {Chapter 8 - {On} the {Behaviour} of {Upwind} {Schemes} in the {Low} {Mach} {Number} {Limit}: {A} {Review}},
	volume = {18},
	shorttitle = {Chapter 8 - {On} the {Behaviour} of {Upwind} {Schemes} in the {Low} {Mach} {Number} {Limit}},
	url = {https://www.sciencedirect.com/science/article/pii/S1570865916300114},
	language = {en},
	urldate = {2021-04-08},
	booktitle = {Handbook of {Numerical} {Analysis}},
	publisher = {Elsevier},
	author = {Guillard, H. and Nkonga, B.},
	editor = {Abgrall, Rémi and Shu, Chi-Wang},
	month = jan,
	year = {2017},
	doi = {10.1016/bs.hna.2016.09.002},
	pages = {203--231},
}

@article{li_mechanism_2013,
	title = {Mechanism of {Roe}-type schemes for all-speed flows and its application},
	volume = {86},
	issn = {0045-7930},
	url = {https://www.sciencedirect.com/science/article/pii/S0045793013002752},
	doi = {10.1016/j.compfluid.2013.07.004},
	language = {en},
	urldate = {2021-04-08},
	journal = {Computers \& Fluids},
	author = {Li, Xue-song and Gu, Chun-wei},
	month = nov,
	year = {2013},
	pages = {56--70},
}

@incollection{muller_low_1999,
	title = {Low {Mach} {Number} {Asymptotics} of the {Navier}-{Stokes} {Equations} and {Numerical} {Implications}},
	booktitle = {30th {Computational} {Fluid} {Dynamics} {Lecture} {Series}},
	publisher = {von Karman Institute for Fluid Dynamics},
	author = {Müller, B},
	year = {1999},
}

@incollection{potsdam_unsteady_2007,
	title = {Unsteady {Low} {Mach} {Preconditioning} with {Application} to {Rotorcraft} {Flows}},
	url = {https://arc.aiaa.org/doi/abs/10.2514/6.2007-4473},
	urldate = {2021-04-08},
	booktitle = {18th {AIAA} {Computational} {Fluid} {Dynamics} {Conference}},
	publisher = {American Institute of Aeronautics and Astronautics},
	author = {Potsdam, Mark and Sankaran, Venkateswaran and Pandya, Shishir},
	year = {2007},
	doi = {10.2514/6.2007-4473},
	note = {\_eprint: https://arc.aiaa.org/doi/pdf/10.2514/6.2007-4473},
}

@incollection{sachdev_improved_2012,
	series = {Fluid {Dynamics} and {Co}-located {Conferences}},
	title = {Improved {Flux} {Formulations} for {Unsteady} {Low} {Mach} {Number} {Flows}},
	url = {https://arc.aiaa.org/doi/10.2514/6.2012-3067},
	urldate = {2021-04-08},
	booktitle = {42nd {AIAA} {Fluid} {Dynamics} {Conference} and {Exhibit}},
	publisher = {American Institute of Aeronautics and Astronautics},
	author = {Sachdev, Jai and Hosangadi, Ashvin and Sankaran, V.},
	month = jun,
	year = {2012},
	doi = {10.2514/6.2012-3067},
}

@incollection{venkateswaran_artficial_2003,
	title = {Artficial {Dissipation} {Control} for {Viscous} and {Unsteady} {Computations}},
	url = {https://arc.aiaa.org/doi/abs/10.2514/6.2003-3695},
	urldate = {2021-04-08},
	booktitle = {16th {AIAA} {Computational} {Fluid} {Dynamics} {Conference}},
	publisher = {American Institute of Aeronautics and Astronautics},
	author = {Venkateswaran, Sankaran and Merkle, Charles},
	year = {2003},
	doi = {10.2514/6.2003-3695},
	note = {\_eprint: https://arc.aiaa.org/doi/pdf/10.2514/6.2003-3695},
}

@article{shima_new_2013,
	series = {International {Workshop} on {Future} of {CFD} and {Aerospace} {Sciences}},
	title = {New approaches for computation of low {Mach} number flows},
	volume = {85},
	issn = {0045-7930},
	url = {https://www.sciencedirect.com/science/article/pii/S0045793012004495},
	doi = {10.1016/j.compfluid.2012.11.017},
	language = {en},
	urldate = {2021-04-08},
	journal = {Computers \& Fluids},
	author = {Shima, Eiji and Kitamura, Keiichi},
	month = oct,
	year = {2013},
	pages = {143--152},
}

@article{shima_parameter-free_2011,
	title = {Parameter-{Free} {Simple} {Low}-{Dissipation} {AUSM}-{Family} {Scheme} for {All} {Speeds}},
	volume = {49},
	issn = {0001-1452},
	url = {https://arc.aiaa.org/doi/10.2514/1.J050905},
	doi = {10.2514/1.J050905},
	number = {8},
	urldate = {2021-04-08},
	journal = {AIAA Journal},
	author = {Shima, Eiji and Kitamura, Keiichi},
	month = aug,
	year = {2011},
	note = {Publisher: American Institute of Aeronautics and Astronautics},
	pages = {1693--1709},
}

@incollection{venkateswaran_dual_1995,
	series = {Aerospace {Sciences} {Meetings}},
	title = {Dual time-stepping and preconditioning for unsteady computations},
	url = {https://arc.aiaa.org/doi/10.2514/6.1995-78},
	urldate = {2021-04-08},
	booktitle = {33rd {Aerospace} {Sciences} {Meeting} and {Exhibit}},
	publisher = {American Institute of Aeronautics and Astronautics},
	author = {Venkateswaran, S and Merkle, Charles},
	month = jan,
	year = {1995},
	doi = {10.2514/6.1995-78},
}

@article{weiss_preconditioning_1995,
	title = {Preconditioning applied to variable and constant density flows},
	volume = {33},
	issn = {0001-1452},
	url = {https://arc.aiaa.org/doi/10.2514/3.12946},
	doi = {10.2514/3.12946},
	number = {11},
	urldate = {2021-04-08},
	journal = {AIAA Journal},
	author = {Weiss, Jonathan M. and Smith, Wayne A.},
	month = nov,
	year = {1995},
	note = {Publisher: American Institute of Aeronautics and Astronautics},
	pages = {2050--2057},
}

@article{liou_new_1993,
	title = {A {New} {Flux} {Splitting} {Scheme}},
	volume = {107},
	issn = {0021-9991},
	url = {https://www.sciencedirect.com/science/article/pii/S0021999183711228},
	doi = {10.1006/jcph.1993.1122},
	language = {en},
	number = {1},
	urldate = {2021-04-08},
	journal = {Journal of Computational Physics},
	author = {Liou, Meng-Sing and Steffen, Christopher J.},
	month = jul,
	year = {1993},
	pages = {23--39},
}

@article{liou_sequel_1996,
	title = {A {Sequel} to {AUSM}: {AUSM}+},
	volume = {129},
	issn = {0021-9991},
	shorttitle = {A {Sequel} to {AUSM}},
	url = {https://www.sciencedirect.com/science/article/pii/S0021999196902569},
	doi = {10.1006/jcph.1996.0256},
	language = {en},
	number = {2},
	urldate = {2021-04-08},
	journal = {Journal of Computational Physics},
	author = {Liou, Meng-Sing},
	month = dec,
	year = {1996},
	pages = {364--382},
}

@article{liou_sequel_2006,
	title = {A sequel to {AUSM}, {Part} {II}: {AUSM}+-up for all speeds},
	volume = {214},
	issn = {0021-9991},
	shorttitle = {A sequel to {AUSM}, {Part} {II}},
	url = {https://www.sciencedirect.com/science/article/pii/S0021999105004274},
	doi = {10.1016/j.jcp.2005.09.020},
	language = {en},
	number = {1},
	urldate = {2021-04-08},
	journal = {Journal of Computational Physics},
	author = {Liou, Meng-Sing},
	month = may,
	year = {2006},
	pages = {137--170},
}

@article{choi_application_1993,
	title = {The {Application} of {Preconditioning} in {Viscous} {Flows}},
	volume = {105},
	issn = {0021-9991},
	url = {https://www.sciencedirect.com/science/article/pii/S0021999183710697},
	doi = {10.1006/jcph.1993.1069},
	language = {en},
	number = {2},
	urldate = {2021-04-08},
	journal = {Journal of Computational Physics},
	author = {Choi, Y. -H. and Merkle, C. L.},
	month = apr,
	year = {1993},
	pages = {207--223},
}

@article{kitamura_reduced_2016,
	title = {Reduced dissipation {AUSM}-family fluxes: {HR}-{SLAU2} and {HR}-{AUSM}+-up for high resolution unsteady flow simulations},
	volume = {126},
	issn = {0045-7930},
	shorttitle = {Reduced dissipation {AUSM}-family fluxes},
	url = {https://www.sciencedirect.com/science/article/pii/S0045793015003850},
	doi = {10.1016/j.compfluid.2015.11.014},
	language = {en},
	urldate = {2021-04-08},
	journal = {Computers \& Fluids},
	author = {Kitamura, Keiichi and Hashimoto, Atsushi},
	month = mar,
	year = {2016},
	pages = {41--57},
}

@article{chorin_numerical_1967,
	title = {A numerical method for solving incompressible viscous flow problems},
	volume = {2},
	issn = {0021-9991},
	url = {https://www.sciencedirect.com/science/article/pii/002199916790037X},
	doi = {10.1016/0021-9991(67)90037-X},
	language = {en},
	number = {1},
	urldate = {2021-04-08},
	journal = {Journal of Computational Physics},
	author = {Chorin, Alexandre Joel},
	month = aug,
	year = {1967},
	pages = {12--26},
}

@article{guillard_behavior_2004,
	title = {On the behavior of upwind schemes in the low {Mach} number limit: {II}. {Godunov} type schemes},
	volume = {33},
	issn = {0045-7930},
	shorttitle = {On the behavior of upwind schemes in the low {Mach} number limit},
	url = {https://www.sciencedirect.com/science/article/pii/S0045793003000781},
	doi = {10.1016/j.compfluid.2003.07.001},
	language = {en},
	number = {4},
	urldate = {2021-04-08},
	journal = {Computers \& Fluids},
	author = {Guillard, Hervé and Murrone, Angelo},
	month = may,
	year = {2004},
	pages = {655--675},
}

@inproceedings{venkateswaran_evaluation_1998,
	address = {Berlin, Heidelberg},
	series = {Lecture {Notes} in {Physics}},
	title = {Evaluation of artificial dissipation models and their relationship to the accuracy of {Euler} and {Navier}-{Stokes} computations},
	isbn = {978-3-540-49540-6},
	doi = {10.1007/BFb0106619},
	language = {en},
	booktitle = {Sixteenth {International} {Conference} on {Numerical} {Methods} in {Fluid} {Dynamics}},
	publisher = {Springer},
	author = {Venkateswaran, S. and Merkle, C. L.},
	editor = {Bruneau, Charles-Henri},
	year = {1998},
	pages = {427--432},
}

@article{turkel_review_1993,
	series = {{SPECIAL} {ISSUE}},
	title = {Review of preconditioning methods for fluid dynamics},
	volume = {12},
	issn = {0168-9274},
	url = {https://www.sciencedirect.com/science/article/pii/0168927493901228},
	doi = {10.1016/0168-9274(93)90122-8},
	language = {en},
	number = {1},
	urldate = {2021-04-09},
	journal = {Applied Numerical Mathematics},
	author = {Turkel, E.},
	month = may,
	year = {1993},
	pages = {257--284},
}

@article{bruel_low_2019,
	title = {A low {Mach} correction able to deal with low {Mach} acoustics},
	volume = {378},
	issn = {0021-9991},
	url = {https://www.sciencedirect.com/science/article/pii/S0021999118307460},
	doi = {10.1016/j.jcp.2018.11.020},
	language = {en},
	urldate = {2021-04-12},
	journal = {Journal of Computational Physics},
	author = {Bruel, Pascal and Delmas, Simon and Jung, Jonathan and Perrier, Vincent},
	month = feb,
	year = {2019},
	pages = {723--759},
}

@article{barsukow_truly_2021,
	title = {Truly multi-dimensional all-speed schemes for the {Euler} equations on {Cartesian} grids},
	volume = {435},
	issn = {0021-9991},
	url = {https://www.sciencedirect.com/science/article/pii/S002199912100111X},
	doi = {10.1016/j.jcp.2021.110216},
	language = {en},
	urldate = {2021-04-12},
	journal = {Journal of Computational Physics},
	author = {Barsukow, Wasilij},
	month = jun,
	year = {2021},
	pages = {110216},
}

@article{lin_density_2018,
	title = {Density enhancement mechanism of upwind schemes for low {Mach} number flows},
	volume = {34},
	issn = {1614-3116},
	url = {https://doi.org/10.1007/s10409-017-0737-9},
	doi = {10.1007/s10409-017-0737-9},
	language = {en},
	number = {3},
	urldate = {2021-04-12},
	journal = {Acta Mechanica Sinica},
	author = {Lin, Bo-Xi and Yan, Chao and Chen, Shu-Sheng},
	month = jun,
	year = {2018},
	pages = {431--445},
}

@article{hughes_new_1986,
	title = {A new finite element formulation for computational fluid dynamics: {V}. {Circumventing} the babuška-brezzi condition: a stable {Petrov}-{Galerkin} formulation of the stokes problem accommodating equal-order interpolations},
	volume = {59},
	issn = {0045-7825},
	shorttitle = {A new finite element formulation for computational fluid dynamics},
	url = {https://www.sciencedirect.com/science/article/pii/0045782586900253},
	doi = {10.1016/0045-7825(86)90025-3},
	language = {en},
	number = {1},
	urldate = {2021-04-20},
	journal = {Computer Methods in Applied Mechanics and Engineering},
	author = {Hughes, Thomas J. R. and Franca, Leopoldo P. and Balestra, Marc},
	month = nov,
	year = {1986},
	pages = {85--99},
}

@article{rhie_numerical_1983,
	title = {Numerical study of the turbulent flow past an airfoil with trailing edge separation},
	volume = {21},
	issn = {0001-1452},
	url = {https://arc.aiaa.org/doi/10.2514/3.8284},
	doi = {10.2514/3.8284},
	number = {11},
	urldate = {2021-04-20},
	journal = {AIAA Journal},
	author = {Rhie, C. M. and Chow, W. L.},
	month = nov,
	year = {1983},
	note = {Publisher: American Institute of Aeronautics and Astronautics},
	pages = {1525--1532},
}

@article{chen_improved_2018,
	title = {An improved entropy-consistent {Euler} flux in low {Mach} number},
	volume = {27},
	issn = {1877-7503},
	url = {https://www.sciencedirect.com/science/article/pii/S1877750318301509},
	doi = {10.1016/j.jocs.2018.06.006},
	language = {en},
	urldate = {2021-04-29},
	journal = {Journal of Computational Science},
	author = {Chen, Shu-sheng and Yan, Chao and Lou, Shuai and Lin, Bo-xi},
	month = jul,
	year = {2018},
	pages = {271--283},
}

@article{matsuyama_performance_2014,
	title = {Performance of all-speed {AUSM}-family schemes for {DNS} of low {Mach} number turbulent channel flow},
	volume = {91},
	issn = {0045-7930},
	url = {https://www.sciencedirect.com/science/article/pii/S0045793013004908},
	doi = {10.1016/j.compfluid.2013.12.010},
	language = {en},
	urldate = {2021-04-29},
	journal = {Computers \& Fluids},
	author = {Matsuyama, Shingo},
	month = mar,
	year = {2014},
	pages = {130--143},
}

@article{dellacherie_analysis_2010,
	title = {Analysis of {Godunov} type schemes applied to the compressible {Euler} system at low {Mach} number},
	volume = {229},
	issn = {0021-9991},
	url = {https://www.sciencedirect.com/science/article/pii/S0021999109005361},
	doi = {10.1016/j.jcp.2009.09.044},
	language = {en},
	number = {4},
	urldate = {2021-04-29},
	journal = {Journal of Computational Physics},
	author = {Dellacherie, Stéphane},
	month = feb,
	year = {2010},
	pages = {978--1016},
}

@article{rieper_low-mach_2011,
	title = {A low-{Mach} number fix for {Roe}’s approximate {Riemann} solver},
	volume = {230},
	issn = {0021-9991},
	url = {https://www.sciencedirect.com/science/article/pii/S0021999111001689},
	doi = {10.1016/j.jcp.2011.03.025},
	language = {en},
	number = {13},
	urldate = {2021-04-29},
	journal = {Journal of Computational Physics},
	author = {Rieper, Felix},
	month = jun,
	year = {2011},
	pages = {5263--5287},
}

@article{thornber_numerical_2008,
	title = {Numerical dissipation of upwind schemes in low {Mach} flow},
	volume = {56},
	copyright = {Copyright © 2007 John Wiley \& Sons, Ltd.},
	issn = {1097-0363},
	url = {https://onlinelibrary.wiley.com/doi/abs/10.1002/fld.1628},
	doi = {https://doi.org/10.1002/fld.1628},
	language = {en},
	number = {8},
	urldate = {2021-04-29},
	journal = {International Journal for Numerical Methods in Fluids},
	author = {Thornber, B. J. R. and Drikakis, D.},
	year = {2008},
	note = {\_eprint: https://onlinelibrary.wiley.com/doi/pdf/10.1002/fld.1628},
	pages = {1535--1541},
}

@article{oswald_l2roe_2016,
	title = {{L2Roe}: a low dissipation version of {Roe}'s approximate {Riemann} solver for low {Mach} numbers},
	volume = {81},
	copyright = {Copyright © 2015 John Wiley \& Sons, Ltd.},
	issn = {1097-0363},
	shorttitle = {{L2Roe}},
	url = {https://onlinelibrary.wiley.com/doi/abs/10.1002/fld.4175},
	doi = {https://doi.org/10.1002/fld.4175},
	language = {en},
	number = {2},
	urldate = {2021-05-12},
	journal = {International Journal for Numerical Methods in Fluids},
	author = {Oßwald, K. and Siegmund, A. and Birken, P. and Hannemann, V. and Meister, A.},
	year = {2016},
	note = {\_eprint: https://onlinelibrary.wiley.com/doi/pdf/10.1002/fld.4175},
	pages = {71--86},
}

@incollection{brezzi_stabilization_1984,
	address = {Wiesbaden},
	series = {Notes on {Numerical} {Fluid} {Mechanics}},
	title = {On the {Stabilization} of {Finite} {Element} {Approximations} of the {Stokes} {Equations}},
	isbn = {978-3-663-14169-3},
	url = {https://doi.org/10.1007/978-3-663-14169-3_2},
	language = {en},
	urldate = {2021-05-14},
	booktitle = {Efficient {Solutions} of {Elliptic} {Systems}: {Proceedings} of a {GAMM}-{Seminar} {Kiel}, {January} 27 to 29, 1984},
	publisher = {Vieweg+Teubner Verlag},
	author = {Brezzi, F. and Pitkäranta, J.},
	editor = {Hackbusch, Wolfgang},
	year = {1984},
	doi = {10.1007/978-3-663-14169-3_2},
	pages = {11--19},
}

@article{eymard_stabilized_2006,
	title = {On a stabilized colocated {Finite} {Volume} scheme for the {Stokes} problem},
	volume = {40},
	copyright = {© EDP Sciences, SMAI, 2006},
	issn = {0764-583X, 1290-3841},
	url = {https://www.esaim-m2an.org/articles/m2an/abs/2006/03/m2an0537/m2an0537.html},
	doi = {10.1051/m2an:2006024},
	language = {en},
	number = {3},
	urldate = {2021-05-14},
	journal = {ESAIM: Mathematical Modelling and Numerical Analysis},
	author = {Eymard, Robert and Herbin, Raphaèle and Latché, Jean Claude},
	month = may,
	year = {2006},
	note = {Number: 3
Publisher: EDP Sciences},
	pages = {501--527},
}

@article{guillard_behavior_2009,
	title = {On the behavior of upwind schemes in the low {Mach} number limit. {IV}: {P0} approximation on triangular and tetrahedral cells},
	volume = {38},
	issn = {0045-7930},
	shorttitle = {On the behavior of upwind schemes in the low {Mach} number limit. {IV}},
	url = {https://www.sciencedirect.com/science/article/pii/S0045793009000838},
	doi = {10.1016/j.compfluid.2009.06.003},
	language = {en},
	number = {10},
	urldate = {2021-05-14},
	journal = {Computers \& Fluids},
	author = {Guillard, Hervé},
	month = dec,
	year = {2009},
	pages = {1969--1972},
}

@inproceedings{dellacherie_checkerboard_2009,
	title = {Checkerboard {Modes} and {Wave} {Equation}},
	booktitle = {Proceedings of {ALGORITMY} 2009},
	author = {Dellacherie, Stéphane},
	year = {2009},
}

@incollection{venkateswaran_efficiency_2000,
	title = {Efficiency and accuracy issues in contemporary {CFD} algorithms},
	url = {https://arc.aiaa.org/doi/abs/10.2514/6.2000-2251},
	urldate = {2021-05-17},
	booktitle = {Fluids 2000 {Conference} and {Exhibit}},
	publisher = {American Institute of Aeronautics and Astronautics},
	author = {Venkateswaran, S. and Merkle, C.},
	year = {2000},
	doi = {10.2514/6.2000-2251},
	note = {\_eprint: https://arc.aiaa.org/doi/pdf/10.2514/6.2000-2251},
}

@article{roe_approximate_1981,
	title = {Approximate {Riemann} solvers, parameter vectors, and difference schemes},
	volume = {43},
	issn = {0021-9991},
	url = {https://www.sciencedirect.com/science/article/pii/0021999181901285},
	doi = {10.1016/0021-9991(81)90128-5},
	language = {en},
	number = {2},
	urldate = {2021-05-17},
	journal = {Journal of Computational Physics},
	author = {Roe, P. L},
	month = oct,
	year = {1981},
	pages = {357--372},
}

@article{thornber_improved_2008,
	title = {An improved reconstruction method for compressible flows with low {Mach} number features},
	volume = {227},
	issn = {0021-9991},
	url = {https://www.sciencedirect.com/science/article/pii/S0021999108000429},
	doi = {10.1016/j.jcp.2008.01.036},
	language = {en},
	number = {10},
	urldate = {2021-05-18},
	journal = {Journal of Computational Physics},
	author = {Thornber, B. and Mosedale, A. and Drikakis, D. and Youngs, D. and Williams, R. J. R.},
	month = may,
	year = {2008},
	pages = {4873--4894},
}

@article{miczek_new_2015,
	title = {New numerical solver for flows at various {Mach} numbers},
	volume = {576},
	copyright = {© ESO, 2015},
	issn = {0004-6361, 1432-0746},
	url = {https://www.aanda.org/articles/aa/abs/2015/04/aa25059-14/aa25059-14.html},
	doi = {10.1051/0004-6361/201425059},
	language = {en},
	urldate = {2021-06-17},
	journal = {Astronomy \& Astrophysics},
	author = {Miczek, F. and Röpke, F. K. and Edelmann, P. V. F.},
	month = apr,
	year = {2015},
	note = {Publisher: EDP Sciences},
	pages = {A50},
}

@incollection{godfrey_preconditioning_1993,
	title = {Preconditioning for the {Navier}-{Stokes} equations with finite-rate chemistry},
	url = {https://arc.aiaa.org/doi/abs/10.2514/6.1993-535},
	urldate = {2021-06-21},
	booktitle = {31st {Aerospace} {Sciences} {Meeting}},
	publisher = {American Institute of Aeronautics and Astronautics},
	author = {Godfrey, Andrew and Walters, Robert and Van Leer, Bram},
	year = {1993},
	doi = {10.2514/6.1993-535},
	note = {\_eprint: https://arc.aiaa.org/doi/pdf/10.2514/6.1993-535},
}

@article{volpe_performance_1993,
	title = {Performance of compressible flow codes at low {Mach} numbers},
	volume = {31},
	issn = {0001-1452},
	url = {https://arc.aiaa.org/doi/10.2514/3.11317},
	doi = {10.2514/3.11317},
	number = {1},
	urldate = {2021-06-22},
	journal = {AIAA Journal},
	author = {Volpe, G.},
	month = jan,
	year = {1993},
	note = {Publisher: American Institute of Aeronautics and Astronautics},
	pages = {49--56},
}

@article{yu_fast_2021,
	title = {A fast transient solver for low-{Mach} number aerodynamics and aeroacoustics},
	volume = {214},
	issn = {0045-7930},
	url = {https://www.sciencedirect.com/science/article/pii/S0045793020303182},
	doi = {10.1016/j.compfluid.2020.104748},
	language = {en},
	urldate = {2021-06-24},
	journal = {Computers \& Fluids},
	author = {Yu, L. and Diasinos, S. and Thornber, B.},
	month = jan,
	year = {2021},
	pages = {104748},
}

@article{wall_semi-implicit_2002,
	title = {A {Semi}-implicit {Method} for {Resolution} of {Acoustic} {Waves} in {Low} {Mach} {Number} {Flows}},
	volume = {181},
	issn = {0021-9991},
	url = {https://www.sciencedirect.com/science/article/pii/S002199910297141X},
	doi = {10.1006/jcph.2002.7141},
	language = {en},
	number = {2},
	urldate = {2021-06-25},
	journal = {Journal of Computational Physics},
	author = {Wall, Clifton and Pierce, Charles D. and Moin, Parviz},
	month = sep,
	year = {2002},
	pages = {545--563},
}

@incollection{merkle_use_1998,
	title = {The use of asymptotic expansions to enhance computational methods},
	url = {https://arc.aiaa.org/doi/abs/10.2514/6.1998-2485},
	urldate = {2021-06-25},
	booktitle = {2nd {AIAA}, {Theoretical} {Fluid} {Mechanics} {Meeting}},
	publisher = {American Institute of Aeronautics and Astronautics},
	author = {Merkle, Charles and Venkateswaran, Sankaran},
	year = {1998},
	doi = {10.2514/6.1998-2485},
	note = {\_eprint: https://arc.aiaa.org/doi/pdf/10.2514/6.1998-2485},
}

@article{dellacherie_construction_2016,
	title = {Construction of modified {Godunov}-type schemes accurate at any {Mach} number for the compressible {Euler} system},
	volume = {26},
	issn = {0218-2025},
	url = {https://www.worldscientific.com/doi/abs/10.1142/S0218202516500603},
	doi = {10.1142/S0218202516500603},
	number = {13},
	urldate = {2021-07-09},
	journal = {Mathematical Models and Methods in Applied Sciences},
	author = {Dellacherie, S. and Jung, J. and Omnes, P. and Raviart, P.-A.},
	month = dec,
	year = {2016},
	note = {Publisher: World Scientific Publishing Co.},
	pages = {2525--2615},
}

@article{thornber_entropy_2008,
	title = {On entropy generation and dissipation of kinetic energy in high-resolution shock-capturing schemes},
	volume = {227},
	issn = {0021-9991},
	url = {https://www.sciencedirect.com/science/article/pii/S0021999108000405},
	doi = {10.1016/j.jcp.2008.01.035},
	language = {en},
	number = {10},
	urldate = {2021-08-05},
	journal = {Journal of Computational Physics},
	author = {Thornber, B. and Drikakis, D. and Williams, R. J. R. and Youngs, D.},
	month = may,
	year = {2008},
	pages = {4853--4872},
}

@article{fillion_flica-ovap_2011,
	series = {13th {International} {Topical} {Meeting} on {Nuclear} {Reactor} {Thermal} {Hydraulics} ({NURETH}-13)},
	title = {{FLICA}-{OVAP}: {A} new platform for core thermal–hydraulic studies},
	volume = {241},
	issn = {0029-5493},
	shorttitle = {{FLICA}-{OVAP}},
	url = {https://www.sciencedirect.com/science/article/pii/S0029549311004006},
	doi = {10.1016/j.nucengdes.2011.04.048},
	language = {en},
	number = {11},
	urldate = {2021-08-05},
	journal = {Nuclear Engineering and Design},
	author = {Fillion, Philippe and Chanoine, Augustin and Dellacherie, Stéphane and Kumbaro, Anela},
	month = nov,
	year = {2011},
	pages = {4348--4358},
}

@incollection{liu_upwind_1989,
	series = {Aerospace {Sciences} {Meetings}},
	title = {Upwind algorithms for general thermo-chemical nonequilibrium flows},
	url = {https://arc.aiaa.org/doi/10.2514/6.1989-201},
	urldate = {2021-09-07},
	booktitle = {27th {Aerospace} {Sciences} {Meeting}},
	publisher = {American Institute of Aeronautics and Astronautics},
	author = {Liu, Yen and Vinokur, Marcel},
	month = jan,
	year = {1989},
	doi = {10.2514/6.1989-201},
}

@inproceedings{rieper_influence_2008,
	title = {Influence of cell geometry on the behaviour of the first-order {Roe} scheme in the low {Mach} number regime},
	booktitle = {Proceedings of the {Fifth} {Conference} on {Finite} {Volumes} for {Complex} {Applications}},
	publisher = {Wiley},
	author = {Rieper, Felix},
	year = {2008},
	pages = {625--632},
}

@article{rieper_influence_2009,
	title = {The influence of cell geometry on the accuracy of upwind schemes in the low mach number regime},
	volume = {228},
	issn = {0021-9991},
	url = {https://www.sciencedirect.com/science/article/pii/S0021999109000096},
	doi = {10.1016/j.jcp.2009.01.002},
	language = {en},
	number = {8},
	urldate = {2021-09-21},
	journal = {Journal of Computational Physics},
	author = {Rieper, Felix and Bader, Georg},
	month = may,
	year = {2009},
	pages = {2918--2933},
}

@article{dellacherie_influence_2010,
	title = {The influence of cell geometry on the {Godunov} scheme applied to the linear wave equation},
	volume = {229},
	issn = {0021-9991},
	url = {https://www.sciencedirect.com/science/article/pii/S0021999110001221},
	doi = {10.1016/j.jcp.2010.03.012},
	language = {en},
	number = {14},
	urldate = {2021-09-21},
	journal = {Journal of Computational Physics},
	author = {Dellacherie, Stéphane and Omnes, Pascal and Rieper, Felix},
	month = jul,
	year = {2010},
	pages = {5315--5338},
}

@article{mary_large_2002,
	title = {Large {Eddy} {Simulation} of {Flow} {Around} an {Airfoil} {Near} {Stall}},
	volume = {40},
	issn = {0001-1452},
	url = {https://arc.aiaa.org/doi/10.2514/2.1763},
	doi = {10.2514/2.1763},
	number = {6},
	urldate = {2021-09-17},
	journal = {AIAA Journal},
	author = {Mary, Ivan and Sagaut, Pierre},
	month = jun,
	year = {2002},
	note = {Publisher: American Institute of Aeronautics and Astronautics},
	pages = {1139--1145},
}

@article{li_momentum_2010,
	title = {The momentum interpolation method based on the time-marching algorithm for {All}-{Speed} flows},
	volume = {229},
	issn = {0021-9991},
	url = {https://www.sciencedirect.com/science/article/pii/S0021999110003621},
	doi = {10.1016/j.jcp.2010.06.039},
	language = {en},
	number = {20},
	urldate = {2021-09-17},
	journal = {Journal of Computational Physics},
	author = {Li, Xue-song and Gu, Chun-wei},
	month = oct,
	year = {2010},
	pages = {7806--7818},
}

@article{li_development_2009,
	title = {Development of {Roe}-type scheme for all-speed flows based on preconditioning method},
	volume = {38},
	issn = {0045-7930},
	url = {https://www.sciencedirect.com/science/article/pii/S0045793008001928},
	doi = {10.1016/j.compfluid.2008.08.002},
	language = {en},
	number = {4},
	urldate = {2021-09-17},
	journal = {Computers \& Fluids},
	author = {Li, Xue-song and Gu, Chun-wei and Xu, Jian-zhong},
	month = apr,
	year = {2009},
	pages = {810--817},
}

@article{li_all-speed_2008,
	title = {An {All}-{Speed} {Roe}-type scheme and its asymptotic analysis of low {Mach} number behaviour},
	volume = {227},
	issn = {0021-9991},
	url = {https://www.sciencedirect.com/science/article/pii/S0021999108000697},
	doi = {10.1016/j.jcp.2008.01.037},
	language = {en},
	number = {10},
	urldate = {2021-09-17},
	journal = {Journal of Computational Physics},
	author = {Li, Xue-song and Gu, Chun-wei},
	month = may,
	year = {2008},
	pages = {5144--5159},
}

@article{liska_comparison_2003,
	title = {Comparison of {Several} {Difference} {Schemes} on {1D} and {2D} {Test} {Problems} for the {Euler} {Equations}},
	volume = {25},
	issn = {1064-8275},
	url = {https://epubs.siam.org/doi/10.1137/S1064827502402120},
	doi = {10.1137/S1064827502402120},
	number = {3},
	urldate = {2021-09-15},
	journal = {SIAM Journal on Scientific Computing},
	author = {Liska, Richard and Wendroff, Burton},
	month = jan,
	year = {2003},
	note = {Publisher: Society for Industrial and Applied Mathematics},
	pages = {995--1017},
}

@article{gresho_theory_1990,
	title = {On the theory of semi-implicit projection methods for viscous incompressible flow and its implementation via a finite element method that also introduces a nearly consistent mass matrix. {Part} 2: {Implementation}},
	volume = {11},
	issn = {1097-0363},
	shorttitle = {On the theory of semi-implicit projection methods for viscous incompressible flow and its implementation via a finite element method that also introduces a nearly consistent mass matrix. {Part} 2},
	url = {https://onlinelibrary.wiley.com/doi/abs/10.1002/fld.1650110510},
	doi = {10.1002/fld.1650110510},
	language = {en},
	number = {5},
	urldate = {2021-09-15},
	journal = {International Journal for Numerical Methods in Fluids},
	author = {Gresho, Philip M. and Chan, Stevens T.},
	year = {1990},
	note = {\_eprint: https://onlinelibrary.wiley.com/doi/pdf/10.1002/fld.1650110510},
	pages = {621--659},
}

@article{klein_semi-implicit_1995,
	title = {Semi-implicit extension of a godunov-type scheme based on low mach number asymptotics {I}: {One}-dimensional flow},
	volume = {121},
	issn = {0021-9991},
	shorttitle = {Semi-implicit extension of a godunov-type scheme based on low mach number asymptotics {I}},
	url = {https://www.sciencedirect.com/science/article/pii/S0021999195900349},
	doi = {10.1016/S0021-9991(95)90034-9},
	language = {en},
	number = {2},
	urldate = {2021-09-27},
	journal = {Journal of Computational Physics},
	author = {Klein, R.},
	month = oct,
	year = {1995},
	pages = {213--237},
}

@article{roller_calculation_2005,
	title = {Calculation of low {Mach} number acoustics: a comparison of {MPV}, {EIF} and linearized {Euler} equations},
	volume = {39},
	copyright = {© EDP Sciences, SMAI, 2005},
	issn = {0764-583X, 1290-3841},
	shorttitle = {Calculation of low {Mach} number acoustics},
	url = {https://www.esaim-m2an.org/articles/m2an/abs/2005/03/m2an07EDP/m2an07EDP.html},
	doi = {10.1051/m2an:2005016},
	language = {en},
	number = {3},
	urldate = {2021-09-27},
	journal = {ESAIM: Mathematical Modelling and Numerical Analysis},
	author = {Roller, Sabine and Schwartzkopff, Thomas and Fortenbach, Roland and Dumbser, Michael and Munz, Claus-Dieter},
	month = may,
	year = {2005},
	note = {Number: 3
Publisher: EDP Sciences},
	pages = {561--576},
}

@article{rehm_equations_1978,
	title = {The equations of motion for thermally driven, buoyant flows},
	volume = {83},
	number = {3},
	journal = {Journal of Research of the National Bureau of Standards},
	author = {Rehm, Ronald G. and Baum, Howard R.},
	year = {1978},
	pages = {297--308},
}

@article{majda_derivation_1985,
	title = {The {Derivation} and {Numerical} {Solution} of the {Equations} for {Zero} {Mach} {Number} {Combustion}},
	volume = {42},
	issn = {0010-2202},
	url = {https://doi.org/10.1080/00102208508960376},
	doi = {10.1080/00102208508960376},
	number = {3-4},
	urldate = {2021-09-27},
	journal = {Combustion Science and Technology},
	author = {MAJDA, ANDFREW and SETHIAN, JAMES},
	month = jan,
	year = {1985},
	note = {Publisher: Taylor \& Francis
\_eprint: https://doi.org/10.1080/00102208508960376},
	pages = {185--205},
}

@article{klainerman_compressible_1982,
	title = {Compressible and incompressible fluids},
	volume = {35},
	issn = {1097-0312},
	url = {https://onlinelibrary.wiley.com/doi/abs/10.1002/cpa.3160350503},
	doi = {10.1002/cpa.3160350503},
	language = {en},
	number = {5},
	urldate = {2021-09-27},
	journal = {Communications on Pure and Applied Mathematics},
	author = {Klainerman, Sergiu and Majda, Andrew},
	year = {1982},
	note = {\_eprint: https://onlinelibrary.wiley.com/doi/pdf/10.1002/cpa.3160350503},
	pages = {629--651},
}

@article{schochet_fast_1994,
	title = {Fast {Singular} {Limits} of {Hyperbolic} {PDEs}},
	volume = {114},
	issn = {0022-0396},
	url = {https://www.sciencedirect.com/science/article/pii/S0022039684711570},
	doi = {10.1006/jdeq.1994.1157},
	language = {en},
	number = {2},
	urldate = {2021-09-27},
	journal = {Journal of Differential Equations},
	author = {Schochet, S.},
	month = dec,
	year = {1994},
	pages = {476--512},
}

@article{edwards_low-diffusion_1998,
	title = {Low-{Diffusion} {Flux}-{Splitting} {Methods} for {Flows} at {All} {Speeds}},
	volume = {36},
	issn = {0001-1452},
	url = {https://arc.aiaa.org/doi/10.2514/2.587},
	doi = {10.2514/2.587},
	number = {9},
	urldate = {2021-09-27},
	journal = {AIAA Journal},
	author = {Edwards, Jack R. and Liou, Meng-Sing},
	month = sep,
	year = {1998},
	note = {Publisher: American Institute of Aeronautics and Astronautics},
	pages = {1610--1617},
}

@incollection{liou_numerical_1999,
	title = {Numerical speed of sound and its application to schemes for all speeds},
	url = {https://arc.aiaa.org/doi/abs/10.2514/6.1999-3268},
	urldate = {2021-09-27},
	booktitle = {14th {Computational} {Fluid} {Dynamics} {Conference}},
	publisher = {American Institute of Aeronautics and Astronautics},
	author = {Liou, Meng-Sing and Edwards, Jack},
	year = {1999},
	doi = {10.2514/6.1999-3268},
	note = {\_eprint: https://arc.aiaa.org/doi/pdf/10.2514/6.1999-3268},
}

@article{toro_flux_2012,
	title = {Flux splitting schemes for the {Euler} equations},
	volume = {70},
	issn = {0045-7930},
	url = {https://www.sciencedirect.com/science/article/pii/S0045793012003398},
	doi = {10.1016/j.compfluid.2012.08.023},
	language = {en},
	urldate = {2021-09-27},
	journal = {Computers \& Fluids},
	author = {Toro, E. F. and Vázquez-Cendón, M. E.},
	month = nov,
	year = {2012},
	pages = {1--12},
}

@article{zha_numerical_1993,
	title = {Numerical solutions of {Euler} equations by using a new flux vector splitting scheme},
	volume = {17},
	issn = {1097-0363},
	url = {https://onlinelibrary.wiley.com/doi/abs/10.1002/fld.1650170203},
	doi = {10.1002/fld.1650170203},
	language = {en},
	number = {2},
	urldate = {2021-09-27},
	journal = {International Journal for Numerical Methods in Fluids},
	author = {Zha, G-C. and Bilgen, E.},
	year = {1993},
	note = {\_eprint: https://onlinelibrary.wiley.com/doi/pdf/10.1002/fld.1650170203},
	pages = {115--144},
}

@book{chassignet_buoyancy-driven_2012,
	address = {Cambridge},
	title = {Buoyancy-{Driven} {Flows}},
	isbn = {978-1-107-00887-8},
	url = {https://www.cambridge.org/core/books/buoyancydriven-flows/E9A065CBA7E7B854DE793E200F71B367},
	urldate = {2021-10-01},
	publisher = {Cambridge University Press},
	editor = {Chassignet, Eric P. and Cenedese, Claudia and Verron, Jacques},
	year = {2012},
	doi = {10.1017/CBO9780511920196},
}

@article{sohel_murshed_critical_2017,
	title = {A critical review of traditional and emerging techniques and fluids for electronics cooling},
	volume = {78},
	issn = {1364-0321},
	url = {https://www.sciencedirect.com/science/article/pii/S1364032117305944},
	doi = {10.1016/j.rser.2017.04.112},
	language = {en},
	urldate = {2021-10-01},
	journal = {Renewable and Sustainable Energy Reviews},
	author = {Sohel Murshed, S. M. and Nieto de Castro, C. A.},
	month = oct,
	year = {2017},
	pages = {821--833},
}

@article{coletti_turbulent_2014,
	title = {Turbulent flow in rib-roughened channel under the effect of {Coriolis} and rotational buoyancy forces},
	volume = {26},
	issn = {1070-6631},
	url = {https://aip.scitation.org/doi/full/10.1063/1.4871019},
	doi = {10.1063/1.4871019},
	number = {4},
	urldate = {2021-10-01},
	journal = {Physics of Fluids},
	author = {Coletti, Filippo and Jacono, David Lo and Cresci, Irene and Arts, Tony},
	month = apr,
	year = {2014},
	note = {Publisher: American Institute of Physics},
	pages = {045111},
}

@article{inoue_sound_2002,
	title = {Sound generation by a two-dimensional circular cylinder in a uniform flow},
	volume = {471},
	issn = {1469-7645, 0022-1120},
	url = {https://www.cambridge.org/core/journals/journal-of-fluid-mechanics/article/sound-generation-by-a-twodimensional-circular-cylinder-in-a-uniform-flow/58647F205BA2C0DE5BDA09CD301FC6FF},
	doi = {10.1017/S0022112002002124},
	language = {en},
	urldate = {2021-10-01},
	journal = {Journal of Fluid Mechanics},
	author = {Inoue, Osamu and Hatakeyama, Nozomu},
	month = nov,
	year = {2002},
	note = {Publisher: Cambridge University Press},
	pages = {285--314},
}

@article{daviller_generalized_2019,
	title = {A generalized non-reflecting inlet boundary condition for steady and forced compressible flows with injection of vortical and acoustic waves},
	volume = {190},
	issn = {0045-7930},
	url = {https://www.sciencedirect.com/science/article/pii/S0045793019301951},
	doi = {10.1016/j.compfluid.2019.06.027},
	language = {en},
	urldate = {2021-10-01},
	journal = {Computers \& Fluids},
	author = {Daviller, G. and Oztarlik, G. and Poinsot, T.},
	month = aug,
	year = {2019},
	pages = {503--513},
}

@article{silva_assessment_2013,
	series = {{LES} of turbulence aeroacoustics and combustion},
	title = {Assessment of combustion noise in a premixed swirled combustor via {Large}-{Eddy} {Simulation}},
	volume = {78},
	issn = {0045-7930},
	url = {https://www.sciencedirect.com/science/article/pii/S0045793010002628},
	doi = {10.1016/j.compfluid.2010.09.034},
	language = {en},
	urldate = {2021-10-01},
	journal = {Computers \& Fluids},
	author = {Silva, Camilo F. and Leyko, Matthieu and Nicoud, Franck and Moreau, Stéphane},
	month = apr,
	year = {2013},
	pages = {1--9},
}

@book{blazek_computational_2015,
	edition = {3rd},
	title = {Computational {Fluid} {Dynamics}: {Principles} and {Applications}},
	isbn = {978-0-08-099995-1},
	publisher = {Butterworth-Heinemann Ltd},
	author = {Blazek, Jiri},
	year = {2015},
}

@incollection{turkel_preconditioning_1994,
	title = {Preconditioning and the limit of the compressible to the incompressible flow equations for finite difference schemes},
	booktitle = {Frontiers of {Computational} {Fluid} {Dynamics} 1994. {Editors} {D}. {A}. {Caughey} \& {M}. {M}. {Hafez}},
	publisher = {John Wiley \& Sons Ltd},
	author = {Turkel, E. and Fiterman, A. and Van Leer, Bram},
	year = {1994},
	pages = {215--234},
}

@article{subbareddy_fully_2009,
	title = {A fully discrete, kinetic energy consistent finite-volume scheme for compressible flows},
	volume = {228},
	issn = {0021-9991},
	url = {https://www.sciencedirect.com/science/article/pii/S0021999108005573},
	doi = {10.1016/j.jcp.2008.10.026},
	language = {en},
	number = {5},
	urldate = {2022-11-12},
	journal = {Journal of Computational Physics},
	author = {Subbareddy, Pramod K. and Candler, Graham V.},
	month = mar,
	year = {2009},
	pages = {1347--1364},
}

\appendix
\appendixpage
\addappheadtotoc

\section{Transonic schemes at low Mach number}\label{app:transonic_scaling}
In this appendix we show, in a very approximate manner, why many numerical schemes designed for transonic flow approach the acoustic diffusion scaling at low Mach number.
Diffusion schemes for the Euler equations can be classified by how many waves they resolve.
For example, scalar Lax-Friedrichs/Rusonov is a 1-wave scheme, HLL is 2-wave scheme, and  Roe, HLLC and AUSM are all 3-wave schemes.
In the characteristic variables, the diffusion Jacobian for the the acoustic waves of a 1- 2- or 3-waves scheme will be close to:
\begin{equation} \label{eq:3wave_diffusion}
    \begin{matrix}
    \begin{pmatrix}
    |u+a| &   0   \\
      0   & |u\pm a| \\
    \end{pmatrix}
    \end{matrix}
\end{equation}
Where the bottom right entry has an addition for 1-wave schemes and a subtraction for 2- and 3- wave schemes.
At low Mach number this Jacobian can be approximated by:
\begin{equation} \label{eq:3wave_diffusion_lowmach}
    |u|
    \begin{matrix}
    \begin{pmatrix}
    M^{\-1}+1 &  0        \\
     0        & M^{\-1}\pm1 \\
    \end{pmatrix}
    \end{matrix}
    \approx
    |u|
    \begin{matrix}
    \begin{pmatrix}
    M^{\-1} &  0      \\
     0      & M^{\-1} \\
    \end{pmatrix}
    \end{matrix}
\end{equation}
The diffusion on the acoustic waves $du \pm dp/\rho a$ is then:
\begin{equation} \label{eq:3wave_acoustic_component}
    (|u|M^{\-1})\uuline{\mathcal{I}}
\end{equation}
which matches the acoustic scaling.
Because the matrix (\ref{eq:3wave_acoustic_component}) is a scalar multiple of the identity matrix, any variable basis for the acoustic system (for example $dp$ and $du$) will also have this diffusion Jacobian at low Mach number.

\section{Non-dimensionalised artificial diffusion for finite-volume schemes}\label{app:fv-diffusive-fluxes}
The non-dimensional diffusion with convective scaling and $\epsilon_{\alpha}=1$ is:
\begin{subequations} \label{eq:nondim_discrete_cdiff}
\begin{alignat}{5}
    \label{eq:density_discrete_cdiff}
    f^d_{\rho} & = \frac{1}{2} \facesum S_{il}\bigg(
        |U|\Delta_{il}\rho &&+
        \rho\bigg(\frac{M^{\-2}\nu_{11}}{\rho|v|}\Delta_{il}p &&+ \frac{U_{il}}{|U_{il}|}\nu_{12}\Delta_{il}U \bigg) \bigg) && && \\
    \label{eq:momentum_discrete_cdiff}
    f^d_{\underline{\rho u}} & = \frac{1}{2} \facesum S_{il}\bigg(
        |U|\Delta_{il}\underline{\rho u} &&+
        \underline{\rho u}\bigg(\frac{M^{\-2}\nu_{11}}{\rho|v|}\Delta_{il}p &&+ \frac{U_{il}}{|U_{il}|}\nu_{12}\Delta_{il}U \bigg) &&+
        \underline{n}_{il}\big(M^{\-2}\frac{U_{il}}{|U_{il}|}\nu_{21}\Delta_{il}p &&+ \rho|v|\nu_{22}\Delta_{il}U\big) \bigg) \\
    \label{eq:energy_discrete_cdiff}
    f^d_{\rho E} & = \frac{1}{2} \facesum S_{il}\bigg(
        |U|\Delta_{il}\rho E &&+
        \rho H\bigg(\frac{M^{\-2}\nu_{11}}{\rho|v|}\Delta_{il}p &&+ \frac{U_{il}}{|U_{il}|}\nu_{12}\Delta_{il}U \bigg) &&+
        UM^{2}\big(M^{\-2}\frac{U_{il}}{|U_{il}|}\nu_{21}\Delta_{il}p &&+ \rho|v|\nu_{22}\Delta_{il}U\big)  \bigg)
\end{alignat}
\end{subequations}

The non-dimensional diffusion with acoustic scaling and $\epsilon_{\alpha}=1$ is:
\begin{subequations} \label{eq:nondim_discrete_adiff}
\begin{alignat}{5}
    \label{eq:density_discrete_adiff}
    f^d_{\rho} & = \frac{1}{2} \facesum S_{il}\bigg(
        |U|\Delta_{il}\rho &&+
        \rho\bigg(\frac{M^{\-1}\nu_{11}}{\rho|v|}\Delta_{il}p &&+ \frac{U_{il}}{|U_{il}|}\nu_{12}\Delta_{il}U \bigg) \bigg) && && \\
    \label{eq:momentum_discrete_adiff}
    f^d_{\underline{\rho u}} & = \frac{1}{2} \facesum S_{il}\bigg(
        |U|\Delta_{il}\underline{\rho u} &&+
        \underline{\rho u}\bigg(\frac{M^{\-1}\nu_{11}}{\rho|v|}\Delta_{il}p &&+ \frac{U_{il}}{|U_{il}|}\nu_{12}\Delta_{il}U \bigg) &&+
        \underline{n}_{il}\big(M^{\-2}\frac{U_{il}}{|U_{il}|}\nu_{21}\Delta_{il}p &&+ M^{\-1}\rho|v|\nu_{22}\Delta_{il}U\big) \bigg) \\
    \label{eq:energy_discrete_adiff}
    f^d_{\rho E} & = \frac{1}{2} \facesum S_{il}\bigg(
        |U|\Delta_{il}\rho E &&+
        \rho H\bigg(\frac{M^{\-1}\nu_{11}}{\rho|v|}\Delta_{il}p &&+ \frac{U_{il}}{|U_{il}|}\nu_{12}\Delta_{il}U \bigg) &&+
        UM^{2}\big(M^{\-2}\frac{U_{il}}{|U_{il}|}\nu_{21}\Delta_{il}p &&+ M^{\-1}\rho|v|\nu_{22}\Delta_{il}U\big)  \bigg)
\end{alignat}
\end{subequations}

The non-dimensional diffusion with blended scaling and $\epsilon_{\alpha}=1$ is:
\begin{subequations} \label{eq:nondim_discrete_bdiff}
\begin{alignat}{5}
    \label{eq:density_discrete_bdiff}
    f^d_{\rho} & = \frac{1}{2} \facesum S_{il}\bigg(
        |U|\Delta_{il}\rho &&+
        \rho\bigg(\frac{M^{\-1}\nu_{11}}{\rho|v|}\Delta_{il}p &&+ \frac{U_{il}}{|U_{il}|}\nu_{12}\Delta_{il}U \bigg) \bigg) && && \\
    \label{eq:momentum_discrete_bdiff}
    f^d_{\underline{\rho u}} & = \frac{1}{2} \facesum S_{il}\bigg(
        |U|\Delta_{il}\underline{\rho u} &&+
        \underline{\rho u}\bigg(\frac{M^{\-1}\nu_{11}}{\rho|v|}\Delta_{il}p &&+ \frac{U_{il}}{|U_{il}|}\nu_{12}\Delta_{il}U \bigg) &&+
        \underline{n}_{il}\big(M^{\-2}\frac{U_{il}}{|U_{il}|}\nu_{21}\Delta_{il}p &&+ \rho|v|\nu_{22}\Delta_{il}U\big) \bigg) \\
    \label{eq:energy_discrete_bdiff}
    f^d_{\rho E} & = \frac{1}{2} \facesum S_{il}\bigg(
        |U|\Delta_{il}\rho E &&+
        \rho H\bigg(\frac{M^{\-1}\nu_{11}}{\rho|v|}\Delta_{il}p &&+ \frac{U_{il}}{|U_{il}|}\nu_{12}\Delta_{il}U \bigg) &&+
        UM^{2}\big(M^{\-2}\frac{U_{il}}{|U_{il}|}\nu_{21}\Delta_{il}p &&+ \rho|v|\nu_{22}\Delta_{il}U\big)  \bigg)
\end{alignat}
\end{subequations}

\end{document}